\documentclass[preprint,aps,prc,tightenlines,showpacs,superscriptaddress,nofootinbib]{revtex4-1}
\usepackage{amsmath,amssymb,amsfonts}
\usepackage[dvips]{graphicx,color}
\usepackage{dcolumn}
\usepackage{epsfig}
\usepackage{bm}
\usepackage{bbm}
\usepackage{multirow}
\usepackage{ulem}
\usepackage{color}

\newcommand{\ket}[1]{|{#1}\rangle}
\newcommand{\bra}[1]{\langle{#1}|}
\newcommand{\inp}[2]{\langle{#1}|{#2}\rangle}

\newcommand{\unit}{\mathbbm{1}}

\def\gtap{\ \raise.3ex\hbox{$>$\kern-.75em\lower1ex\hbox{$\sim$}}\ }
\def\ltap{\ \raise.3ex\hbox{$<$\kern-.75em\lower1ex\hbox{$\sim$}}\ }

\usepackage[dvipdfmx]{hyperref}

\begin{document}

\title{Coupled-channel analysis of $D^+\to K^- \pi^+\pi^+$ decay
}
\author{Satoshi X. Nakamura}
\email{nakamura@kern.phys.sci.osaka-u.ac.jp}
\affiliation{Department of Physics, Osaka University, Toyonaka, Osaka 560-0043, Japan}

\begin{abstract}

We perform a coupled-channel analysis of 
pseudodata for the $D^+\to K^-\pi^+\pi^+$ Dalitz plot.
The pseudodata are generated from the isobar model of the E791
Collaboration, and are reasonably realistic.
We demonstrate that it is feasible to analyze the
high-quality data within a coupled-channel framework that
describes
the final state interaction of $D^+\to K^-\pi^+\pi^+$ 
as multiple rescatterings of three pseudoscalar mesons through
two-pseudoscalar-meson interactions in accordance with the two-body and
three-body unitarity.
The two-pseudoscalar-meson interactions are designed to reproduce
empirical $\pi\pi$ and $\pi \bar K$ scattering amplitudes.
Furthermore, 
we also include mechanisms that are beyond simple iterations
of the two-body interactions, i.e., a three-meson force, 
derived from the hidden local symmetry model.
A picture of hadronic dynamics in $D^+\to K^-\pi^+\pi^+$ 
described by our coupled-channel model
is found to be quite different from those of the previous isobar-type
analyses.
For example, 
we find that 
the $D^+\to {K}^-\pi^+\pi^+$ decay width can get almost triplicated
when the rescattering mechanisms are turned on.
Among the rescattering mechanisms, 
those associated with the
$\rho(770)\bar{K}^0$ channel,
which contribute to $D^+\to {K}^-\pi^+\pi^+$ 
only through a channel coupling,
give a large contribution,
and significantly improve the quality of the fits. 
The $K^-\pi^+$ $s$-wave amplitude from our analysis is reasonably
consistent with those extracted from the E791 
model independent partial-wave
analysis;
the hadronic rescattering and the coupling to the 
$\rho(770)\bar{K}^0$ channel play a major role here.
We also find that 
the dressed $D^+$ decay vertices have
phases, induced by the strong rescatterings,
that strongly depend on the momenta of the final
pseudoscalar mesons.
Although the conventional isobar-type analyses have assumed the phases to be
constant, this common assumption is not supported by our more microscopic
viewpoint.

\end{abstract}

% 13.25.-k	Hadronic decays of mesons
% 14.40.Rt	Exotic mesons
% 11.80.Jy	Many-body scattering and Faddeev equation
% 11.80.Et	Partial-wave analysis
% 11.80.Gw	Multichannel scattering
% 11.80.La	Multiple scattering

% 14.40.-n	Mesons (for leptonic decays of mesons, see 13.20.-v; for hadronic decays of mesons, see 13.25.-k)
% 14.40.Be	Light mesons (S=C=B=0)
% 14.40.Df	Strange mesons (|S|>0, C=B=0)
% 14.40.Lb	Charmed mesons (|C|>0, B=0)
% 13.25.Ft	Decays of charmed mesons
% 13.75.Lb	Meson-meson interactions
\maketitle

\section{Introduction}

With the advent of charm and B-factories, a large amount of data for
charmed-meson decays have been accumulated in the last decades. 
Among a number of physical interests,
one appealing aspect of studying these charmed-meson decays is that
we can gain information about
interactions between light mesons and resonances.
This was particularly highlighted by the E791 Collaboration's report on
their identification of the $\sigma$ meson in 
the Dalitz plot of the $D^+\to\pi^-\pi^+\pi^+$ decay~\cite{e791-sigma}.
A similar analysis was also made for the
$D^+\to K^-\pi^+\pi^+$ decay to identify the $\kappa$
resonance~\cite{e791-prl}.
These findings triggered further analyses of 
the $D^+\to K^-\pi^+\pi^+$ Dalitz plot data,
paying special attention to 
the $K^-\pi^+$ $s$-wave amplitude, as follows.
Oller~\cite{oller} analyzed the E791 data~\cite{e791-prl} using the 
$\bar{K}\pi$ $I$=1/2 ($I$: total isospin) $s$-wave amplitude
based on the chiral unitary approach~\cite{oller2}, instead of
Breit-Wigner functions
for the $K^-\pi^+$ $s$-wave 
used in the E791 analysis,
and obtained a reasonable fit.
The $I$=3/2 $s$-wave was not considered in his analysis.
The E791 Collaboration reanalyzed their data without assuming a
particular functional form for the $K^-\pi^+$ $s$-wave amplitude.
Rather, they determined it bin by bin, which they call
model independent partial-wave analysis (MIPWA)~\cite{e791}.
An interesting finding in the MIPWA was that 
the obtained $K^-\pi^+$ $s$-wave amplitude has the phase that depends on
the $K^-\pi^+$ energy in a manner
significantly different from what is expected from the Watson theorem
combined with the LASS empirical amplitude~\cite{lass},
assuming the $I$=1/2 dominance.
Edera et al. suggested that this difference can be understood as a
substantial mixture of the $\bar{K}\pi$ $I$=1/2
and 3/2 $s$-wave amplitudes~\cite{pik_I=3/2_model}.
This idea was implemented in the Dalitz plot analysis done by the
FOCUS Collaboration~\cite{focus}. 
They parametrized the $K^-\pi^+$ $s$-wave amplitude in terms of 
the $K$-matrix of $I$=1/2 and 3/2 that had been fitted to the
LASS amplitudes~\cite{lass,pik_I=3/2}.
They found that the
$D^+\to K^-\pi^+\pi^+$ Dalitz plot can be well fitted with 
the $K^-\pi^+$ $s$-wave amplitude in which 
the $I$=1/2 and 3/2 components interfere with each other in a rather
destructive manner.
The FOCUS Collaboration also has done a MIPWA in a subsequent
work~\cite{focus2009} to find results similar to those of the E791 MIPWA. 
The quasi-MIPWA has also been done by the CLEO
Collaboration~\cite{cleo}.
Their new finding was that $I$=2 $\pi\pi$ nonresonant amplitude can
give a non-negligible contribution.
Meanwhile, an analogous decay, $D^+\to K_S^0\pi^0\pi^+$,
has been analyzed by the BESIII Collaboration~\cite{bes3}.
An interesting finding was that the
$\rho(770)\bar K$ channel gives by far the dominant contribution.
This implies that,
although the $\rho(770)\bar K$ contribution is not directly observed in 
the $D^+\to K^-\pi^+\pi^+$ Dalitz plot, 
it can play a substantial role
in the hadronic final state interactions (FSI) of
the $D^+\to K^-\pi^+\pi^+$ decay,
considering that the two $D^+$ decays share the same
hadronic dynamics to a large extent.

Many of the previous analyses of the $D^+\to K^-\pi^+\pi^+$ Dalitz plot
have been done with the so-called isobar
model in which a $D^+$-meson decays into 
an excited state $R$ ($\bar\kappa, \bar K^*, \bar K^*_2$, etc.)
and a pseudoscalar meson.
The $R$ subsequently decays into a pair of
light pseudoscalar mesons, while
the third pseudoscalar meson is treated as a spectator.
The propagation of $R$ is commonly parametrized by a Breit-Wigner
function.
The total decay amplitude is given by a coherent sum of these isobar
amplitudes supplemented by a flat interfering background.
On the other hand, as mentioned in the above paragraph,
some analysis groups
modified the conventional isobar model to
use the $K$-matrix or chiral unitary model for
the $K^-\pi^+$ $s$-wave amplitude so that
the consistency with the LASS data for
$K^-\pi^+\to K^-\pi^+$ and with the two-body
unitarity is maintained. 
Meanwhile, in (Q)MIPWA,
the $K^-\pi^+$ $s$-wave amplitude is solely determined by 
the $D^+\to K^-\pi^+\pi^+$ Dalitz plot data.
We will refer to these models, which do not explicitly consider
three-meson-rescattering required by the three-body unitarity,
as isobar-type models.
The basic assumptions common to all of these models are that
the spectator pseudoscalar meson interacts with the others very weakly,
and/or $D^+\to Rc$ ($c$: spectator meson) vertices 
with complex coupling constants can absorb such effects.

Although each analysis group has obtained a reasonable fit to its own
data,
their results are not necessarily in conformity with theoretical
expectations.
For example, the E791 Collaboration reported that the phase of 
the $I$=1/2 $\bar K\pi$ $p$-wave amplitude used in their MIPWA is,
according to the Watson theorem, not consistent
with that of the LASS analysis in the elastic region~\cite{e791}.
This may be originated from either or both of two possible reasons below, each of
which signals a serious problem in the basic assumptions underlying in
the analysis model.
One possible reason is that
the $I$=1/2 $\bar K\pi$ $p$-wave amplitude in the E791 analysis
is given by a coherent sum of
Breit-Wigner functions for $\bar K^*(892)$ and $\bar K^*(1680)$,
and it is not consistent with the LASS data to the
required precision.
Or the amplitude does not satisfy the two-body unitarity not only formally
but also quantitatively.
Another possibility is that the (neglected) rescattering of the
spectator meson with the other mesons actually plays a substantial role to
generate an energy-dependent
phase so that the Watson theorem does not hold in reality, and the
analysis model tried to fit it. 
Not only the E791 analysis but also all the previous analyses mentioned
above would share the same problem,
and this seems to indicate a need for going beyond the conventional
isobar-type analysis; a unitary coupled-channel approach.
In order to extract from data a right physics, 
e.g., $K^-\pi^+$ $s$-wave amplitude from 
$D^+\to K^-\pi^+\pi^+$ Dalitz plot data,
one needs to use a theoretically sound model so that unnecessary
model artifacts do not come into play.

Recently, we have developed a unitary coupled-channel framework 
for describing a heavy-meson decay into three light mesons~\cite{3pi-1};
both the two-body and three-body unitarity are maintained.
In the reference, we studied the extent to which 
the isobar-type
description of heavy-meson (or excited meson) decays
is valid by analyzing simple pseudodata.
We found a significant effect of the channel couplings and
multiple rescattering on the Dalitz plot distributions.
This study has been extended to an analysis of pseudodata for
excited meson photoproductions~\cite{3pi-2}.
In the present work, we apply the formalism of Ref.~\cite{3pi-1} with some
modifications to a realistic case.
Thus, we will perform a coupled-channel analysis of 
the $D^+\to K^-\pi^+\pi^+$ Dalitz plot pseudodata
generated from the E791 isobar model~\cite{e791}.
To the best of our knowledge,
this is the first coupled-channel Dalitz plot analysis of a $D$-meson decay into three
pseudoscalar mesons.
We will demonstrate that a quantitative coupled-channel partial-wave
analysis of 
the $D^+\to K^-\pi^+\pi^+$ Dalitz plot is feasible.
Then we will examine the hadronic dynamics in the FSI of the
$D^+\to K^-\pi^+\pi^+$ decay within the coupled-channel model.
We will study how 
the partial-wave amplitudes and their fit fractions
are different between an isobar-type model and a model that
includes the three-body scattering.
We also examine
contributions from the rescattering and channel couplings
to the Dalitz plot distribution.
Through these investigations, we address
the validity of the above-mentioned
basic assumptions of the isobar-type model 
from this more microscopic viewpoint. 

So far, the three-body FSI for the $D^+\to K^-\pi^+\pi^+$ decay
has been explored by Magalh\~aes et al.~\cite{usp}
(see also Guimar\~aes et al.~\cite{LF}).
They were concerned with whether 
the difference in the phase between the
$K^-\pi^+$ $s$-wave amplitudes from MIPWA (E791~\cite{e791} and FOCUS~\cite{focus2009})
and the $\bar K\pi$ $I$=1/2 $s$-wave amplitude from the LASS analysis
can be understood as a result of the FSI.
They calculated the 
$K^-\pi^+$ $s$-wave amplitude in the $D^+\to K^-\pi^+\pi^+$ decay
with only the $\bar K\pi$ $I$=1/2 $s$-wave scattering amplitude
of the chiral unitary model
that had been fitted to the LASS amplitude in the elastic region;
other partial waves as well as inelasticities were not taken into
account. 
An interesting finding in their work was that 
the $K^-\pi^+$ $s$-wave amplitude originated from
the weak vector current and the subsequent $s$-wave rescattering 
is in a fairly good
agreement with the phases from the MIPWA in the elastic region;
the qualitative feature of 
the modulus of the MIPWA amplitude is also described.
This finding was further confirmed in their subsequent works~\cite{usp2}
where $\rho(770)$ contribution was also partly taken into account.
Although their results are interesting and suggestive, they should be looked with a
caution for the reasons below.
First, their calculated amplitudes are still qualitatively in agreement
with those from the MIPWA in the elastic region, and more refinements
are needed particularly in the inelastic region. 
There will be a delicate interplay between 
new mechanisms to be included for the refinements and the
already existing mechanisms, and thus it is not clear if their current
findings persist after the refinements. 
Second, the authors of Refs.~\cite{usp,usp2} treated the 
$K^-\pi^+$ $s$-wave amplitudes from the MIPWA as data, and 
did not analyze the Dalitz plot directly.
However, we note that 
the MIPWA $s$-wave amplitude
was obtained under the assumption that all the other partial-wave amplitudes are
basically not changed by the FSI.
In case the $s$-wave amplitude is significantly modified by the FSI, 
the other partial-wave amplitudes are also probably modified,
and the MIPWA $s$-wave amplitude may no longer be compatible with them.
In order to fully examine the FSI effects on each of the partial-wave
amplitudes for the $D^+\to K^-\pi^+\pi^+$ decay,
it is first necessary to analyze the Dalitz plot
with a model that takes account of all relevant partial waves and the FSI,
thereby extracting the partial-wave amplitudes,
as we will do in this work.
Then, the FSI effects on the extracted amplitudes
can be studied.

In our analysis, we use two-pseudoscalar-meson interactions that generate
unitary amplitudes for
$\bar K\pi$ and  $\pi\pi$ scatterings, and consider resonances 
($\bar\kappa, \bar K^*, \rho$, etc.)
as poles in the amplitudes.
The two-body interactions are fitted to the LASS and CERN-Munich~\cite{grayer,hyams,na48}
data.
With the two-body interactions,
we solve the Faddeev equation for a three-pseudoscalar-meson scattering to
obtain an amplitude that respects the three-body unitarity and thus channel couplings.
This amplitude is used to describe the FSI of the $D^+$ decay.
The pseudodata are fitted by adjusting the strengths and phases of (bare)
$D^+\to Rc$ vertices, while the two-pseudoscalar-meson interactions are fixed as those
obtained by fitting the two-body scattering data.
In this way, 
we will examine the extent to which we can fit the $D^+$-decay
pseudodata, keeping the consistency with the
two-body scattering data for {\it all} partial waves considered in our model.
This is in contrast with the previous
$D^+\to K^-\pi^+\pi^+$ analyses where some of the resonance parameters
were also adjusted along with the $D^+\to Rc$ vertices.
We consider the $I$=1/2 $\bar K\pi$ $s$-, $p$-, and $d$-waves as
commonly included in the previous analyses.
We also explicitly include the $I$=3/2 $\bar K\pi$ $s$-wave
($I$=2 $\pi\pi$ $s$-wave)
as has been done so in the FOCUS~\cite{focus} (CLEO~\cite{cleo}) analysis.
We also consider the $I$=1 $\pi\pi$ $p$-wave where the
$\rho(770)$ plays a major role.
This partial wave has not been considered in the previous Dalitz plot analyses because 
it does not directly decay into the $K^-\pi^+\pi^+$ final state.
However, this partial wave can still contribute to
$D^+\to K^-\pi^+\pi^+$ through the channel couplings.
Considering the BESIII analysis mentioned above, the $\rho(770)\bar K$
channel is expected to play a substantial role also here, and we will
see that this is indeed the case, at least within our analysis.

Now we discuss the last piece of our model.
The FSI of the $D^+\to K^-\pi^+\pi^+$ decay is a 
three-pseudoscalar-meson scattering.
Generally in a three-body scattering, there can exist a mechanism that
cannot be expressed by a combination of two-body mechanisms, i.e., a
three-body force. 
Based on the hidden local symmetry (HLS) model~\cite{hls}, in which vector and
pseudoscalar mesons are implemented together in a chiral Lagrangian, 
we can actually derive a ``three-meson force'' essentially in a
parameter-free fashion, up to form factors we include.
Thus we consider some of the HLS-based three-meson force acting on
important channels, and examine how they play a role in the 
$D^+\to K^-\pi^+\pi^+$ decay.
If two-pseudoscalar-meson interactions are
well determined by precise two-body scattering data,
the $D\to \bar K \pi\pi$ decays 
and also other decay modes such as $D\to \pi\pi\pi$
could serve as a ground to study the
three-meson force.

The organization of this paper is as follows.
In Sec.~\ref{sec:formalism},
we discuss our coupled-channel model and present formulas to calculate 
the $D^+\to K^-\pi^+\pi^+$ decay amplitude.
Then in Sec.~\ref{sec:results},
we present numerical results from our analyses 
of the two-pseudoscalar-meson scattering data and of
the $D^+\to K^-\pi^+\pi^+$ decay Dalitz plot pseudodata.
Finally, we give a summary and future prospects in Sec.~\ref{sec:conclusion}.
A derivation and the resulting expressions for the three-meson force,
and model parameters are presented in the Appendixes.

\section{Formulation}
\label{sec:formalism}

We have already developed a formalism to describe a heavy-meson decay
into three light mesons in Ref.~\cite{3pi-1}.
In the present work, we basically use the same formalism with some modifications.
Thus, here we just present expressions that are needed in the following
discussions, specifying the modifications we make for this work.
For derivations of most of the expressions, see Sec.~II of Ref.~\cite{3pi-1}.
Our formalism can be regarded as a three-dimensional reduction of a
fully relativistic formulation.
Because we deal with scatterings of light particles, i.e. pions, one
may question the validity of this approximation. 
Although a legitimate concern,
we made an argument on this 
along with improvements needed in the future
in Sec.~V of Ref.~\cite{3pi-1}.
In our formalism, we first construct a two-light-meson ($\pi$, $K$) interaction model that
is subsequently applied to three-light-meson scattering.
The following presentations are also given in this order.

\subsection{Two-light-meson scattering model}

We describe two-light-meson scatterings with
a unitary coupled-channel model.
For example, we consider $\pi\pi$ and $K\bar K$ channels for a $\pi\pi$
scattering, and $\pi\bar K$ and $\eta' \bar K$ channels for
$I$=1/2 $s$-wave $\pi\bar K$ scattering. 
We will specify channels for each partial wave later.
We model the two-light-meson interactions with
resonance($R$)-excitation mechanisms or contact interactions or both.
We choose this parametrization of the two-meson interactions so that it can
be easily applied to a three-meson scattering.
A difference from Ref.~\cite{3pi-1} is that 
we include the contact interactions here but we did not in Ref.~\cite{3pi-1}.
This is because we include partial-wave amplitudes that do not have resonances
in analyzing the $D^+\to K^-\pi^+\pi^+$ decay Dalitz plot.
Besides,
we use a parametrization for $R\to ab$ ($a,b$: pseudoscalar mesons) vertex functions that are
different from those used in the previous work and,
with this parametrization,
we need to include contact interactions in addition to
$R$-excitation mechanisms
in order to obtain reasonable fits
to some empirical partial-wave amplitudes.

\begin{figure}[t]
 \includegraphics[width=12cm]{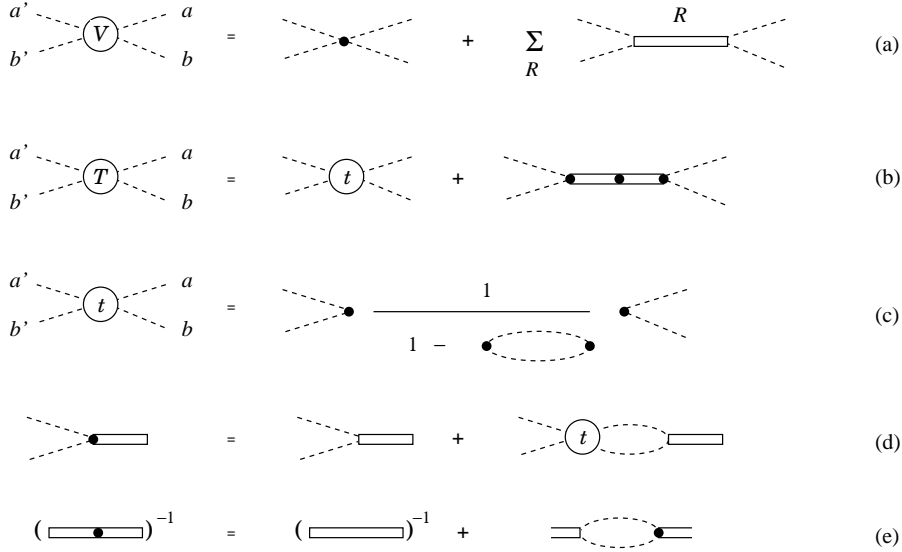}
\caption{\label{fig:two-body}
Diagrammatic representation of two-light-meson partial-wave
scattering, $ab\to a'b'$.
(a) two-meson interaction potentials; 
(b) two-meson partial-wave scattering amplitude;
(c) two-meson partial-wave scattering amplitude from separable contact interactions;
(d) dressed $R\to ab$ decay vertex; 
(e) the inverse of dressed $R$ Green function.
}
\end{figure}
We consider a partial-wave $ab\to a'b'$ scattering 
(see Fig.~\ref{fig:two-body} for a diagrammatic representation)
with total energy $E$,
total angular momentum $L$, total isospin $I$.
We will also denote the incoming and outgoing momenta by $q$ an $q'$,
respectively; $q=|\bm q|$ throughout this paper, except for the Appendix.
When $q$ is on-shell, it is related to $E$ by
$E=E_a(q)+E_b(q)$ and 
$E_a(q)=\sqrt{{q}^2+m_a^2}$; 
$m_a$ being the mass of $a$.
First let us consider a $ab\to a'b'$ scattering
that is not accompanied by a
resonance excitation.
In this case, we use a separable two-light-meson interaction potential
[first term of rhs of Fig.~\ref{fig:two-body}(a)]
as follows:
\begin{eqnarray}
v^{LI}_{a'b',ab} (q',q) = 
w^{LI}_{a'b'}(q') h^{LI}_{a'b',ab}\; w^{LI}_{ab}(q) ,
\label{eq:cont-ptl}
\end{eqnarray}
where $h^{LI}_{a'b',ab}$ is a coupling constant, and $w^{LI}_{ab}(q)$ is a
vertex function. We use the following parametrization for $w^{LI}_{ab}(q)$:
\begin{eqnarray}
w^{LI}_{ab}(q) = {1\over \sqrt{E_a(q)E_b(q)}} 
\left[{1\over 1 +(q/b^{LI}_{ab})^2}\right]^{2+L/2} 
\left(q\over m_\pi\right)^L \ ,
\label{eq:vf-cont}
\end{eqnarray}
where $b^{LI}_{ab}$ is a cutoff parameter. 
Meanwhile,
we allow an exception for $L$=0, $I$=2 $\pi\pi$ scattering for which we
use a different parametrization for the vertex function:
\begin{eqnarray}
w^{LI}_{\pi\pi}(q) = {1+h'(q/m_\pi)^2\over E_\pi(q)} 
\left[{1\over 1 +(q/b^{LI}_{\pi\pi})^2}\right]^{3}
\ ,
\label{eq:vf-cont2}
\end{eqnarray}
where an additional coupling constant $h'$ has been introduced to obtain a
reasonable fit to data.
For a later convenience, let us define $\tilde w^{LI}_{ab}$ by
\begin{equation}
w^{LI}_{ab}(q) =
\left\{
\begin{array}{ll}
\displaystyle
\frac{1}{\sqrt{2}}\tilde w^{LI}_{ab}(q)
& \text{(if $a$ and $b$ are identical particles)}, \\
\displaystyle
\tilde w^{LI}_{ab}(q)
& \text{(otherwise)} .
\end{array}
\right.
\label{eq:tilde-w}
\end{equation}
With the above interaction potential, 
the partial-wave amplitude is given as follows
[see Fig.~\ref{fig:two-body}(c) for a diagrammatic representation]:
\begin{eqnarray}
t^{LI}_{a'b',ab} (q',q; E) = \sum_{a''b''}
w^{LI}_{a'b'}(q') \tau^{LI}_{a'b',a''b''}(E)\;
h^{LI}_{a''b'',ab}\; w^{LI}_{ab}(q) ,
\label{eq:pw-2body-cont}
\end{eqnarray}
with
\begin{eqnarray}
\left[(\tau^{LI}(E))^{-1}\right]_{a'b',ab} &=& 
\delta_{a'b',ab} -  \sigma^{LI}_{a'b',ab}(E)
\ , \\
\sigma^{LI}_{a'b',ab}(E) &=& 
\int_0^\infty dq\; q^2  
{ h^{LI}_{a'b',ab}\left[w^{LI}_{ab}(q)\right]^2
\over E-E_a(q)-E_b(q)+i\epsilon}
\ .
\label{eq:pw-cont-self}
\end{eqnarray}
The amplitude of Eq.~(\ref{eq:pw-2body-cont}) can contain
resonance pole(s), in general.

Now we extend the model to include explicit degrees of freedom for
excitations of spin-$L$ isospin-$I$ resonances that contribute
to the $ab\to a'b'$ scattering. 
In this case, a two-light-meson interaction potential 
includes bare $R$-excitation mechanisms in addition to the contact
potential $v^{LI}_{a'b',ab}$ defined in Eq.~(\ref{eq:cont-ptl})
as follows [see also Fig.~\ref{fig:two-body}(a)]:
\begin{eqnarray}
V^{LI}_{a'b',ab} (q',q; E) =
v^{LI}_{a'b',ab} (q',q) 
+ \sum_{R}
\bar{f}^{LI}_{a'b',R}(q') {1\over E-m_R} \bar{f}^{LI}_{R,ab}(q) 
\ ,
\label{eq:pw-2body-v}
\end{eqnarray}
where $m_R$ is the bare mass of $R$;
$\bar{f}^{LI}_{ab,R}(q)$ denotes 
a bare $R\to ab$ vertex function and
$\bar f^{LI}_{R,ab}(q)= \bar f^{{LI}\ast}_{ab,R}(q)$.
Following Ref.~\cite{3pi-1}, we introduce $\tilde f^{LI}_{ab,R}(q)$ that
is related to $\bar f^{LI}_{ab,R}(q)$ as follows:
\begin{equation}
\bar{f}^{LI}_{ab,R}(q) = 
\left\{
\begin{array}{ll}
\displaystyle
\frac{1}{\sqrt{2}}\tilde{f}^{LI}_{ab,R}(q) & \text{(if $a$ and $b$ are identical particles)}, \\
\displaystyle
\tilde{f}^{LI}_{ab,R}(q) & \text{(otherwise)} .
\end{array}
\right.
\label{eq:tilde-f}
\end{equation}
Then we employ the following parametrization for the bare vertex function:
\begin{eqnarray}
\tilde{f}^{LI}_{ab,R}(q)=
g_{ab,R}
{m_\pi\over \sqrt{m_R E_a(q) E_b(q)}}
\left[\frac{1}{1+(q/c_{ab,R})^2}\right]^{1+(L/2)}
\left( \frac{q}{m_\pi} \right)^L  ,
\label{eq:pipi-vertex}
\end{eqnarray}
where $g_{ab,R}$ and $c_{ab,R}$ are coupling constant and cutoff,
respectively.
This parametrization is different from that used in Eq.~(35) of Ref.~\cite{3pi-1},
and has a proper kinematical factor.
With the potential in Eq.~(\ref{eq:pw-2body-v}),
the resulting scattering amplitude is given by
[Fig.~\ref{fig:two-body}(b)]:
\begin{eqnarray}
T^{LI}_{a'b',ab} (q',q; E) =
t^{LI}_{a'b',ab} (q',q; E) 
+ \sum_{R',R}
\bar{\bar f}^{LI}_{a'b',R'}(q';E) \tau^{LI}_{R',R}(E) \bar{\bar f}^{LI}_{R,ab}(q;E) 
\ ,
\label{eq:pw-2body-t}
\end{eqnarray}
where the first term is the scattering amplitude from the contact
interactions only, as has been defined in Eq.~(\ref{eq:pw-2body-cont}).
The symbol $\bar{\bar f}_{ab,R}$ denotes
the dressed vertex that describes the bare $R\to ab$ decay
followed by $ab$ rescattering through the contact interactions.
Expressions for $\bar{\bar f}_{ab,R}$ and $\bar{\bar f}_{R,ab}$ are
[Fig.~\ref{fig:two-body}(d)]:
\begin{eqnarray}
\bar{\bar f}^{LI}_{ab,R}(q;E) &=& 
{\bar f}^{LI}_{ab,R}(q) + \sum_{a'b'}
\int_0^\infty dq' q'^2\; 
{t^{LI}_{ab,a'b'} (q,q'; E)\; {\bar f}^{LI}_{a'b',R}(q')
\over E-E_{a'}(q')-E_{b'}(q')+ i\epsilon}  \ ,\\
\bar{\bar f}^{LI}_{R,ab}(q;E) &=& 
{\bar f}^{LI}_{R,ab}(q) + \sum_{a'b'}
\int_0^\infty dq' q'^2\; 
{{\bar f}^{LI}_{R,a'b'}(q')\; t^{LI}_{a'b',ab} (q',q; E)
\over E-E_{a'}(q')-E_{b'}(q')+ i\epsilon} 
\ .
\label{eq:dressed-vertex}
\end{eqnarray}
In Eq.~(\ref{eq:pw-2body-t}), 
the dressed Green function for $R$, $\tau^{LI}_{R',R}(E)$, 
has been introduced, and is given by
[Fig.~\ref{fig:two-body}(e)]:
\begin{eqnarray}
[(\tau^{LI}(E))^{-1}]_{R'R} = (E-m_R)\delta_{R',R}
 -\bar\Sigma^{LI}_{R',R}(E) ,
\end{eqnarray}
where $\bar\Sigma^{LI}_{R',R}(E)$ is
the self-energy of $R$, and is 
defined by
\begin{eqnarray}
\bar\Sigma^{LI}_{R',R}(E) = \sum_{ab}
\int_0^\infty q^2 dq 
\frac{\bar{f}^{LI}_{R',ab}(q) \bar{\bar f}^{LI}_{ab,R}(q;E)} {E - E_a(q) - E_b(q) + i\epsilon} .
\label{eq:pipi-sigma}
\end{eqnarray}
In case $ab\to a'b'$ interaction is given by only resonant mechanisms 
[no first term in Eq.~(\ref{eq:pw-2body-v})],
which is the case in Ref.~\cite{3pi-1},
the corresponding scattering amplitude is obtained from Eq.~(\ref{eq:pw-2body-t})
by dropping the first term, and replacing the dressed
vertex ($\bar{\bar f}$) with the bare one ($\bar{ f}$) in Eqs.~(\ref{eq:pw-2body-t})
and (\ref{eq:pipi-sigma}).

The partial-wave amplitude, $T^{LI}_{a'b',ab}$ in Eq.~(\ref{eq:pw-2body-t}),
 is related to the $S$-matrix by
\begin{eqnarray}
s^{LI}_{ab,ab} (E)
 = \eta_{LI}\, e^{2i\delta_{LI}} = 
1 - 2\pi i\rho_{ab}\, T^{LI}_{ab,ab} (q_o,q_o; E) \ ,
\label{eq:s-matrix}
\end{eqnarray}
where $q_o$ is the on-shell momentum that satisfies $E=E_a(q_o)+E_b(q_o)$,
and $\rho_{ab}= q_o E_a(q_o)E_b(q_o)/E$ is the phase-space factor.
The phase shift and inelasticity are denoted by $\delta_{LI}$ and
$\eta_{LI}$, respectively.

The parameters contained in the two-light-meson potentials such as
$m_R$, $g_{ab,R}$, $c_{ab,R}$, $h^{LI}_{a'b',ab}$, and
$b^{LI}_{ab}$ are determined by fitting experimental data.
A particular choice of the potential, such as the number of
$R$ and contact interactions included,
will be specified later for each 
partial wave and for each of $\pi\pi$ and $\bar K\pi$ interactions.

\subsection{Three-light-meson scattering model}
\label{sec:three-meson}

\begin{figure}[t]
\begin{minipage}[t]{105mm}
 \includegraphics[width=10.5cm]{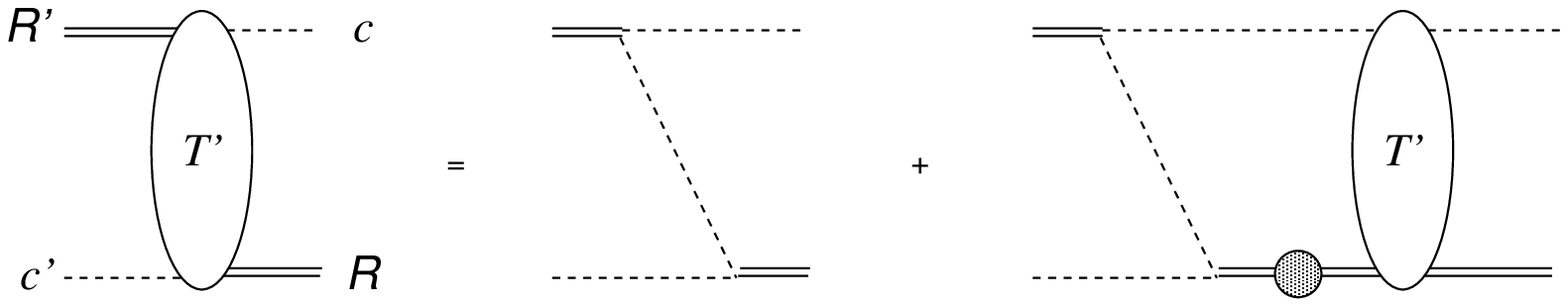}
\caption{\label{fig:3body-t}
Diagrammatic 
representation of scattering equation 
of Eq.~(\ref{eq:pw-tcr})
for $c{\cal R}\to c'{\cal R}'$ scattering.
The gray blob represents the dressed ${\cal R}$ Green function in 
Eqs.~(\ref{eq:green-Rc})-(\ref{eq:green-Rc4}).
}
\end{minipage}
\hspace{5mm}
\begin{minipage}[t]{50mm}
 \includegraphics[width=25mm]{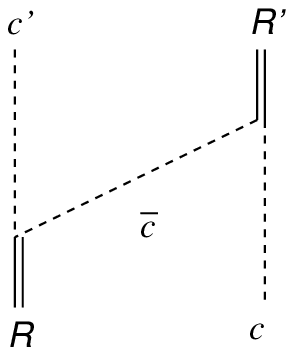}
\caption{\label{fig:cr-int}
Z-diagram
for $c{\cal R}\to c'{\cal R}'$ process
via $\bar c$-exchange.
}
\end{minipage}
\end{figure}

We now consider a case where three light-mesons are scattering. 
First, let us assume that the three mesons interact with each other only
through the two-meson interactions discussed in the previous subsection.
Because our two-meson interaction potential is given in a separable
form,
we can cast the Faddeev equation into a two-body like
scattering equation (the so-called Alt-Grassberger-Sandhas (AGS) equation~\cite{AGS})
for a $c{\cal R}\to c'{\cal R}'$ scattering.
Here, ${\cal R}$ stands for either $R$ or $r_{ab}$, and $r_{ab}$ is
a spurious ``state'' that is supposed to live within a contact
interaction in a very short time, and decays (going to the left in
equations) into the two light-mesons, $ab$.
This degree of freedom is introduced merely for
extending the AGS-type $cR\to c'R'$ scattering equation
used in Ref.~\cite{3pi-1} 
to include the contact two-meson interactions.
Thus, the scattering equation for
a partial-wave amplitude,
$T'^{JPT}_{(c'{\cal R}')_{l'},(c{\cal R})_{l}} (p',p; E)$,
is given as (see Fig.~\ref{fig:3body-t} for a diagrammatic representation)
\begin{eqnarray}
T'^{JPT}_{(c'{\cal R}')_{l'},(c{\cal R})_{l}} (p',p; E)
&=&
Z^{\bar{c},JPT}_{(c'{\cal R}')_{l'},(c{\cal R})_{l}} (p',p; E)
\nonumber\\
&&
+ \sum_{c'',{\cal R}'',{\cal R}''',l''}
\int^\infty_0 q^2dq\, Z^{\bar{c}'',JPT}_{(c'{\cal R}')_{l'},(c''{\cal R}''')_{l''}}(p',q;E) 
\nonumber\\
&&
\qquad \times G_{c''{\cal R}''',c''{\cal R}''}(q,E) 
T'^{JPT}_{(c''{\cal R}'')_{l''},(c{\cal R})_{l}}(q,p;E) \ ,
\label{eq:pw-tcr}
\end{eqnarray}
where $JPT$ are the total angular momentum, parity, and the total
isospin of the $c{\cal R}$ system and they are conserved in the scattering.
The $c{\cal R}$ state with the relative orbital angular momentum $l$ is denoted by $(c{\cal R})_{l}$; 
the allowed range for $l$ is determined by $JP$ and the spin-parity of ${\cal R}$.
The magnitude of the incoming (outgoing) relative momentum of the $c{\cal R}$
($c'{\cal R}'$) state is denoted by $p$ ($p'$).
The driving term of the scattering is a partial-wave form of the
so-called Z-diagram, $Z^{\bar{c},JPT}_{(c'{\cal R}')_{l'},(c{\cal R})_{l}} (p',p; E)$.
The Z-diagram is a process in which 
${\cal R}\to c'\bar{c}$ decay is followed by $\bar{c}c\to {\cal R}'$-formation,
as illustrated in Fig.~\ref{fig:cr-int}.
For a more explicit definition as well as the partial-wave expansion of the
Z-diagram, we refer the readers to Appendix~C of Ref.~\cite{3pi-1};
in particular, Eqs.~(C10)-(C12) of Ref.~\cite{3pi-1} give an explicit
expression for $Z^{\bar{c},JPT}_{(c'R')_{l'},(cR)_{l}} (p',p; E)$
(note ${\cal R}$=$R$ and ${\cal R}'$=$R'$ in Ref.~\cite{3pi-1}) in which
$\tilde{f}^{LI}_{c'\bar{c},R}$ and
$\tilde{f}^{LI}_{\bar{c}c,R'}$ 
defined in Eq.~(\ref{eq:pipi-vertex}) can be directly inserted.
When ${\cal R}=r_{c'\bar{c}}$ (${\cal R'}=r'_{a'b'}$), the corresponding expression for the Z-diagram can be
practically obtained by replacing
$\tilde{f}^{LI}_{c'\bar{c},R}$ ($\tilde{f}^{LI}_{\bar{c}c,R'}$) in
$Z^{\bar{c},JPT}_{(c'R')_{l'},(cR)_{l}} (p',p; E)$
with $\tilde w^{LI}_{c'\bar{c}}$ ($h^{LI}_{a'b',\bar{c}c} \tilde w^{LI}_{\bar{c}c}$)
defined in Eq.~(\ref{eq:tilde-w}).
The Z-diagrams are known to have the moon-shape singularity~\cite{msl07} that
prevents us from solving Eq.~(\ref{eq:pw-tcr}) with the standard
subtraction method. 
Here we employ the spline method (see Ref.~\cite{msl07} for detailed explanations) to
obtain numerical solutions from Eq.~(\ref{eq:pw-tcr}).

In Eq.~(\ref{eq:pw-tcr}), we have also used
the Green function,
$G_{c{\cal R}',c{\cal R}}(q,E)$,
for ${\cal R}$ and ${\cal R}'$ which can be coupled through
${\cal R}\to ab\to {\cal R}'$.
It is given by
\begin{eqnarray}
[ G^{-1}(q,E) ]_{cR',cR}
&=& 
[ E - E_{c}(q) - E_{R}(q) ] \delta_{R',R} 
- \Sigma^{LI}_{R',R}\left(q, E-E_{c}(q)\right)
\ ,
\label{eq:green-Rc}
\end{eqnarray}
for $({\cal R},{\cal R}') = (R,R')$, and 
\begin{eqnarray}
\left[ G^{-1}(q,E) \right]_{cr'_{a'b'},cR} &=&
- \sigma^{LI}_{r'_{a'b'},R}\left(q, E-E_{c}(q)\right) 
\qquad {\rm for\ }
({\cal R},{\cal R}') = (R,r'_{a'b'})
\ , \\
\left[G^{-1}(q,E)\right]_{cR',cr_{ab}} 
&=&
- \sigma^{LI}_{R',r_{ab}}\left(q, E-E_{c}(q)\right)
\qquad {\rm for\ }
({\cal R},{\cal R}') = (r_{ab},R')\ , \\
\left[G^{-1}(q,E)\right]_{cr'_{a'b'},cr_{ab}} &=&
\delta_{r_{ab},r'_{a'b'}} - \sigma^{LI}_{r'_{a'b'},r_{ab}}\left(q, E-E_{c}(q)\right)
\quad {\rm for\ }
({\cal R},{\cal R}') = (r_{ab},r'_{a'b'})\ ,
\label{eq:green-Rc4}
\end{eqnarray}
where we have introduced the self-energies
$\Sigma^{LI}_{R', R}\left(q, E\right)$, 
and also quantities 
$\sigma^{LI}_{{\cal R'},{\cal R}}\left(q, E\right)$, 
which is either dimensionless or dimension of square-root of the energy,
as defined by
\begin{eqnarray}
\Sigma^{LI}_{R',R}(p,E) &=& 
\sum_{ab} \sqrt{\frac{m_{R'}m_{R}}{E_{R'}(p)E_{R}(p)}}
\int_0^\infty q^2 dq
{M_{ab}(q)\over \sqrt{M^2_{ab}(q) + p^2}}
\frac{ \bar{f}^{LI}_{R', ab}(q) \bar{f}^{LI}_{ab,R}(q)}
{E - \sqrt{M^2_{ab}(q) + p^2} + i\epsilon} ,
\label{eq:RR-self}
%\nonumber
\\
\sigma^{LI}_{r'_{a'b'},R}(p,E) &=& 
\sum_{ab} \sqrt{\frac{m_{R}}{E_{R}(p)}}
\int_0^\infty q^2 dq
{M_{ab}(q)\over \sqrt{M^2_{ab}(q) + p^2}}
\frac{h^{LI}_{a'b',ab} w^{LI}_{ab}(q) \bar{f}^{LI}_{ab,R}(q)}
{E - \sqrt{M^2_{ab}(q) + p^2} + i\epsilon} ,
%\nonumber
\\
\sigma^{LI}_{R',r_{ab}}(p,E) &=& 
\sqrt{\frac{m_{R'}}{E_{R'}(p)}}
\int_0^\infty q^2 dq
{M_{ab}(q)\over \sqrt{M^2_{ab}(q) + p^2}}
\frac{ \bar{f}^{LI}_{R', ab}(q) w^{LI}_{ab}(q)}
{E - \sqrt{M^2_{ab}(q) + p^2} + i\epsilon} ,
%\nonumber
\\
\sigma^{LI}_{r'_{a'b'},r_{ab}}(p,E) &=& 
\int_0^\infty q^2 dq
{M_{ab}(q)\over \sqrt{M^2_{ab}(q) + p^2}}
\frac{h^{LI}_{a'b',ab} [w^{LI}_{ab}(q)]^2}
{E - \sqrt{M^2_{ab}(q) + p^2} + i\epsilon} ,
\label{eq:R-self}
\end{eqnarray}
with $M_{ab}(q) =  E_{a}(q)+E_{b}(q)$, and $\sum_{ab}$ runs over
all two-meson states from $R\to ab$ decays.
The kinematical factors in the expressions are from the Lorentz
transformation to boost the ${\cal R}$-at-rest frame to the $c{\cal R}$
center-of-mass frame.

So far, we have considered the three-meson scattering 
due to multiple iterations of the two-meson interactions.
Given the two-meson interactions, this is a necessary consequence of the
three-body unitarity.
In a three-meson system, however, there may be a room for a new
mechanism that is absent in a two-meson system to play a role.
We will refer to such mechanisms as a ``three-meson force'' hereafter.
\begin{figure}[t]
\begin{minipage}[t]{77mm}
 \includegraphics[width=7cm]{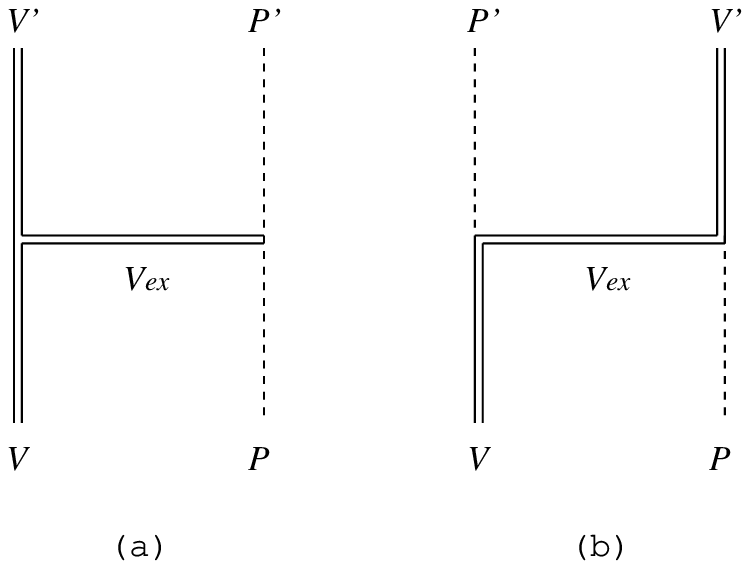}
\caption{\label{fig:vecpot} 
Vector ($V$) and pseudoscalar ($P$) mesons interaction potentials based on the hidden
 local symmetry model~\cite{hls}. Vector mesons ($V_{ex}$) are exchanged.}
%\end{figure}
\end{minipage}
\hspace{3mm}
\begin{minipage}[t]{77mm}
%\begin{figure}[t]
 \includegraphics[width=7cm]{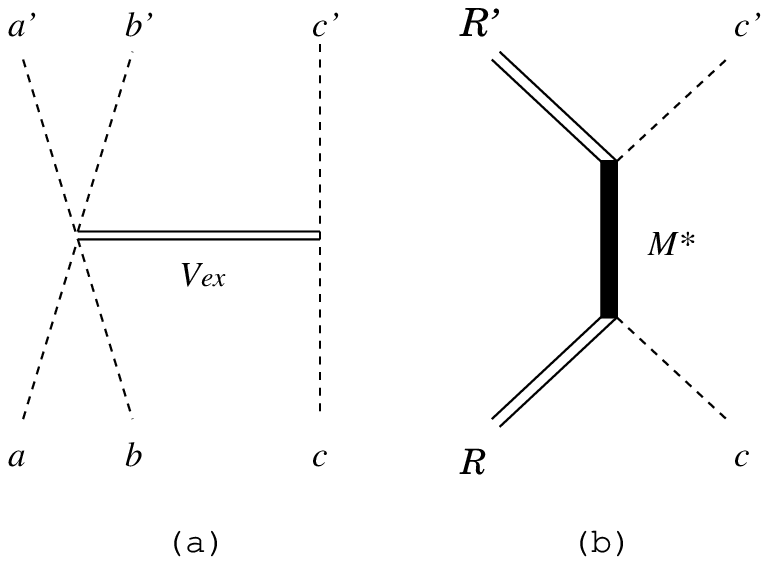}
\caption{\label{fig:3meson} 
Three-meson force {\it not} considered in this work.
(a) Vector-meson exchange between
a pseudoscalar-meson pair ($ab$) in $s$-wave and 
a pseudoscalar meson ($c$).
(b) $Rc$ interaction via meson-resonance ($M^*$) excitation.
}
\end{minipage}
\end{figure}
Diagrams shown in Fig.~\ref{fig:vecpot} can work as a
three-meson force.
These are interactions 
between a vector-meson and a pseudoscalar
meson via a vector-meson exchange; 
for our particular application to $D^+\to K^-\pi^+\pi^+$, 
they are  bare $\rho$-$\bar K$,  
and bare $\bar K^*$-$\pi$ interactions.
These mechanisms are based on the hidden local symmetry (HLS) model~\cite{hls}
in which vector and pseudoscalar mesons are implemented in a
Lagrangian that has a symmetry under nonlinear chiral transformations.
Expressions for the Lagrangian and the resulting interaction potentials
of Fig.~\ref{fig:vecpot} are presented in Appendix~\ref{app:lag}.
These mechanisms in Fig.~\ref{fig:vecpot} along with the Z-diagram in
Fig.~\ref{fig:cr-int} have been studied by 
Jansen et al.~\cite{pirho1,pirho2,pirho3}
to examine the $\pi$-$\rho$ correlation and its relevance to
a soft $\pi NN$ form factor in a $NN$ potential.
There are also other possible mechanisms that can work as a three-meson force. 
We show some diagrams in Fig.~\ref{fig:3meson} as examples.
The diagram in Fig.~\ref{fig:3meson}(a) describes an interaction between
a pseudoscalar-meson-pair ($ab$) in $s$-wave and another
pseudoscalar meson ($c$) via a vector-meson exchange;
this mechanism is also from the HLS Lagrangian.
Meanwhile, in the diagram of Fig.~\ref{fig:3meson}(b),
an $R$ interacts with a pseudoscalar meson to form a resonance ($M^*$),
which is followed by a decay into an $R'$ and a pseudoscalar meson.
This is a familiar mechanism and often assumed in partial-wave analyses for
meson spectroscopy. 

In this work, 
we consider the vector-pseudoscalar interactions shown in 
Fig.~\ref{fig:vecpot} in our analysis of
the $D^+\to K^-\pi^+\pi^+$ decay to study their relevance.
Thus, the scattering equation in Eq.~(\ref{eq:pw-tcr}) is modified by
adding the new mechanisms to the Z-diagrams:
\begin{eqnarray}
Z^{\bar{c},JPT}_{(c'R')_{l'},(cR)_{l}} (p',p; E)
\to
Z^{\bar{c},JPT}_{(c'R')_{l'},(cR)_{l}} (p',p; E)
+V^{JPT}_{(c'R')_{l'},(cR)_{l}} (p',p)
\quad {\rm in\ Eq.~(\ref{eq:pw-tcr})}
\ ,
\label{eq:addition}
\end{eqnarray}
where the added term,
$V^{JPT}_{(c'R')_{l'},(cR)_{l}} (p',p)$,
is in the partial-wave form for which we give explicit
expressions in Eqs.(\ref{eq:vvv-pot}), (\ref{eq:vvp-pot}), and
(\ref{eq:partialv}).
In Eq.~(\ref{eq:addition}),
$R$ and $R'$ are the lightest spin-1 bare states of either
$(I,S[{\rm strangeness}])=(1,0)$ or $(I,S)=(1/2,-1)$ which we denote ``$\rho$'' and
``$\bar{K}^{*}$'', respectively.
For our particular application to
the $D^+\to K^-\pi^+\pi^+$ decay,
we include $(VP,V'P',V_{ex})$=
(``$\rho$''$\bar{K}$,``$\rho$''$\bar{K}$,$\rho$),
(``$\rho$''$\bar{K}$,``$\bar{K}^*$''$\pi$,$K^*$),
(``$\bar{K}^*$''$\pi$,``$\bar{K}^*$''$\pi$,$\rho$)
for the diagram Fig.~\ref{fig:vecpot}(a), and
$(VP,V'P',V_{ex})$=
(``$\rho$''$\bar{K}$,``$\rho$''$\bar{K}$,$K^*$),
(``$\rho$''$\bar{K}$,``$\bar{K}^*$''$\pi$,$\omega$),
(``$\bar{K}^*$''$\pi$,``$\bar{K}^*$''$\pi$,$\bar{K}^*$)
for the diagram Fig.~\ref{fig:vecpot}(b).
On the other hand, we leave examination of mechanisms such as those shown in
Fig.~\ref{fig:3meson} to future work
for the following reasons.
As we emphasized in the introduction, 
even effects of multiple scattering due
to the two-meson force on the $D$-decay
has still not been studied in a realistic setting.
In this situation, 
studying the relevance of the three-meson force to the $D$-decay is
indeed in an exploratory level, and thus 
a reasonable starting point would be to
include it in the most important channel.
As we will see, 
the vector-pseudoscalar ($\rho$-$\bar K$) channel plays a very
important role in the rescattering process,
and therefore it would be good to study 
 the new mechanisms of Fig.~\ref{fig:vecpot} 
in this channel at first.
Also, regarding the mechanism in Fig.~\ref{fig:3meson}(b), 
no relevant meson-resonance of spin-0, $I$=3/2, $S$=$-1$
is known in the $D$-meson mass region, 
and thus we do not need to include it for the moment.

\subsection{$D^+\to K^-\pi^+\pi^+$ decay amplitude}
\label{sec:decay-amp}

\begin{figure}[t]
 \includegraphics[width=11cm]{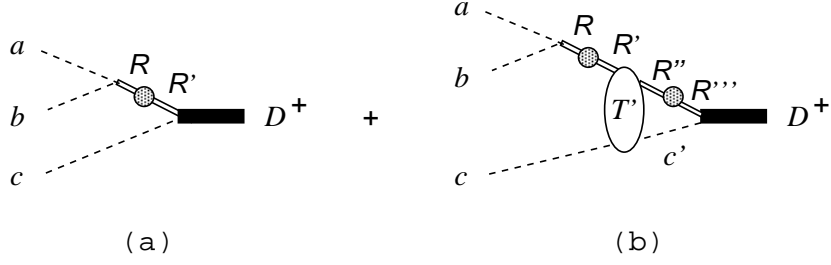}
\caption{\label{fig:d-decay}
Diagrammatic representation of $D^+$-decay into three pseudoscalar
 mesons $a,b,c$ in our coupled-channel model.
(a) The isobar-type diagram.
(b) The rescattering diagram.
The amplitude $T'$ is from the scattering equation represented by Fig.~\ref{fig:3body-t}.
The gray blob represents the dressed ${\cal R}$ Green function.
}
\end{figure}
In our coupled-channel model, 
the decay amplitude for $D^+\to K^-\pi^+\pi^+$ is given by
\begin{eqnarray}
\label{eq:cyclic-sum}
T_{K^-\pi^+\pi^+, D^+} 
(\bm{p}_{K^-},\bm{p}_{\pi_1^+},\bm{p}_{\pi_2^+};E\!=\!m_{D^+})
=\sum_{(abc)}^{\rm cyclic}
T_{(ab)c, D^+} (\bm{p}_a,\bm{p}_b,\bm{p}_c; E) \ ,
\end{eqnarray}
where we have introduced the cyclic summation that takes the sum over
$abc$ = $K^-\pi_1^+\pi_2^+$, $\pi_1^+\pi_2^+K^-$, $\pi_2^+K^-\pi_1^+$,
and
\begin{eqnarray}
\label{eq:decay-amp}
T_{(ab)c, D^+} (\bm{p}_a,\bm{p}_b,\bm{p}_c; E) &=&
\sum_{{\cal R},{\cal R}',s_{\cal R}^z} f^{s_{\cal R}^z}_{ab,{\cal R}}(\bm{p}_a,\bm{p}_b)  
G_{c{\cal R},c{\cal R}'}(p_c,E)  
\bar{\Gamma}^{s_{\cal R}^z}_{c{\cal R}',D^+}(\bm{p}_c,E) ,
\end{eqnarray}
where $s_{\cal R}^z$ is the $z$-component of the spin of ${\cal R}$, and
$G_{c{\cal R},c{\cal R}'}$ is the Green function that has
been defined in Eqs.~(\ref{eq:green-Rc})-(\ref{eq:green-Rc4}).
This decay amplitude in Eq.~(\ref{eq:decay-amp})
 is diagrammatically represented in Fig.~\ref{fig:d-decay}.
The symbol $f^{s_{\cal R}^z}_{ab,{\cal R}}$ denotes a ${\cal R}\to ab$ decay vertex
function which is explicitly given as
\begin{eqnarray}
f^{L^z}_{ab,{\cal R}} (\bm{p}_a,\bm{p}_b) &=&
\sqrt{ \frac{m_{R} E_a(q) E_b(q)}{E_{R}(p_R) E_a(p_a) E_b(p_b)} }
\inp{t_a t^z_{a} t_b t^z_{b}}{I, t^z_a+t^z_b} Y_{L,L^z}(\hat q)
\tilde{f}^{LI}_{ab,R}(q) \quad {\rm for}\ {\cal R}=R, \\
f^{L^z}_{ab,{\cal R}} (\bm{p}_a,\bm{p}_b) &=&
\sqrt{ \frac{E_a(q) E_b(q)}{E_a(p_a) E_b(p_b)} }
\inp{t_a t^z_{a} t_b t^z_{b}}{I, t^z_a+t^z_b} Y_{L,L^z}(\hat q)
\tilde w^{LI}_{ab}(q) \qquad {\rm for}\ {\cal R}=r_{ab} \ , 
\label{eq:R-ab}
\end{eqnarray}
where $\tilde{f}^{LI}_{ab,R}$ and $\tilde w^{LI}_{ab}$ have been defined in 
Eqs.~(\ref{eq:tilde-f}) and (\ref{eq:tilde-w}), respectively;
$\langle j_1 m_1 j_2 m_2 |JM\rangle$ is the Clebsch-Gordan coefficient,
and $t_a$ is the isospin of meson $a$ and $t^z_a$ is its $z$ component.
The kinematical factors in the equations are from the Lorentz
transformation to boost the $ab$-pair center-of-mass (CM) frame to
the total CM frame. 
The momentum $\bm{q}$ is the relative momentum of the $ab$-pair in their
CM frame.
The dressed $D^+\to {\cal R'}c$ vertex,
$\bar{\Gamma}^{s_{\cal R'}^z}_{c{\cal R}',D^+}$,
has also been introduced in Eq.~(\ref{eq:decay-amp}), 
and it is explicitly given by
\begin{eqnarray}
\bar{\Gamma}^{s_{\cal R'}^z}_{c{\cal R'},D^+} (\bm{p}_c,E)
&=&
\sum_{PT}
\sum_{ l, l^z}
\inp{l l^z s_{\cal R'} s^z_{\cal R'}}{S_{D}\; S_{D}^z}
\inp{t_{\cal R'}\; t^z_a\!+\!t^z_b\; t_c t^z_c}{T\; t^z_a\!+\!t^z_b\!+\!t^z_c}
\nonumber\\
&&\times Y_{l,l^z}(\hat p_c) \bar F^{S_{D}PT}_{(c{\cal R'})_l,D^+}(p_c,E),
\label{eq:dressed_mstar}
\end{eqnarray}
where $S_{D}(=0)$ is the $D$-meson spin,
and $t^z_a\!+\!t^z_b\!+\!t^z_c=3/2$ for $D^+\to K^-\pi^+\pi^+$.
We sum over the parity ($P$) and total isospin $(T)$ of the final
hadronic states
because the weak $D$-decay does not conserve them.
The last factor in the above equation is given by
\begin{eqnarray}
\label{eq:dressed-g}
\bar F^{JPT}_{(c{\cal R'})_l ,D^+}(p_c ,E)&=&
F^{JPT}_{(c{\cal R'})_l, D^+}(p_c) 
+ \sum_{{\cal R}'',{\cal R}''',c',l'} 
\int_0^\infty\!\! dp_{c'}\, p_{c'}^2 \;
T'^{JPT}_{(c{\cal R'})_{l},(c'{\cal R}'')_{l'}}(p_c,p_{c'};E) 
\nonumber\\
&&
\qquad\qquad \qquad\qquad \qquad\qquad
\times
G_{c'{\cal R}'',c'{\cal R}'''}(p_{c'},E) F^{JPT}_{(c'{\cal R}''')_{l'},D^+}(p_{c'}) \ ,
\end{eqnarray}
where $T'^{JPT}_{(c{\cal R'})_{l},(c'{\cal R}'')_{l'}}$
is the partial-wave amplitude for 
$c'{\cal R}''\to c{\cal R'}$ scattering
obtained by solving the coupled-channel scattering equation, Eq.~(\ref{eq:pw-tcr}).
The first term on the rhs corresponds to the isobar-type contribution 
[Fig.~\ref{fig:d-decay}(a)] while the second term is the contribution from
the rescattering [Fig.~\ref{fig:d-decay}(b)].
The quantity $F^{JPT}_{(c{\cal R})_l,D^+}(p)$ is
the bare $D^+\to ({\cal R}c)_l$ vertex function for which we choose a parametrization,
\begin{eqnarray}
F^{JPT}_{(c{\cal R})_l,D^+}(p) &=& 
\frac{A_{\cal R}}{(2\pi)^{3/2}} 
\frac{C^{JPT}_{(c{\cal R})_l}\; \exp\left[{i\,\phi^{JPT}_{(c{\cal R})_l}}\right]}
{\sqrt{2 E_c(p)}}
\left(
\frac{(\Lambda^{JPT}_{(c{\cal R})_l})^2 }{p^2 + (\Lambda^{JPT}_{(c{\cal R})_l})^2}
\right)^{2+(l/2)} 
\left(\frac{p}{m_\pi}\right)^{l} ,
\label{eq:bare_mstar}
\end{eqnarray}
with
\begin{eqnarray}
A_{\cal R} &=& \sqrt{m_R\over 2 E_R(p)} \quad {\rm for}\ {\cal R}=R \ ,
\qquad A_{\cal R} = 1 \quad {\rm for}\ {\cal R}=r_{ab} \ .
\label{eq:ar}
\end{eqnarray}
In Eq.~(\ref{eq:bare_mstar}),
$C^{JPT}_{(c{\cal R})_l}$, $\phi^{JPT}_{(c{\cal R})_l}$, 
and $\Lambda^{JPT}_{(c{\cal R})_l}$ are the
coupling, phase, and cutoff, respectively,
and they will be determined by fitting Dalitz plot distribution data.
The couplings $C^{JPT}_{(c{\cal R})_l}$ are nonzero only when
$|S_{D}-s_{\cal R}|\leq l \leq S_{D} + s_{\cal R}$.
The parametrization used in this work [Eq.~(\ref{eq:bare_mstar})]
is different from the one used in Ref.~\cite{3pi-1} in choosing the
kinematical factor.

The isobar-model-type amplitude [Fig.~\ref{fig:d-decay}(a)] for
the $D^+\to K^-\pi^+\pi^+$  decay,
$T^{\rm Isobar}_{(ab)c, D^+}$, 
 is obtained from the above equations~(\ref{eq:cyclic-sum})-(\ref{eq:ar})
by just dropping the second term of the rhs of Eq.~(\ref{eq:dressed-g}).

The procedure to calculate the Dalitz plot distribution from the decay
amplitude of Eq.~(\ref{eq:cyclic-sum}) is explained in detail in
Appendix~B of Ref.~\cite{3pi-1}, and we do not repeat it here.

\section{Analysis and Results}
\label{sec:results}

Now we apply the coupled-channel formalism discussed in the previous section to 
analyses of data.
First we determine parameters in the two-pseudoscalar-meson
scattering model by analyzing experimental data for $\pi\pi$ and
$\pi\bar K$ scatterings.
Then we extract resonance parameters from amplitudes of the two-meson
interaction model.
This two-meson interaction model is a basic ingredient for the three-meson
scattering model.
In the subsequent subsection, 
we analyze 
the $D^+\to K^-\pi^+\pi^+$ decay in a realistic setting.

\subsection{Two-pseudoscalar-meson scattering}
\label{sec:two-meson}

For studying the $D^+\to K^-\pi^+\pi^+$ decay in our coupled-channel
framework, the $\pi\pi$ and $\pi\bar{K}$ scatterings of $E\ltap 2$~GeV
are relevant.
We will determine the model parameters
of our $\pi\pi$ and $\pi\bar{K}$ scattering models, i.e.,
$h^{LI}_{a'b',ab}$, $b^{LI}_{ab}$, $m_R$, $g_{ab,R}$, and $c_{ab,R}$ 
in Eqs.~(\ref{eq:cont-ptl})-(\ref{eq:vf-cont2}), (\ref{eq:pw-2body-v}),
and (\ref{eq:pipi-vertex}),
by fitting empirical scattering amplitudes for $E\ltap 2$~GeV.

\subsubsection{$\pi\bar{K}$ scattering}

We analyze the $\pi\bar{K}$ scattering amplitudes from the LASS
experiment~\cite{lass,pik_I=3/2}.
For our application to the $D^+\to K^-\pi^+\pi^+$ decay 
in the next section, 
we determine the model parameters for 
$\{L,I\}$ = \{0,1/2\}, \{0,3/2\}, \{1,1/2\}, \{2,1/2\}
partial waves.
We explain details of our $\pi\bar{K}$ scattering model for each partial
wave. 
For the $\{L,I\}$=\{0,1/2\} wave,
we consider $\pi\bar{K}$-$\eta'\bar{K}$ coupled channels
because the $\eta'\bar{K}$ channel is known to play a significant role
while $\eta\bar{K}$ does not in this partial wave.
We include two bare $R$ states supplemented by
a contact $\pi\bar{K}$$\to$$\pi\bar{K}$ interaction.
For the $\{L,I\}$=\{0,3/2\} wave,
we consider a contact $\pi\bar{K}$$\to$$\pi\bar{K}$ interaction only.
For the $\{L,I\}$=\{1,1/2\} and \{2,1/2\} waves,
we consider coupling of $\pi\bar{K}$ and effective inelastic
channels; masses of the two ``particles''
in the inelastic channel, denoted by $m^{LI}_{1}$ and
$m^{LI}_{2}$, are also fitted to the data.
We include three bare $R$ states for $\{L,I\}$=\{1,1/2\} while 
a single bare $R$ state for $\{2,1/2\}$.
We present
the $\pi\bar{K}$ model parameters determined by the fits
in Table~\ref{tab:kpi} of Appendix~\ref{sec:para}.

\begin{figure}[t]
\includegraphics[width=0.33\textwidth]{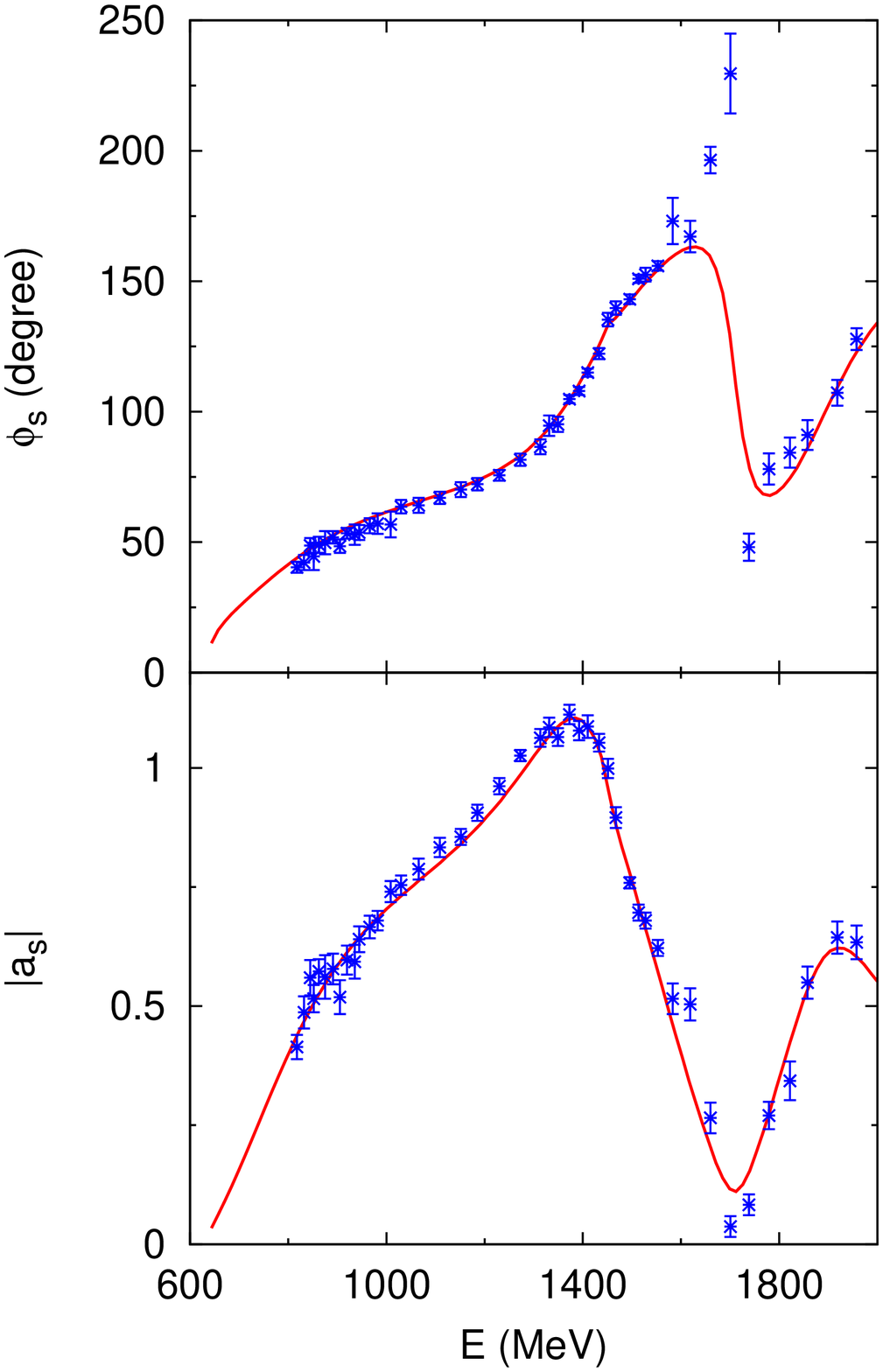}
\hspace{-2mm}
\includegraphics[width=0.33\textwidth]{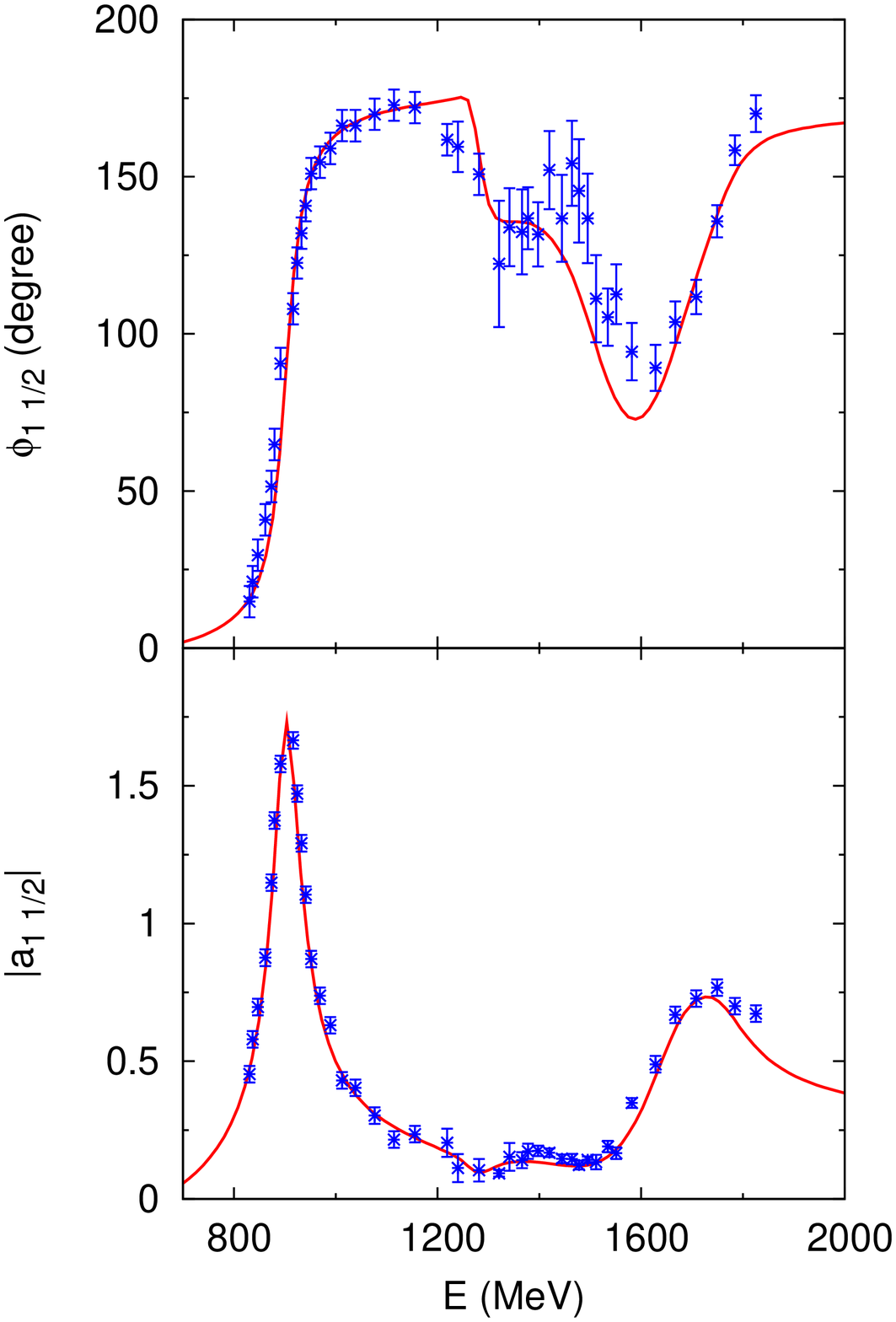}
\hspace{-2mm}
\includegraphics[width=0.33\textwidth]{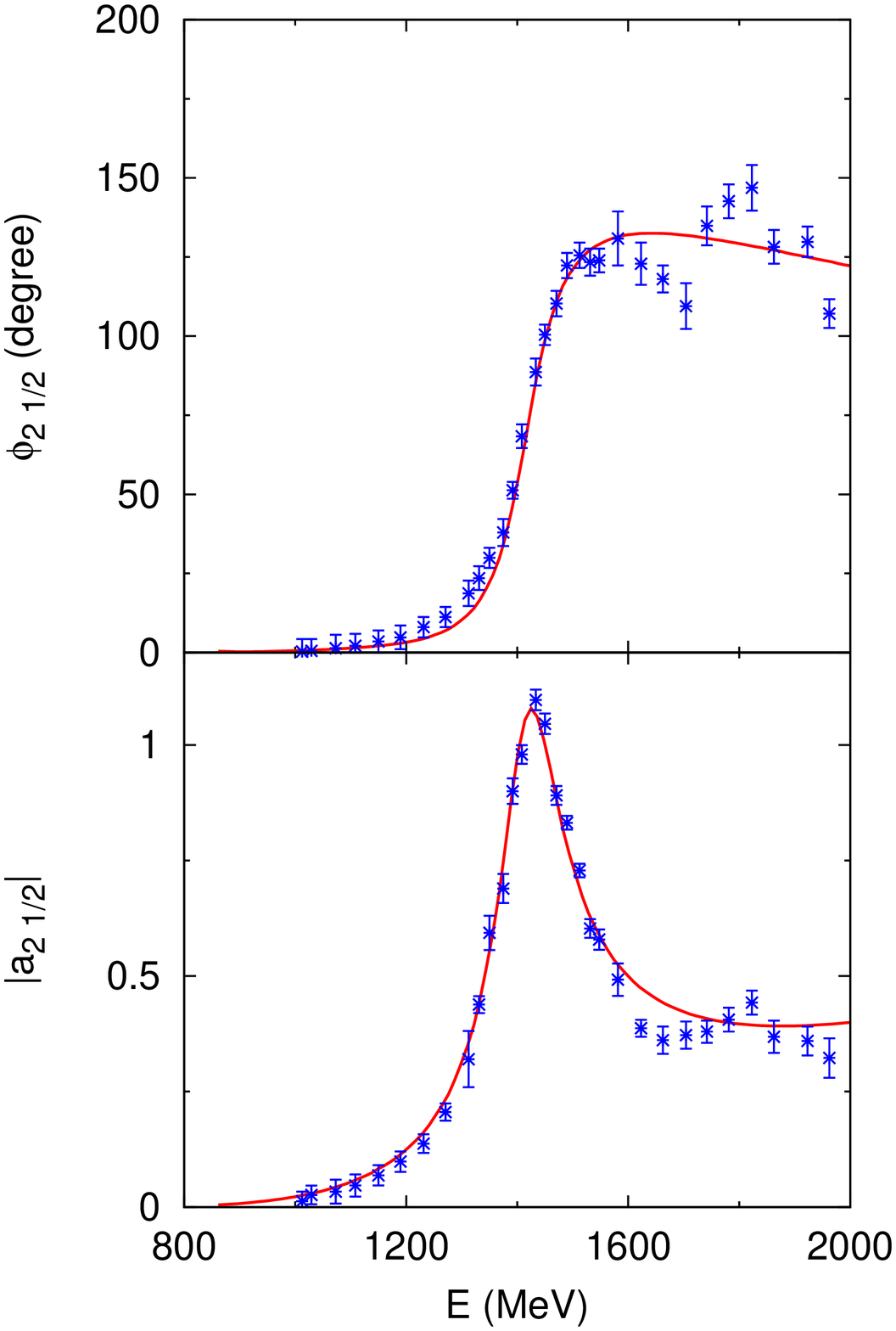}
\caption{\label{fig:pik} 
(Color online) 
Phase (upper) and modulus (lower) of the amplitudes 
for the $\pi\bar{K}$ scatterings:
(Left) $L$=0 for $\pi^+K^-$ ; (Center) $\{L,I\}$=\{1,1/2\}; 
(Right) $\{L,I\}$=\{2,1/2\}.
Data are taken from Ref.~\cite{lass}.}
\end{figure}
\begin{figure}[t]
\includegraphics[width=0.35\textwidth]{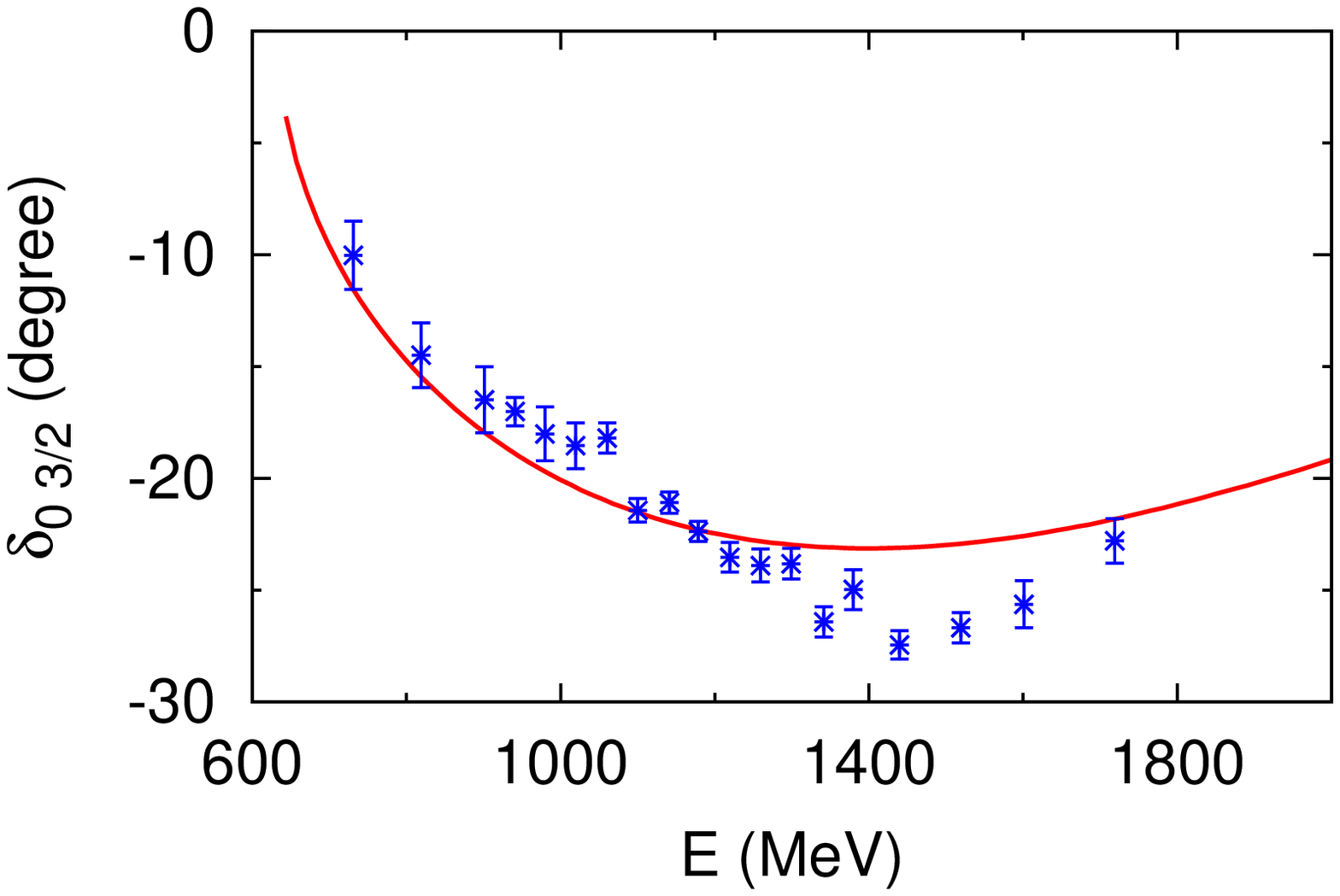}
\includegraphics[width=0.35\textwidth]{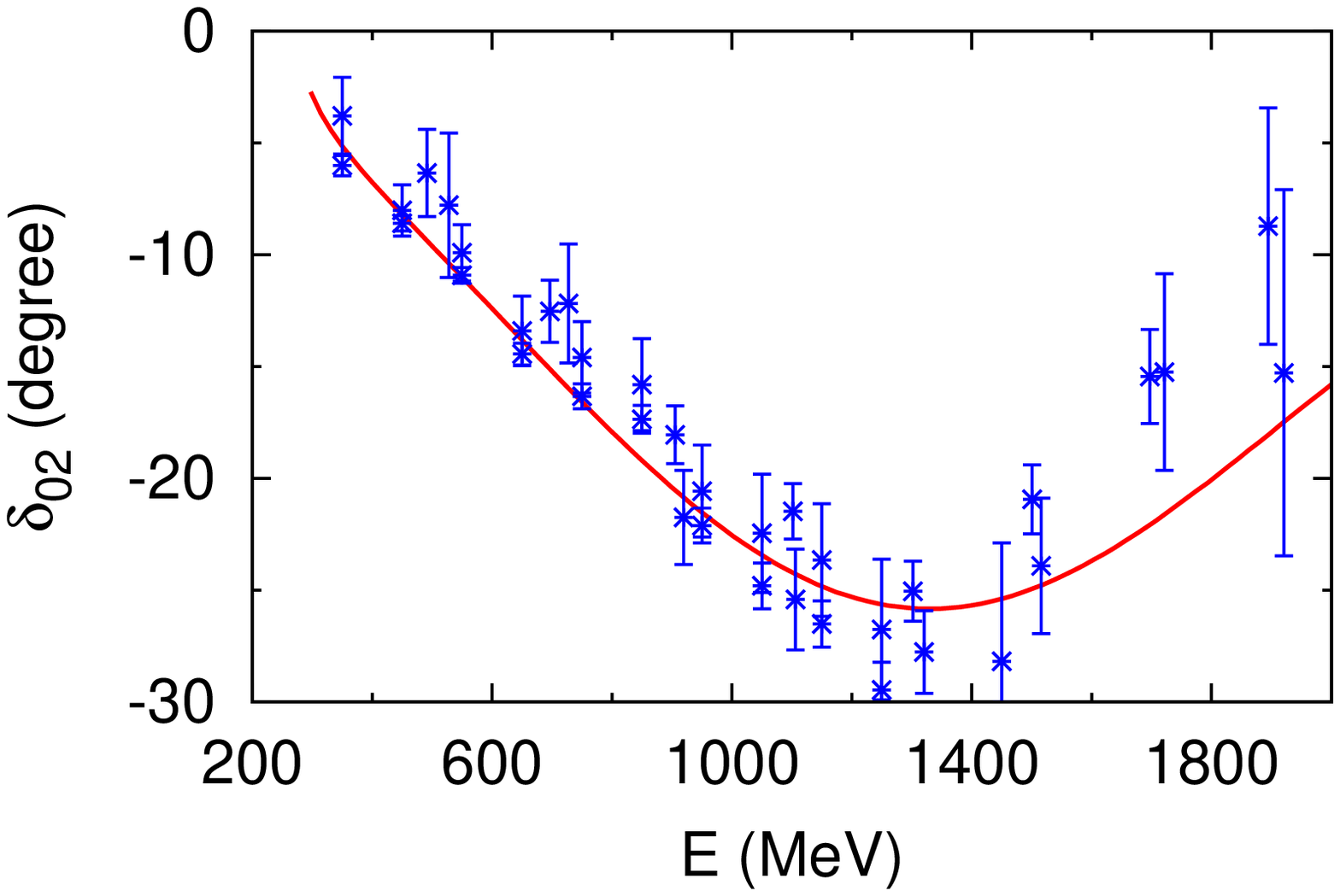}
\caption{\label{fig:pipi-pik} 
(Color online) 
Phase shifts ($\delta_{LI}$) of the $\pi K$ and $\pi\pi$ scatterings:
(Left) $\pi K$ 
phase shifts for $\{L,I\}$=\{0,3/2\} partial wave;
data are from Ref.~\cite{pik_I=3/2}.
(Right) $\pi\pi$ phase shifts for $\{L,I\}$=\{0,2\} partial wave;
data are from Refs.~\cite{pipi_I=2_1,pipi_I=2_2}.
}
\end{figure}
We present the quality of the fits to the empirical partial-wave
amplitude~\cite{lass} of $\pi^+K^-$ $L$=0 partial wave and 
$\{L,I\}$=\{1,1/2\},\{2,1/2\} partial waves in Fig.~\ref{fig:pik} where phases 
(upper panels) and modulus (lower panels) of the amplitudes are shown.
Also the elastic scattering phase shifts for the $\{L,I\}$=\{0,3/2\} partial wave
calculated with our model are compared with the data~\cite{pik_I=3/2}
in Fig.~\ref{fig:pipi-pik}(left).
The $\chi^2$ values obtained in the fits are tabulated in Table~\ref{tab:chi2-2body}.
\begin{table}[t]
\caption{\label{tab:chi2-2body}
$\chi^2$ values for the fits to empirical $\pi\bar K$ and $\pi\pi$
 partial-wave amplitudes. 
Each entry in the row indicated by ``\# of parameters'' is the number of
model parameters adjusted to fit the data.
}
\begin{ruledtabular}
\begin{tabular}{ccccccccc}
& \multicolumn{4}{c}{$\pi\bar K$}
& \multicolumn{4}{c}{$\pi\pi$}\\
$\{L,I\}$& \{0,1/2\}& \{0,3/2\}& \{1,1/2\}& \{2,1/2\}& \{0,0\}& \{0,2\}& \{1,1\}& \{2,0\} \\
\cline{1-1}
\cline{2-5}
\cline{6-9}
$\chi^2$          &304& 158&344&183&274&114&291& 276 \\
\# of data points & 84& 19 & 84& 64&148& 42&130& 130 \\
\# of parameters  & 12&  2 & 17&  7& 12&  3& 10&   5 \\
$\chi^2$/d.o.f    &4.2& 9.3&5.1&3.2&2.0&2.9&2.4& 2.2 \\
\end{tabular}
\end{ruledtabular}
\end{table}
The $s$-wave $\pi^+K^-$ amplitude is calculated by linearly combining
the $\{L,I\}$=\{0,1/2\}, \{0,3/2\} partial-wave amplitudes.
Overall, as seen in the figures, we obtain a reasonable description of
the data included in the fit.
However, one notices that the model has a sudden change 
and deviation from the data in the phase for the
$\{L,I\}$=\{1,1/2\} partial wave at $E\sim 1.3$~GeV.
This is perhaps an artifact of our model that has the threshold where
the effective inelastic channel opens.
Fortunately, the magnitude of the amplitude is rather small
around this energy so that the deviation in the phase will not give a significant
impact on observables calculated with this model.

\begin{table}[t]
\caption{\label{tab:pik-pole}
Pole positions of the $\pi\bar K$ partial-wave amplitudes in the complex-energy plane. 
A partial wave is specified by the orbital angular momentum ($L$) and the isospin ($I$).
Poles below $\text{Re}[E]\le 2$~GeV and $|\text{Im}[E]|\le 0.25$~GeV are presented.
Roman numerals in the square brackets specify the Riemann sheet on which
each of the poles exist.
We use the convention defined in, e.g., Ref.~\cite{sheet},
to specify each of the Riemann sheets, I--IV.
Each of the states is identified with the corresponding particle 
name used in the PDG listings~\cite{pdg2014}.
}
\begin{ruledtabular}
\begin{tabular}{ccccl|ccl|ccl}
$L$ & $I$ & \multicolumn{9}{c}{Pole positions (GeV), [Riemann sheet], Name}  \\\hline
0&1/2&$0.71-0.23i$ &[II] &$\kappa$     &$1.44-0.14i$ &[II] &$K^*_0(1430)$&$1.88-0.13i$ &[III]&$K^*_0(1950)$\\
 &   &             &     &             &$1.46-0.25i$ &[III]&$K^*_0(1430)$&             &     &             \\
1&1/2&$0.90-0.025i$&[II] &$K^*(892)$   &$1.28-0.058i$&[III]&$K^*(1410)$  &$1.66-0.088i$&[III]&$K^*(1680)$  \\ 
2&1/2&$1.42-0.055i$&[III]&$K^*_2(1430)$& ---         &     &             &---          &     &             \\
\end{tabular}
\end{ruledtabular}
\end{table}
From the $\pi\bar{K}$ amplitudes of the model determined above, we extract resonance
pole positions by the analytic continuation~\cite{analytic-cont,analytic-cont2}
as shown in Table~\ref{tab:pik-pole}.
We present poles below $\text{Re}[E]\le 2$~GeV and $|\text{Im}[E]|\le 0.25$~GeV.
We can consistently identify most of the extracted poles with 
the corresponding particles listed by the Particle Data Group
(PDG)~\cite{pdg2014} as shown in the table.
For the $\{L,I\}$=\{0,1/2\} partial wave,
our model has a pole at $0.71-0.23i$~GeV
that corresponds to the so-called $\kappa$ whose 
mass is $682\pm 29$~MeV, and width $547\pm 24$~MeV in the PDG listing.
Also we find two poles at ${\rm Re}[E]\sim$1.4~GeV on different Riemann sheets.
These two poles are associated with a single resonance [$K^*_0(1430)$]
that is split by coupling to $\eta'\bar{K}$ channel.
For the $\{L,I\}$=\{1,1/2\} partial wave,
our model has the well-established $K^*(892)$.
Also in the same partial wave, there is a pole at 
$1.28-0.058i$~GeV that is a bit off the $K^*(1410)$ resonance parameters
from the PDG.

\subsubsection{$\pi\pi$ scattering}

We perform an analysis of $\pi\pi$ scattering data with our
coupled-channel model in a way similar to the analysis of $\pi\bar{K}$ data in
the previous section.
Although we only need a $\pi\pi$ model for the $\{L,I\}$=\{1,1\},\{0,2\} partial
waves for our coupled-channel analysis of the $D^+\to K^-\pi^+\pi^+$ decay, 
we present here our $\pi\pi$ model for all major partial waves 
for a future reference. 
We consider $\pi\pi$-$K\bar{K}$ coupled channels for all partial waves
except for $\{L,I\}$=\{0,2\} where only the elastic
$\pi\pi$ channel is taken into account.
Regarding details of our model for each partial wave, 
we include two bare $R$ states supplemented by
a contact $\pi\pi$$\to$$\pi\pi$ interaction
for the $\{L,I\}$=\{0,0\} wave.
For the $\{L,I\}$=\{1,1\} and \{2,0\} waves,
we include two bare $R$ states and 
a single bare $R$ state, respectively.
Finally for the $\{L,I\}$=\{0,2\} wave,
we consider a contact $\pi\pi$$\to$$\pi\pi$ interaction only.
We present
the $\pi\pi$ interaction model parameters determined by the fits
in Table~\ref{tab:pipi} of Appendix~\ref{sec:para}.

\begin{figure}[t]
\includegraphics[width=0.33\textwidth]{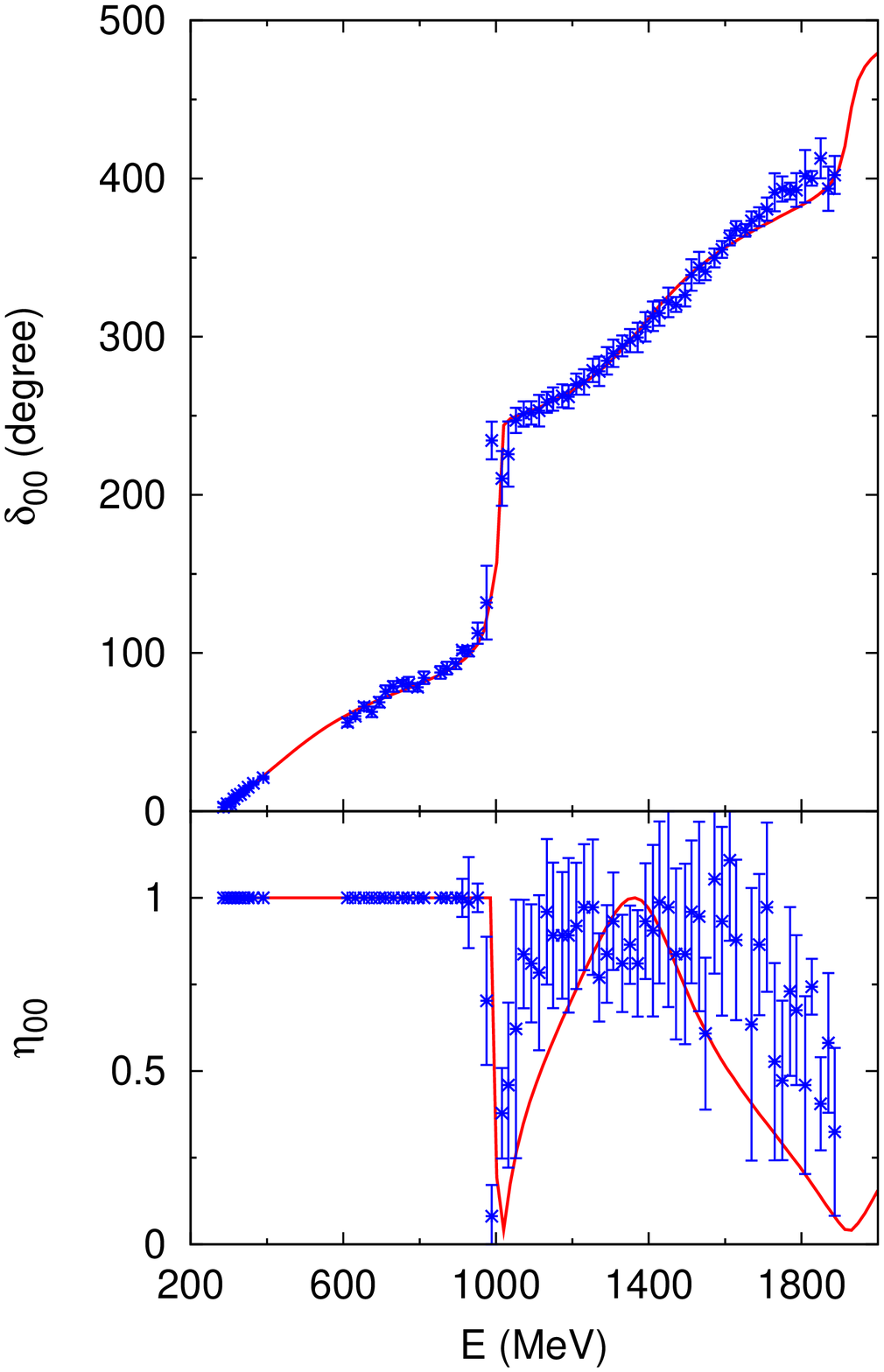}
\hspace{-2mm}
\includegraphics[width=0.33\textwidth]{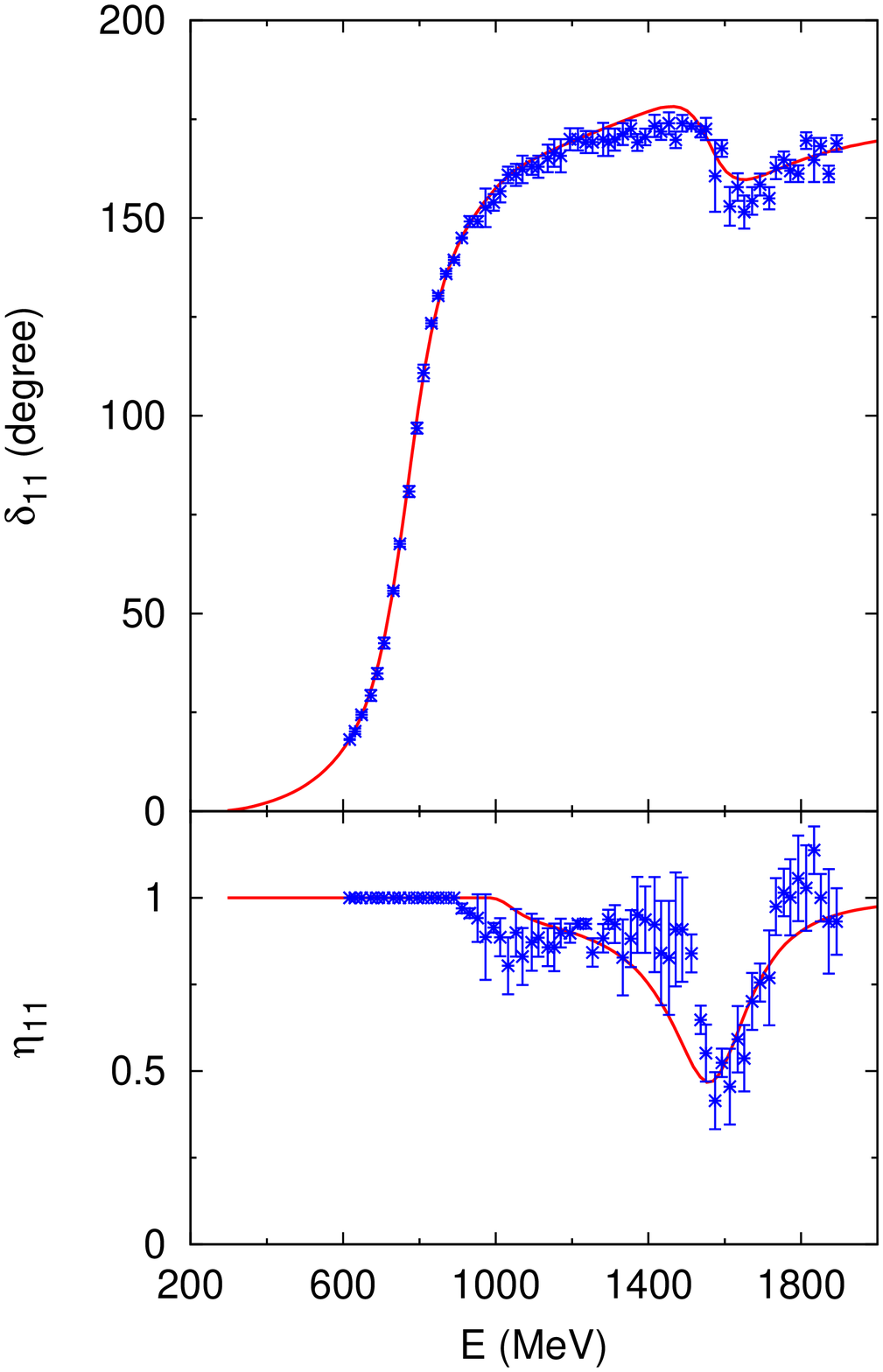}
\hspace{-2mm}
\includegraphics[width=0.33\textwidth]{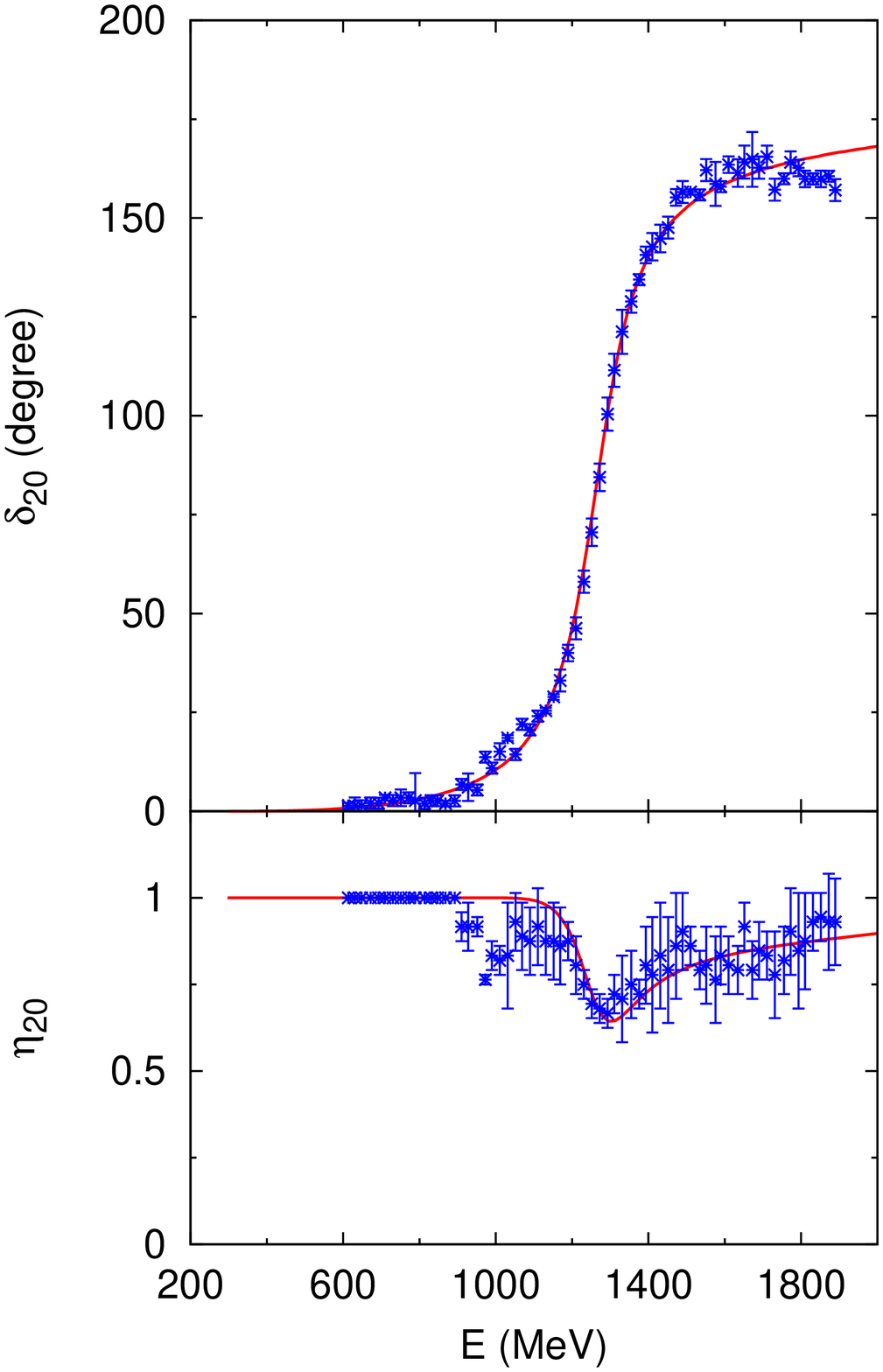}
\caption{\label{fig:pipi} 
(Color online) 
Phase shifts (upper) and inelasticities (lower) for the $\pi\pi$ scattering:
(Left) $\{L,I\}$=\{0,0\}. (Center) $\{L,I\}$=\{1,1\}. (Right) $\{L,I\}$=\{2,0\}.
Data are taken from Ref.~\cite{grayer,hyams,na48}.}
\end{figure}
We present the quality of the fits in 
Figs.~\ref{fig:pipi} and \ref{fig:pipi-pik}(right)
where phase shifts and inelasticities are shown.
The $\chi^2$ values obtained in the fits are tabulated in Table~\ref{tab:chi2-2body}.
As seen in the figures,
we obtain reasonable fits to the data from Refs.~\cite{grayer,hyams,na48}.
Although the quality of the fits is not much improved from those we
obtained in Ref.~\cite{3pi-1}, the purpose of updating 
the $\pi\pi$ scattering model is to take account of the proper
kinematical factor, as mentioned below Eq.~(\ref{eq:pipi-vertex}).
The nonzero inelasticities from our model
are due to the coupling to the $K\bar{K}$
channel of our coupled-channel model.
We note that 
the $K\bar K$ channel in our model effectively simulates all inelastic
channels in which the true $K\bar K$ channel is a major component, 
because we did not include $\pi\pi\to K\bar K$ and 
$K\bar K\to K\bar K$ data in our analysis, 
and also because we did not include the 4$\pi$
channel in the model.
\begin{table}[t]
\caption{\label{tab:pipi-pole}
Pole positions of the $\pi\pi$ partial-wave amplitudes in the complex-energy plane. 
The other features are the same as those in Table~\ref{tab:pik-pole}.
}
\begin{ruledtabular}
\begin{tabular}{ccccl|ccl|ccl}
$L$ & $I$ & \multicolumn{9}{c}{Pole positions (GeV), [Riemann sheet], Name}  \\\hline
0&0&$0.41-0.24i $&[II] &$\sigma$   &$1.02 - 0.001i$&[II] &$f_0(980)$&$1.38-0.15i$&[III]&$f_0(1370)$\\
1&1&$0.77-0.079i$&[II] &$\rho(770)$&$1.01 - 0.083i$&[III]&$\rho(1450)$&$1.58-0.16i$&[III]&$\rho(1700)$\\ 
2&0&$1.25-0.099i$&[III]&$f_2(1270)$&---            &     &          & --- 	 &     &
\end{tabular}
\end{ruledtabular}
\end{table}
From the determined $\pi\pi$ partial-wave amplitudes of our model, we extract resonance
poles as presented in Table~\ref{tab:pipi-pole}.
Most of the extracted poles are consistently identified with the
counterparts in the PDG listings, as shown in Table~\ref{tab:pipi-pole}.
A difference from the PDG value is found for the width of $f_0(980)$; our model
has a rather small width ($\sim$2~MeV) in comparison with the PDG
average (40-100~MeV). 
This difference was also found in the model used in Ref.~\cite{3pi-1}, and possible
sources of the difference were discussed there.
Also the second resonance in the $\{L,I\}$=\{1,1\} partial wave perhaps does
not correspond to $\rho(1450)$. 
However, an effect of this resonance pole on the amplitude seems to be
very small, and our model reproduces the empirical amplitude very well.
These differences 
from the PDG listings in the pole positions
could be due to the fact that we
employ the simple parametrization for the two-pseudoscalar-meson
interactions in order to make our three-meson scattering model
relatively easy to handle.
The behavior of the amplitudes on the unphysical 
(complex energy) region may be different from those of more
sophisticated two-meson interaction models cited in the PDG.
Thus, we would not mean to claim that
the pole positions presented on Tables~\ref{tab:pik-pole} and
\ref{tab:pipi-pole} have
a comparable reliability to those from the more sophisticated analyses.
The tables are just for showing the properties
of the amplitudes used in our analysis of the $D^+ \to K^- \pi^+ \pi^+$
decay, and thus we do not quote errors for the pole positions.
Yet, the $\pi\pi$ and $\pi\bar K$ amplitudes our model generates are
reasonable on the real physical energy axis, and should be good enough
for our purpose, that is, 
a coupled-channel analysis of the $D^+ \to K^- \pi^+ \pi^+$
decay with realistic $\pi\pi$ and $\pi\bar K$
amplitudes.

\subsection{Analysis of $D^+\to K^-\pi^+\pi^+$ Dalitz plot}
\label{sec:D-decay}

Now we will perform a partial-wave analysis of 
pseudodata for $D^+\to K^-\pi^+\pi^+$ Dalitz plot distribution
using our coupled-channel model.
In what follows,
we explain setups of our models used in the analysis.
Then we discuss how we prepare pseudodata
and our analysis procedure, which is followed by numerical results.

\subsubsection{Model setup}
\label{sec:setup}

In our coupled-channel framework,
$D^+$-meson decays into ${\cal R}c$ channels, followed by multiple scatterings due
to the hadronic dynamics, leading to the final $K^-\pi^+\pi^+$ state.
This process is expressed by 
Eqs.~(\ref{eq:decay-amp}), (\ref{eq:dressed_mstar}), and
(\ref{eq:dressed-g});
with the symmetrization, the decay amplitude is given by Eq.~(\ref{eq:cyclic-sum}).
We consider the following 11 ${\cal R}c$ coupled channels in our full calculation:
\begin{eqnarray}
\{{\cal R}c\} = \left\{
R^{01}_1\pi,
R^{01}_2\pi,
r^{01}_{\pi\bar K}\pi,
R^{11}_1\pi,
R^{11}_2\pi,
R^{11}_3\pi,
R^{21}_1\pi,
R^{12}_1\bar{K},
R^{12}_2\bar{K},
r^{03}_{\pi\bar K}\pi,
r^{04}_{\pi\pi}\bar{K}
\right\} \ ,
\label{eq:cc}
\end{eqnarray}
where 
$R^{L,2I}_i$ stands for the $i$-th bare $R$ state with the spin $L$ and the
 isospin $I$; when $I$ is an integer (half-integer), it is understood
 that this $R$ state has the strangeness $S$=0 ($S$=$-1$) in this paper.
Thus, $R^{01}_i$, $R^{11}_i$, $R^{21}_i$, $R^{12}_i$ are seeds of
$\bar K^*_0$, $\bar K^*$, $\bar K^*_2$, $\rho$ resonances, respectively.
In our model, these resonances are included as poles in the unitary
scattering amplitudes. 
$r^{L,2I}_{ab}$ is a ``state'' associated with a contact
interaction in a partial wave of $L$ and $I$, as has been introduced in 
Sec.~\ref{sec:three-meson}.
Most of the partial waves associated with
these channels have been considered in the previous 
Dalitz plot analyses of the
$D^+\to K^-\pi^+\pi^+$ decay~\cite{e791-prl,oller,e791,focus,focus2009,cleo}.
However, the $\{L,I\}$=\{0,2\} partial wave associated with the
$r^{04}_{\pi\pi}\bar{K}$ channel was considered 
only in the CLEO analysis~\cite{cleo}.
Also, the $\{L,I\}$=\{0,3/2\} partial wave associated with
the $r^{03}_{\pi\bar K}\pi$ channel was explicitly considered 
only in the FOCUS analysis~\cite{focus}, but other MIPWA can implicitly take
account of this partial wave.
The $\{L,I\}$=\{1,1\} partial wave associated with the
$R^{12}_i\bar{K}$ channel has not been considered in the previous
analyses.
This channel can contribute to 
$D^+\to K^-\pi^+\pi^+$ only through the coupled-channel dynamics, 
and therefore it does not show up in isobar-type models.
A possible important role of this channel was hinted by the Brazilian
group~\cite{usp,usp2}, as stated in the introduction.
We note that, unlike most of the previous isobar-type analyses, we do not
include a flat interfering background amplitude.
With the coupled channels considered in this work [Eq.~(\ref{eq:cc})],
the final hadronic system has 
the total spin $J$=0, parity $P$=+1, total isospin $T$=3/2, and 
$l$=$s_{\cal R}$.
We fit Dalitz plot pseudodata for $D^+\to K^-\pi^+\pi^+$ by 
adjusting parameters associated with 
the $D^+\to {\cal R}c$ vertex function in Eq.~(\ref{eq:bare_mstar}).
Among the parameters, 
$C^{JPT}_{(c{\cal R})_l}$, $\phi^{JPT}_{(c{\cal R})_l}$, and
$\Lambda^{JPT}_{(c{\cal R})_l}$,
we fix $\Lambda^{JPT}_{(c{\cal R})_l}$=5~GeV for all $D^+\to (c{\cal R})_l$
vertices, and set $C^{JPT}_{(\pi R^{11}_1)_1}$=1, $\phi^{JPT}_{(\pi R^{11}_1)_1}$=0.
Then, we fit the pseudodata by adjusting 
the other $C^{JPT}_{(c{\cal R})_l}$ and $\phi^{JPT}_{(c{\cal R})_l}$
under a requirement that 
all ${\cal R}^{L,2I}$, which belong to the same partial wave characterized
by $\{L,I\}$, share the same phase $\phi^{JPT}_{(c{\cal R})_l}$.
With this requirement, when the hadronic rescattering is turned off,
the Watson theorem is satisfied, up to a slight violation due to
the fact that our model takes account of the center-of-mass motion of
the scattering two mesons.
More specifically, the $p_c$-dependence of
the phase of $G_{c{\cal R},c{\cal R}'}$
in Eq.~(\ref{eq:decay-amp}) leads to the slight violation.

The hadronic rescattering processes are described by the scattering
amplitude, $T'^{JPT}_{(c'{\cal R}')_{l'},(c{\cal R})_{l}}$,
defined in Eq.~(\ref{eq:pw-tcr}), and also by the ${\cal R}c$ Green
function defined in Eqs.~(\ref{eq:green-Rc})-(\ref{eq:green-Rc4}).
The main driving force of the rescattering processes is the
two-pseudoscalar-meson interactions that have been fixed 
in the previous section by fitting the
empirical $\pi\bar{K}$ and $\pi\pi$ scattering amplitudes. 
The two-pseudoscalar-meson interactions enter the scattering equation
[Eq.~(\ref{eq:pw-tcr})] as the Z-diagrams 
and also through the $R$ self-energies and 
$\sigma^{LI}_{{\cal R}',{\cal R}}$ in 
Eqs.~(\ref{eq:RR-self})-(\ref{eq:R-self}).
Given the coupled channels specified above and the two-meson
interactions, we can find all the contributing Z-diagrams that induce
channel couplings.
In our analysis,
we consider all of the contributing Z-diagrams that contain
$\pi\pi\bar{K}$ and $K\bar{K}\bar{K}$ intermediate states.
In addition, we include the three-meson force based on the HLS model and, 
as mentioned below Eq.~(\ref{eq:addition}),
we have totally six diagrams of this type.
As described in Appendix~\ref{app:lag}, 
once two coupling constants are fixed by the $\rho\to\pi\pi$
and $\omega\to\pi\rho$ decay widths, 
all the other couplings are fixed by SU(3) and the HLS.
We use the same form factor [Eq.~(\ref{eq:ff-3mf})] for
all the different diagrams of the three-meson force.
The cutoff, $\Lambda_{\rm 3MF}$, in the form factor is determined by
fitting the pseudodata.

In our analysis of $D^+\to K^-\pi^+\pi^+$ Dalitz plot pseudodata,
we will basically use three models.
The first one is the ``Full model'' that contains all the dynamical contents
described above.
The second one is the ``Z model'' for which the rescattering mechanism is
solely due to multiple iteration of the two-pseudoscalar-meson interactions 
in the form of the Z diagrams and ${\cal R}$ propagators.
Thus the Full model is obtained by adding the three-meson force to the Z model.
The third model is the ``Isobar model'' that does not explicitly contain
the rescattering process.
The decay amplitude for the Isobar model has been described at the end
of Sec.~\ref{sec:decay-amp}.
This Isobar model is still different from most of the isobar models used
in the previous Dalitz plot analyses of $D^+\to K^-\pi^+\pi^+$ in that all
two-pseudoscalar partial-wave amplitudes are unitary, and fit well the
empirical amplitudes in the relevant energy region; 
the Watson theorem is also maintained in the sense discussed in the
previous paragraph.
Finally we remark that the two-pseudoscalar-meson interactions, that have
been determined in the previous sections, will not be adjusted to fit
the $D^+$-decay pseudodata.
This is in contrast with most of the previous analyses where some of
Breit-Wigner parameters were also adjusted along with $D^+\to Rc$ vertices.

\subsubsection{Data and analysis method}
\label{sec:data}

\begin{figure}[t]
\includegraphics[width=0.45\textwidth]{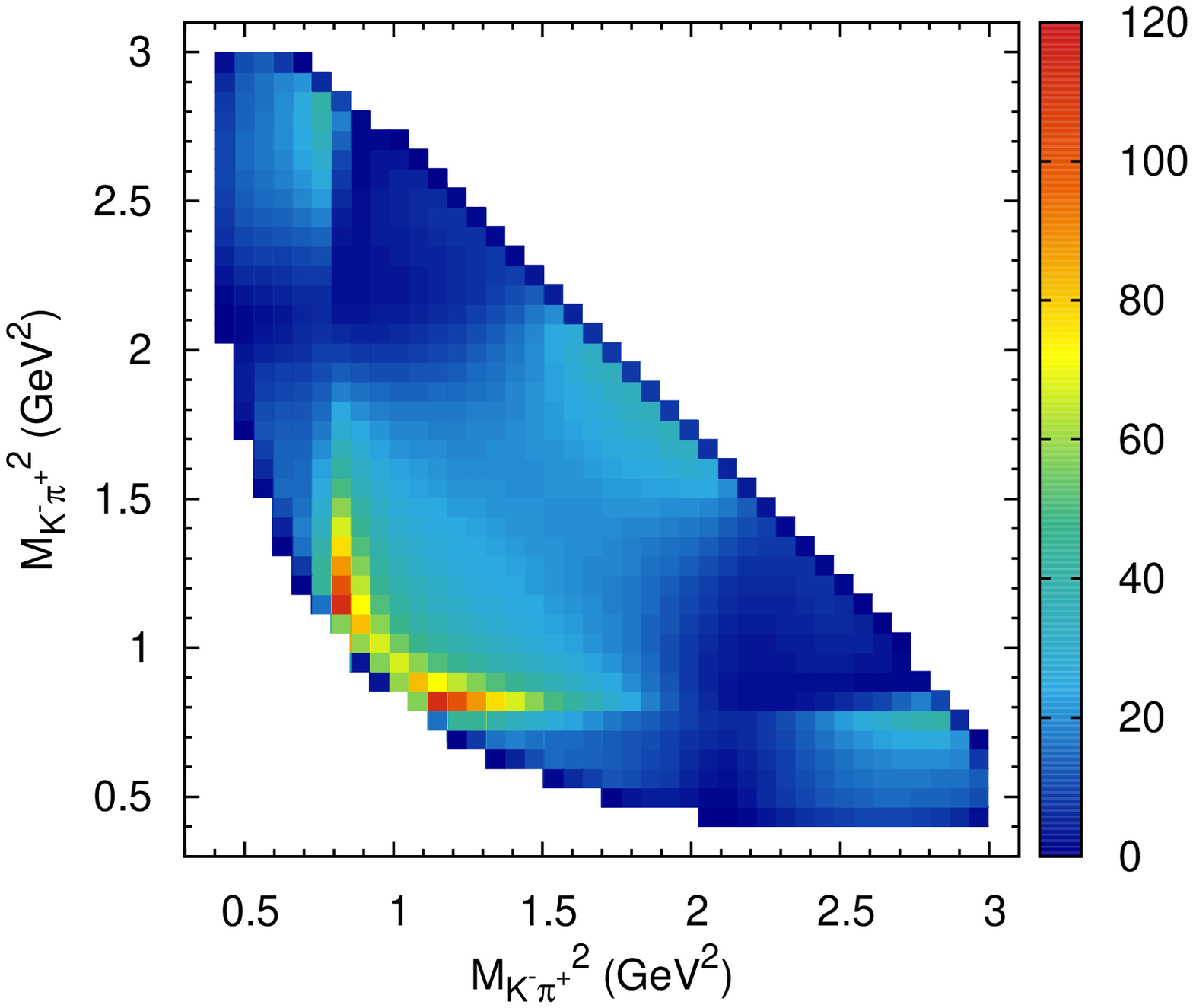}
\hspace{-3mm}
\includegraphics[width=0.45\textwidth]{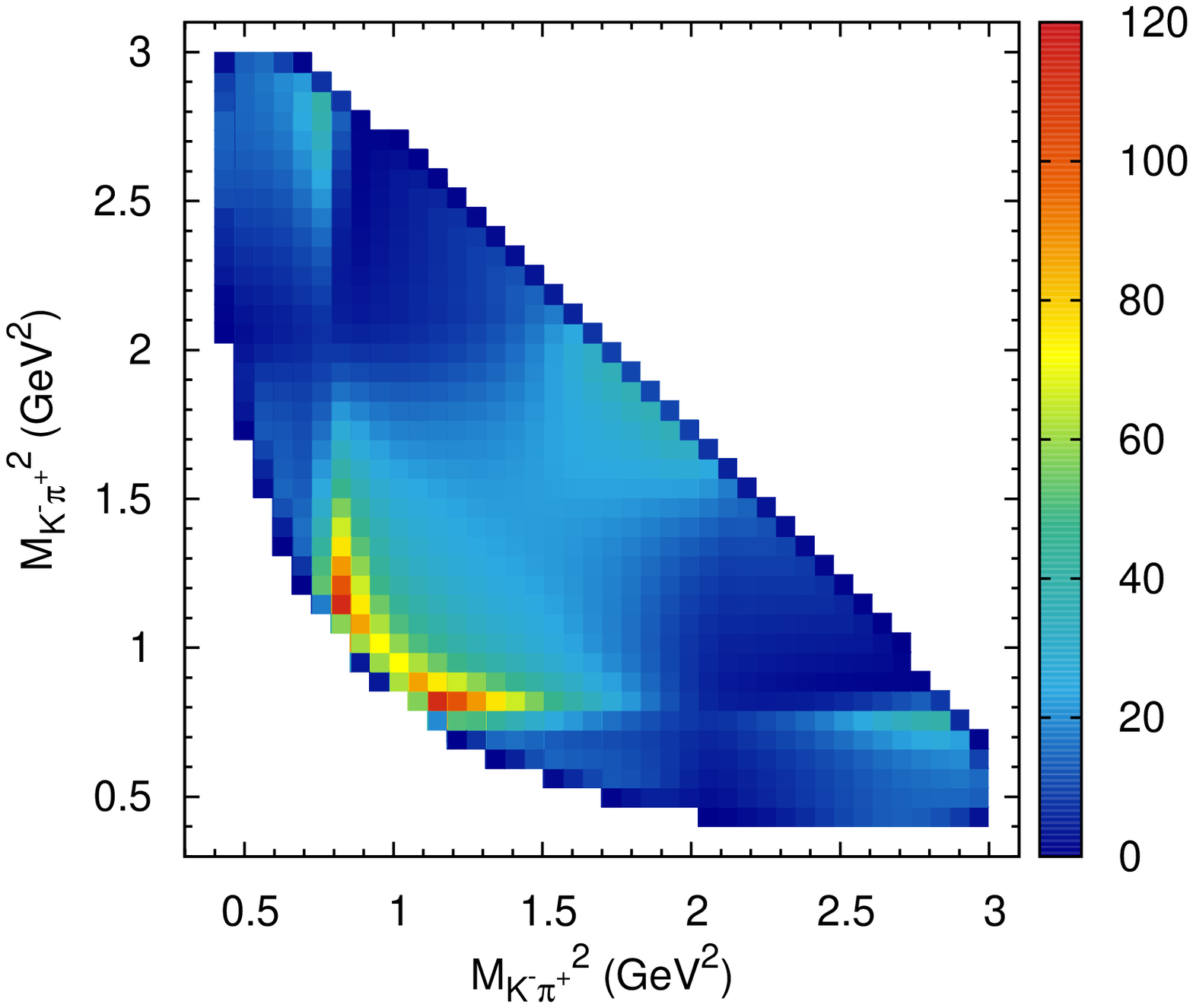}
\caption{\label{fig:dalitz}
(Color online) 
The Dalitz plot distribution for $D^+\to K^-\pi^+\pi^+$
in 40$\times$40 cells.
The left panel shows our pseudodata generated with the isobar model of
the E791 Collaboration~\cite{e791}.
The right panel is the counterpart 
from the Full model that has been fitted to the pseudodata.
An explanation for how the pseudodata are generated is found in 
Sec.~\ref{sec:data}.
}
\end{figure}

We create reasonably realistic pseudodata of 
the $D^+\to K^-\pi^+\pi^+$ Dalitz plot
from the isobar model of
the E791 Collaboration~\cite{e791}.
The E791 Collaboration obtained the isobar model
through their partial-wave analysis of 
the $D^+\to K^-\pi^+\pi^+$ Dalitz plot of 15,079 events, among which
94.4\% were determined to be signals.
In generating pseudodata, we take a procedure similar to
Sec.~V of Ref.~\cite{qcdf} where Dedonder et al. created
pseudodata for $D^0\to K_S^0\pi^+\pi^-$ using the isobar model of the
BABAR Collaboration.
We start with a grid 400$\times$400 squared cells covering 
all kinematical region of the Dalitz plot distribution with 
$M^2_{K^-\pi^+}$ as $x$-axis and another $M^2_{K^-\pi^+}$ as $y$-axis;
$M^2_{K^-\pi^+}$ denotes the squared invariant mass of the $K^-\pi^+$ pair.
Each cell is given by the E791 isobar model a value of the Dalitz plot
distribution at the center of the cell. 
Then, the values of the Dalitz plot distribution in 10$\times$10 adjacent
cells are summed to obtain 40$\times$40 cells, each of which has 
the value of the partially integrated Dalitz plot distribution. 
The width of each cell is 0.0649~GeV$^2$.
In the E791 analysis~\cite{e791},
40$\times$40 cells were used to perform
their MIPWA and thus are a reasonable size also in our analysis. 
The Dalitz plot distribution value in each cell is 
multiplied by a common normalization constant and then
is round off to be an integer;
the common normalization constant is chosen so that 
the sum of the round-off values of all the cells 
coincides with 15,079 $\times$ 94.4\% $\sim$14,234 , 
the number of signals of the E791 experiment.
In this way, we have generated pseudodata for 
the $D^+\to K^-\pi^+\pi^+$ Dalitz plot, as presented in the left panel
of Fig.~\ref{fig:dalitz}.

Next task is to analyze the above pseudodata with our model.
Again, Ref.~\cite{qcdf} serves as a useful reference to find an 
analysis procedure.
We calculate the Dalitz plot distribution using the decay amplitude of 
Eqs.~(\ref{eq:cyclic-sum})
and the formulas given in 
Appendix~B of Ref.~\cite{3pi-1}.
In each cell of the 40$\times$40 grid, we integrate the Dalitz plot
distribution from our model;
the overall normalization is chosen so that the integral over all the
kinematical region of the Dalitz plot distribution becomes equal to
14,234, the number of signals for the E791 data.
In this way, we have the number of events (a real number) in each of the
cells, and can compare it with the counterpart in the pseudodata.
If a given cell of the pseudodata has
the number of events less than 5,
then the cell is merged with the adjacent cell in the same $x$-axis to
be a larger cell. This grouping is repeated until the cell
contains more than or equal to 5 events. 
The same grouping is also applied to the event samples from our model.
Thus, the total number of effective cells, each of which contains more than or
equal to 5 events of the pseudodata, is 725.

The quality of the fit can be quantified by calculating 
$\chi^2$ defined by
\begin{eqnarray}
\chi^2 = \sum_j \chi^2_j
 = \sum_j \left( {N_j^{th} - N_j^{exp} \over \Delta
		 N_j^{exp}}\right)^2 \ ,
\label{eq:chi2}
\end{eqnarray}
where $j$ labels each cell, and $N_j^{th}$ ($N_j^{exp}$) is the number of
events in the $j$-th cell from our model (pseudodata).
The error of the pseudodata in each cell is assigned by 
$\Delta N_j^{exp} = \sqrt{N_j^{exp}}$.

We will perform the least $\chi^2$-fit to the pseudodata with
the CERN program library \textsc{MINUIT}.
Errors for the fitting parameters are estimated in a standard manner as follows.
First we calculate Hessian matrix, $H_{ij}$, defined by 
\begin{eqnarray}
 H_{ij} = {1\over 2}{\partial^2 \chi^2 \over \partial \theta_i\partial
  \theta_j} \bigg|_{\{\theta\}=\{\bar\theta\}} \ ,
\end{eqnarray}
where $\chi^2$ is given in Eq.~(\ref{eq:chi2});
$\theta_i$ is one of the fitting parameters and 
$\{\bar\theta\}$ is a set of the fitting parameters at the minimum of
$\chi^2$.
Then the error matrix is given by the inverse of the Hessian,
$E_{ij}=(H^{-1})_{ij}$, and the error for the parameter $\theta_i$ is
assigned by $\sqrt{E_{ii}}$.
An error $\delta X$ for a quantity $X$ such as a fit fraction is
estimated by the error propagation formula:
\begin{eqnarray}
[\delta X]^2 = 
\sum_{i,j}
{\partial X \over \partial\theta_i}\bigg|_{\{\theta\}=\{\bar\theta\}} \, 
E_{ij} \ 
{\partial X \over \partial\theta_j}\bigg|_{\{\theta\}=\{\bar\theta\}} \ .
\end{eqnarray}

\subsubsection{Numerical results and discussions}

\begin{table}[t]
\caption{\label{tab:chi2}
The total $\chi^2$-values and $\chi^2/{\rm d.o.f.}$
from the Full, Z and Isobar models obtained by fitting 
the $D^+\to K^-\pi^+\pi^+$ Dalitz plot pseudodata.
The Z model without couplings to
$(\pi^+\pi^0)_P^{I=1}\bar{K}^0$ is labeled by 
``Z (without $\rho$)''.
}    
\begin{ruledtabular}
\begin{tabular}{lcccc}
                             & Full   & Z   &Isobar& Z (without $\rho$)\\\hline
$\chi^2$                     & 157.   & 119.& 303. &216. \\ 	 
$\chi^2/{\rm d.o.f.}$        & 0.22   & 0.17&0.42  & 0.30
\end{tabular}
\end{ruledtabular}
\end{table}
We performed the least $\chi^2$-fit to the pseudodata following the procedure
explained in the previous subsection.
We used the three models; the Full, Z, and Isobar models.
All the parameters 
and their statistical errors
determined by the fits are tabulated for each of the models in 
Table~\ref{tab:param}.
The Dalitz plot distribution from the Full model is shown in 
the right panel of Fig.~\ref{fig:dalitz}.
Comparing with the left panel of the pseudodata, a difference is hardly
discernible to the eye.
The situation is the same for the Z and Isobar models.
The quality of the fit can be quantified by calculating the
$\chi^2$-values of Eq.~(\ref{eq:chi2}) that are
presented in the second row of Table~\ref{tab:chi2}.
The $\chi^2$-value of the Z model
is the smallest, and the Full model comes in second.
These models are significantly better than the Isobar
model in the fit quality.
In the third row of the table, we also show $\chi^2/{\rm d.o.f.}$ to
assess if the better $\chi^2$ is simply due to more degrees of freedom
in the fits.
The number of the fitting parameters for the Full, Z, and Isobar models
are 16, 15, and 12, respectively, as can be found in
Table~\ref{tab:param}.
The number of bins at which $\chi^2$ is calculated is 725.
Thus we obtain $\chi^2/{\rm d.o.f.}$ as shown in Table~\ref{tab:chi2}.
Therefore, the ranking of the fit quality is still in the same order as
far as $\chi^2$ is concerned.

\begin{figure}[t]
\includegraphics[width=0.33\textwidth]{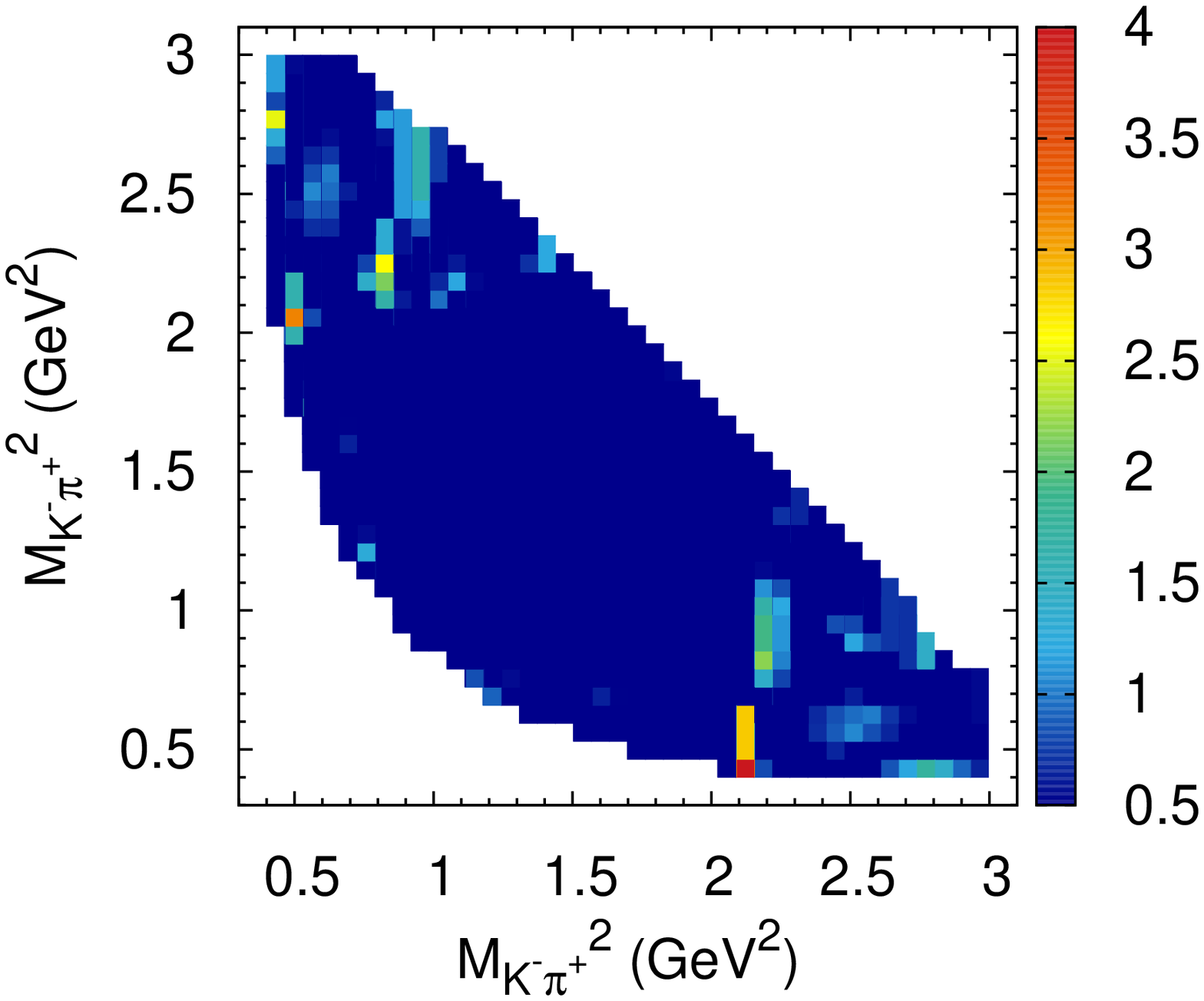}
\hspace{-3mm}
\includegraphics[width=0.33\textwidth]{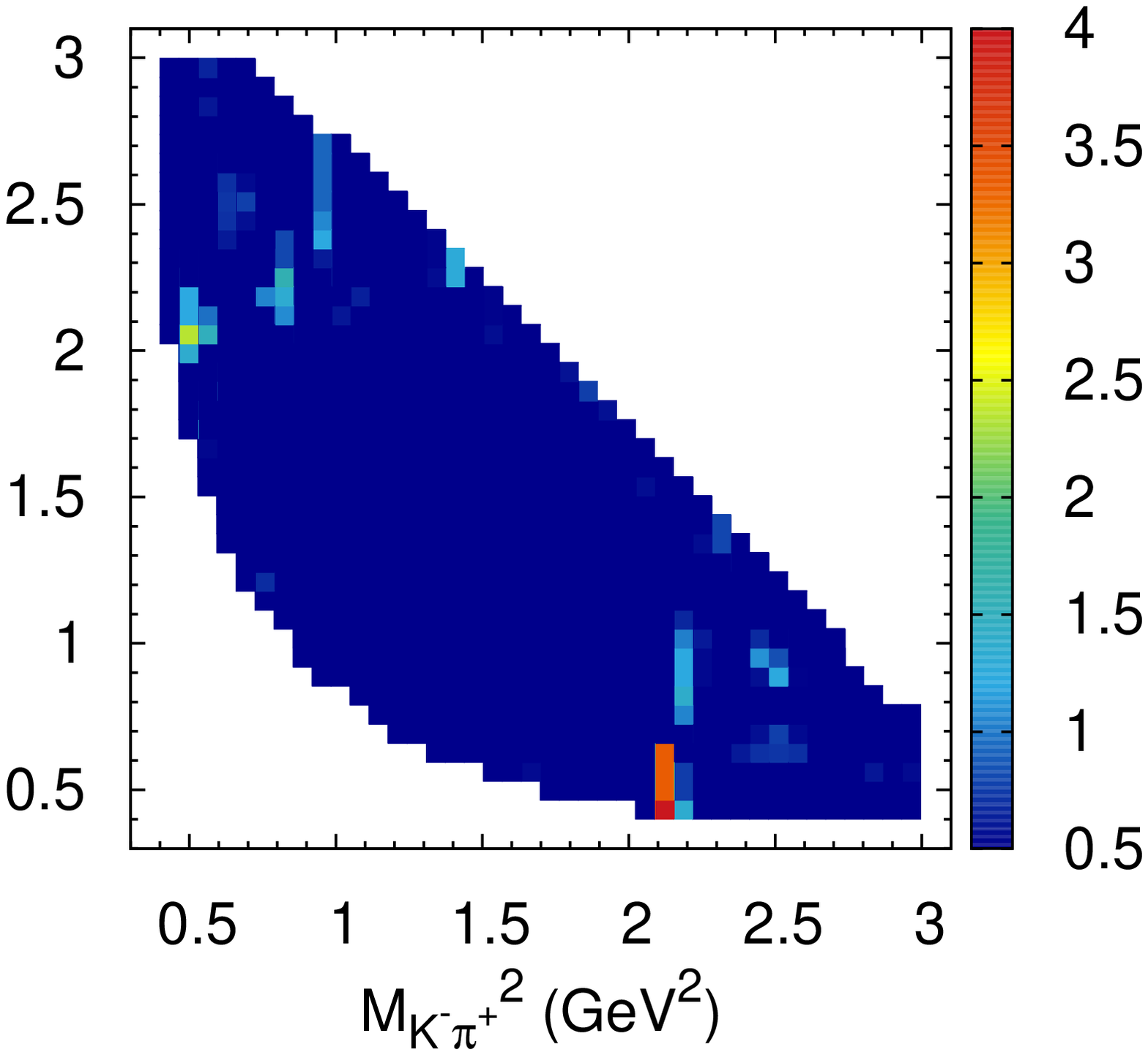}
\hspace{-3mm}
\includegraphics[width=0.33\textwidth]{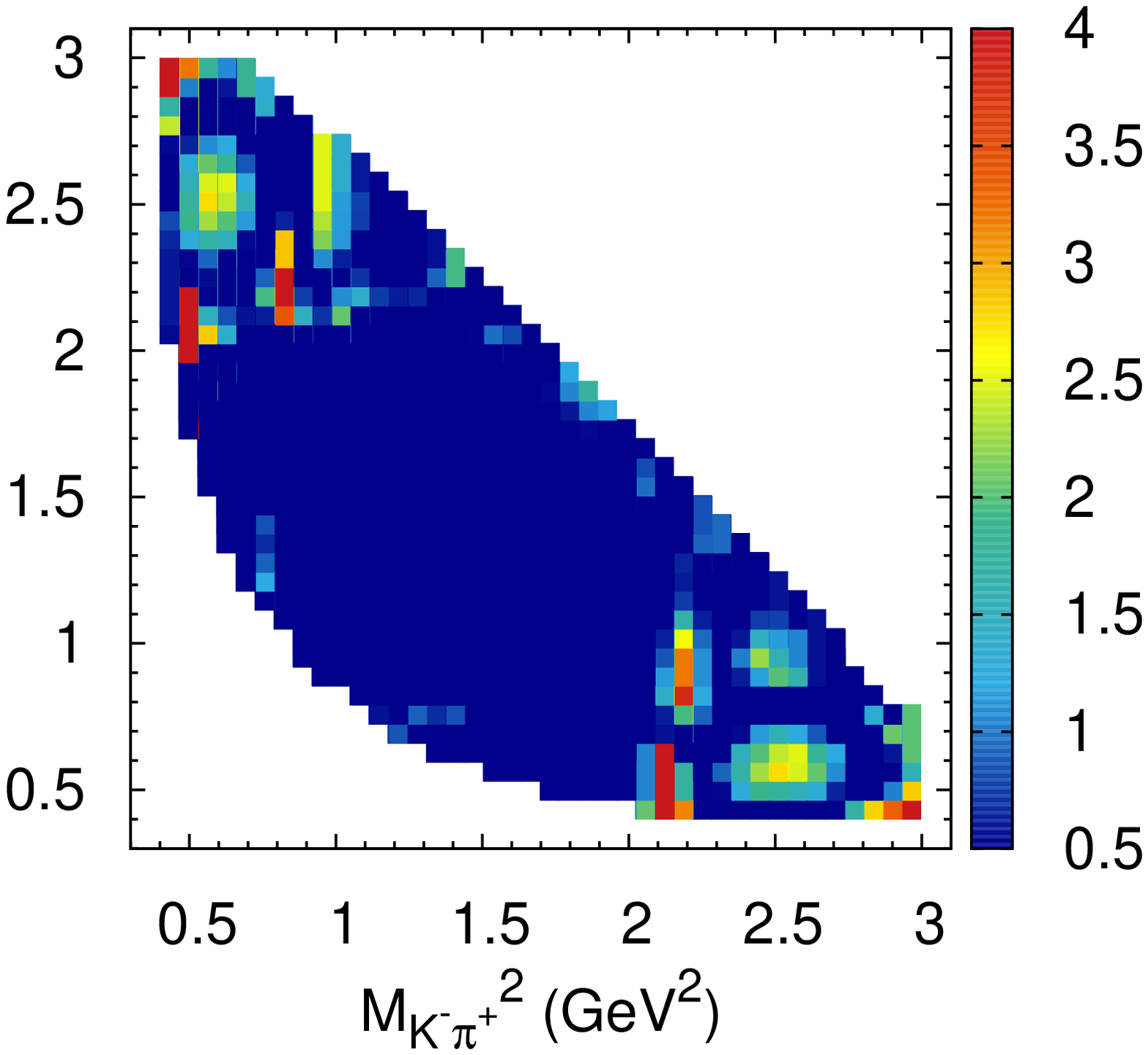}
\caption{\label{fig:error}
(Color online) 
The $\chi^2_j$ distributions.
The left, middle, right panels are from the Full, Z, and Isobar models,
 respectively.
The bins with $\chi^2_j\ge 4$ are all given the same color (red); the same applies to
the bins with $\chi^2_j\le 0.5$ (dark blue).
}
\end{figure}
\begin{figure}[t]
\includegraphics[width=0.45\textwidth]{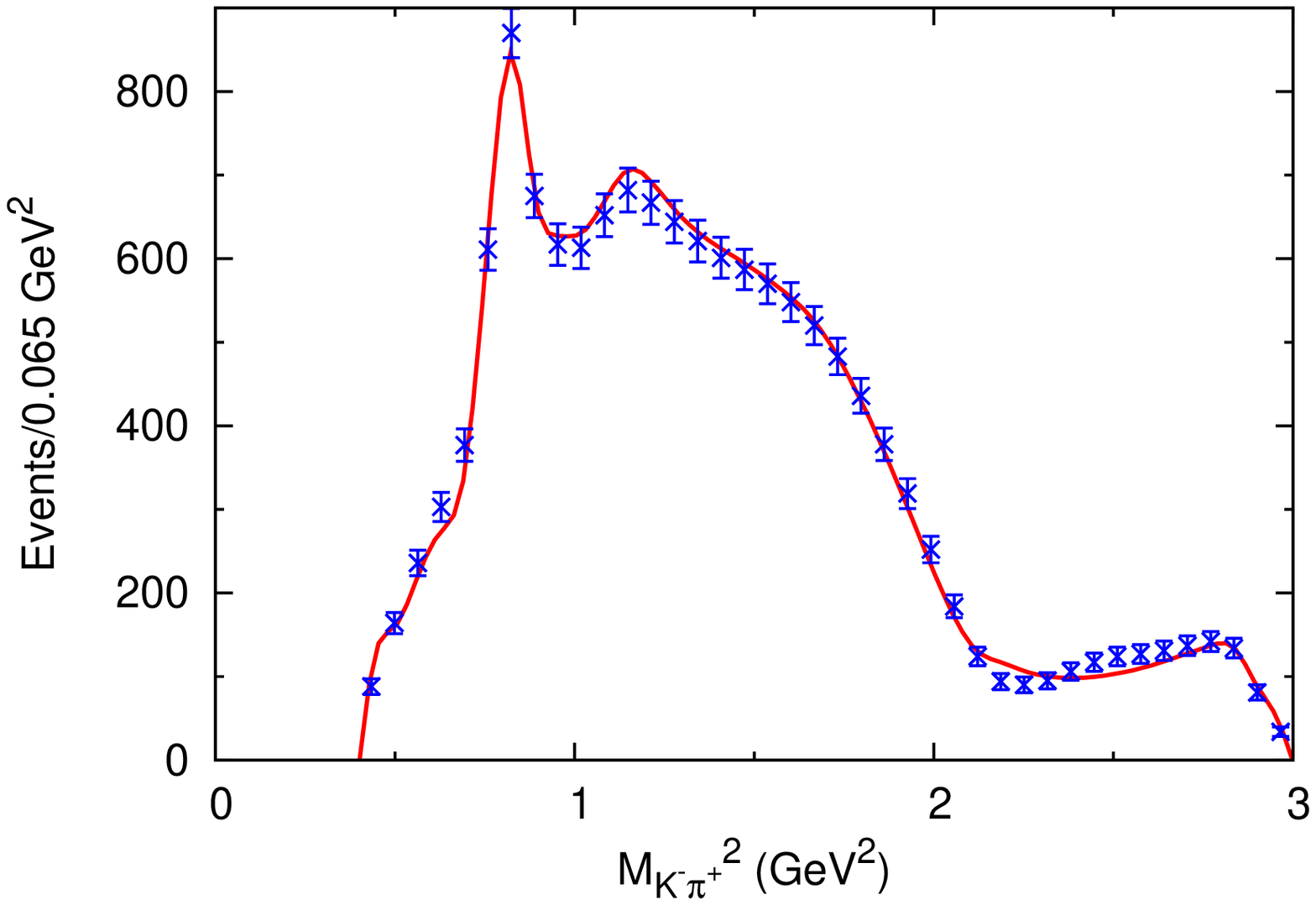}
\hspace{5mm}
\includegraphics[width=0.45\textwidth]{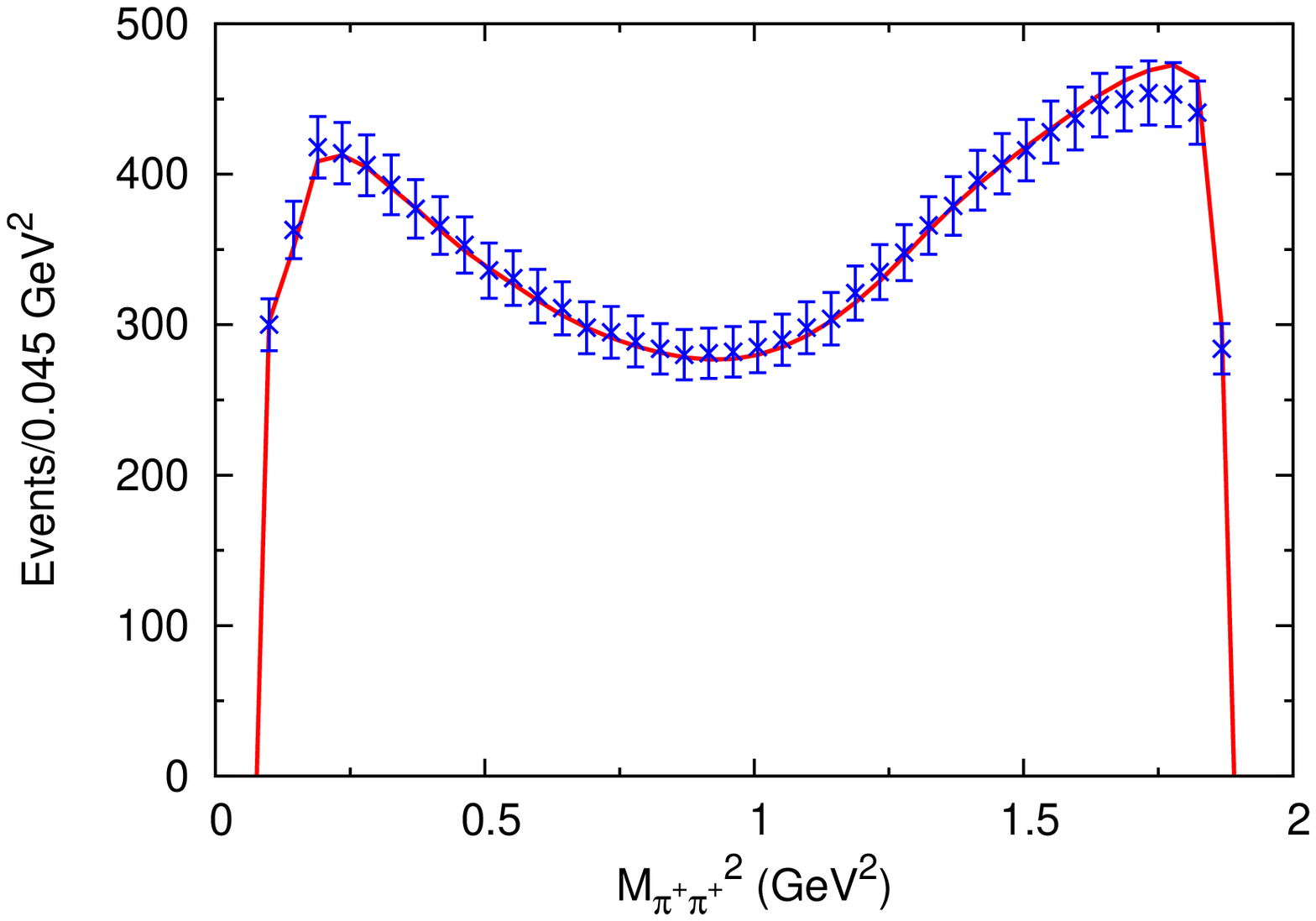}
\caption{\label{fig:proj-full}
(Color online) 
The $K^-\pi^+$ ($\pi^+\pi^+$) squared invariant mass spectrum of the Full model compared
with the pseudodata from the E791 isobar model in the left (right) panel.
}
\end{figure}
In order to see the quality of the fits more clearly, we show the
$\chi^2_j$ distribution at all of the bins in Fig.~\ref{fig:error}.
These figures show that all of our models fit the pseudodata rather
precisely.
Most of the bins are fitted with $\chi^2_j<0.5$, and 
$\chi^2_j$ exceeds 1 at only a small number of the bins.
Yet again, the Full and Z models clearly show a
better performance in the fit than the Isobar model does.
The quality of the fit can also be shown in the projection of the Dalitz
plot distribution onto the $M^2_{K^-\pi^+}$ or $M^2_{\pi^+\pi^+}$
distributions, as presented in Figs.~\ref{fig:proj-full}
for the Full model as a representative.

What is the main reason that the Full and Z models fit the pseudodata better
than the Isobar model ?
Because only the former models have 
the $R_i^{12}\bar{K}$ (i.e., $\rho^+ \bar K^0$) channels
and the hadronic FSI, 
one may guess either or both of these dynamical contents are
responsible.
To address this point, we introduce 
the ``Z (without $\rho$)'' model, that is, the Z-model
with couplings to the $R_i^{12}\bar{K}$ channel turned off.
By fitting the pseudodata with the Z (without $\rho$) model, 
we examine if the rescattering effects can improve the $\chi^2$.
The $\chi^2$-value is shown in Table~\ref{tab:chi2}, and 
is significantly better than that of the Isobar model.
We note that the Isobar model and the Z(without $\rho$) model 
have the same number of adjustable parameters in the fits.
Thus, the quality of the fit is improved just by 
including the rescattering due to the Z-diagrams, 
and the inclusion of the $R_i^{12}\bar{K}$ channels
further improves the fit substantially.

We remark that we obtained the reasonable fits to the pseudodata 
without adjusting the parameters associated with the
two-pseudoscalar-meson interactions.
On the other hand,
as mentioned already, most of the previous analyses varied 
some Breit-Wigner parameters in fitting their $D^+\to K^-\pi^+\pi^+$
Dalitz plot data.
A common claim in those analyses~\cite{e791-prl,e791,focus,cleo}
was that the width of $\bar K^*_0(1430)$ obtained in their fits, 
$\sim$170~MeV, was significantly smaller than those from the PDG and the
LASS analysis, $\sim$270~MeV.
In our model, there are two poles associated with 
$\bar K^*_0(1430)$ as shown in Table~\ref{tab:pik-pole},
and they are on different Riemann sheets, the branch point of which
is the $\eta' \bar K$ threshold.
This two-pole structure is a coupled-channel effect.
One of the poles is close to the PDG
value, while the other one has a rather broad width.
Thus, in our analysis, we did not need 
$K^*_0(1430)$ with a narrower width to obtain the reasonable fits to the 
$D^+\to K^-\pi^+\pi^+$ Dalitz plot pseudodata.

\begin{figure}[t]
\includegraphics[width=0.33\textwidth]{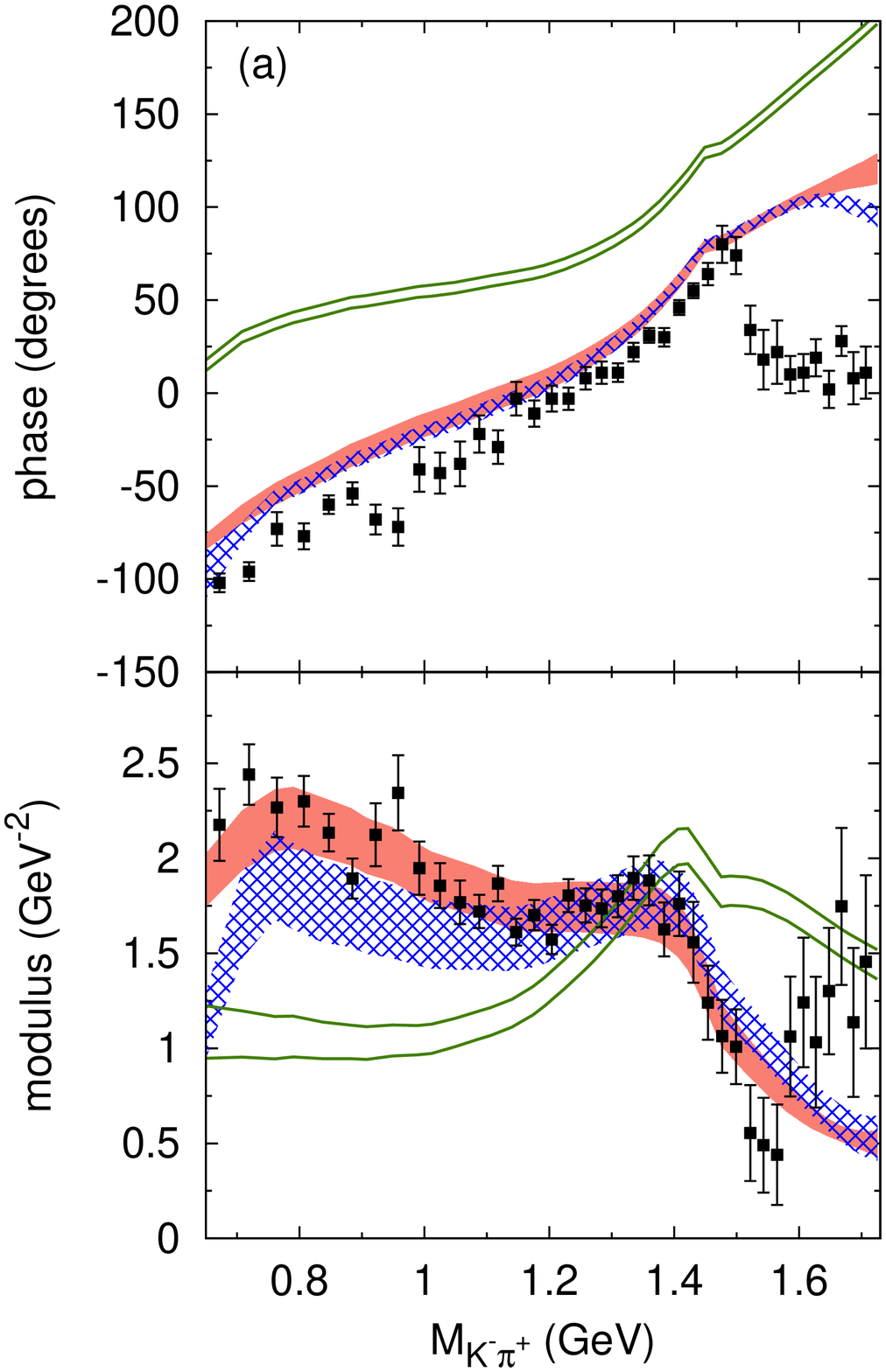}
\hspace{-3mm}
\includegraphics[width=0.33\textwidth]{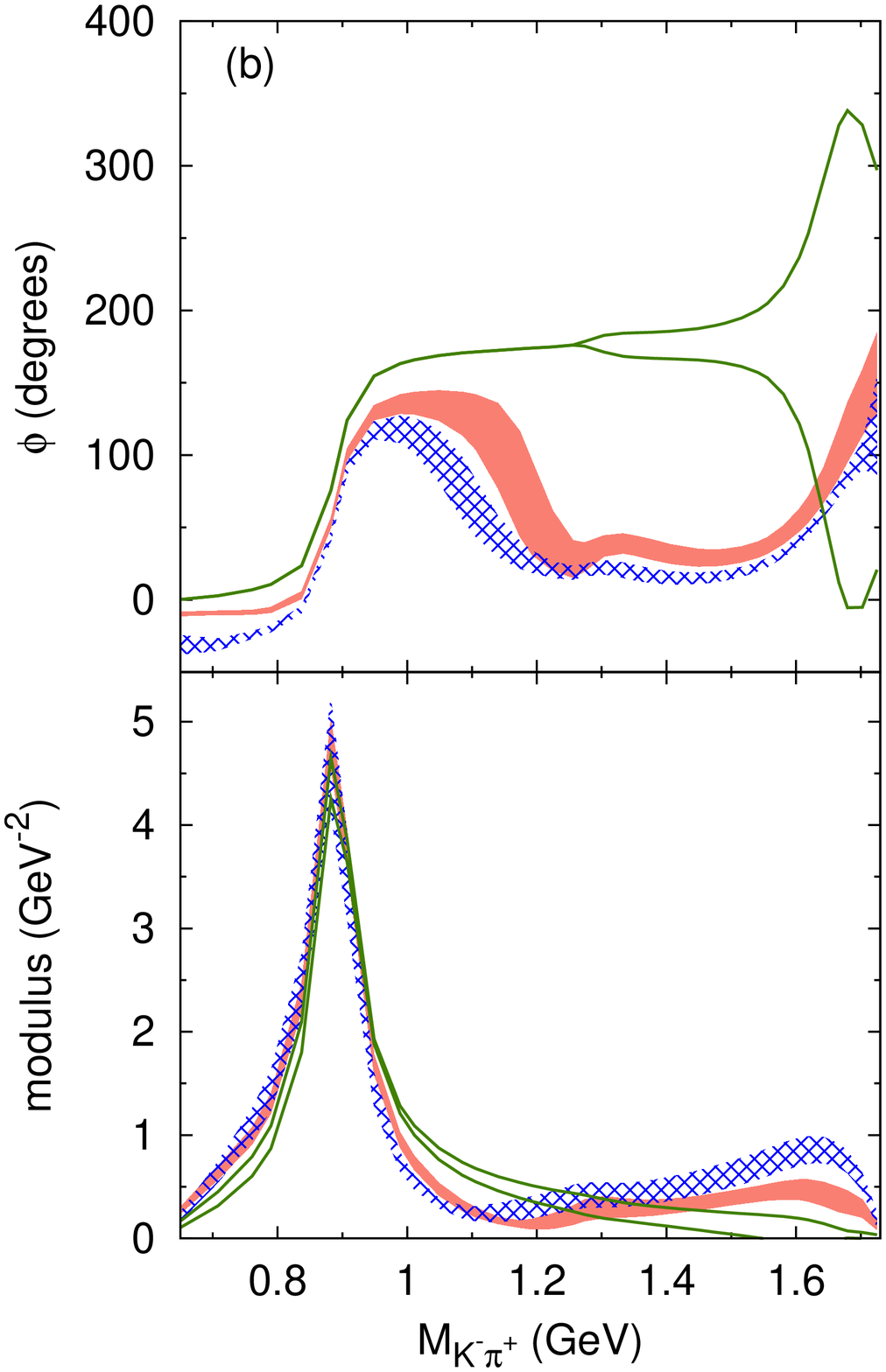}
\hspace{-3mm}
\includegraphics[width=0.33\textwidth]{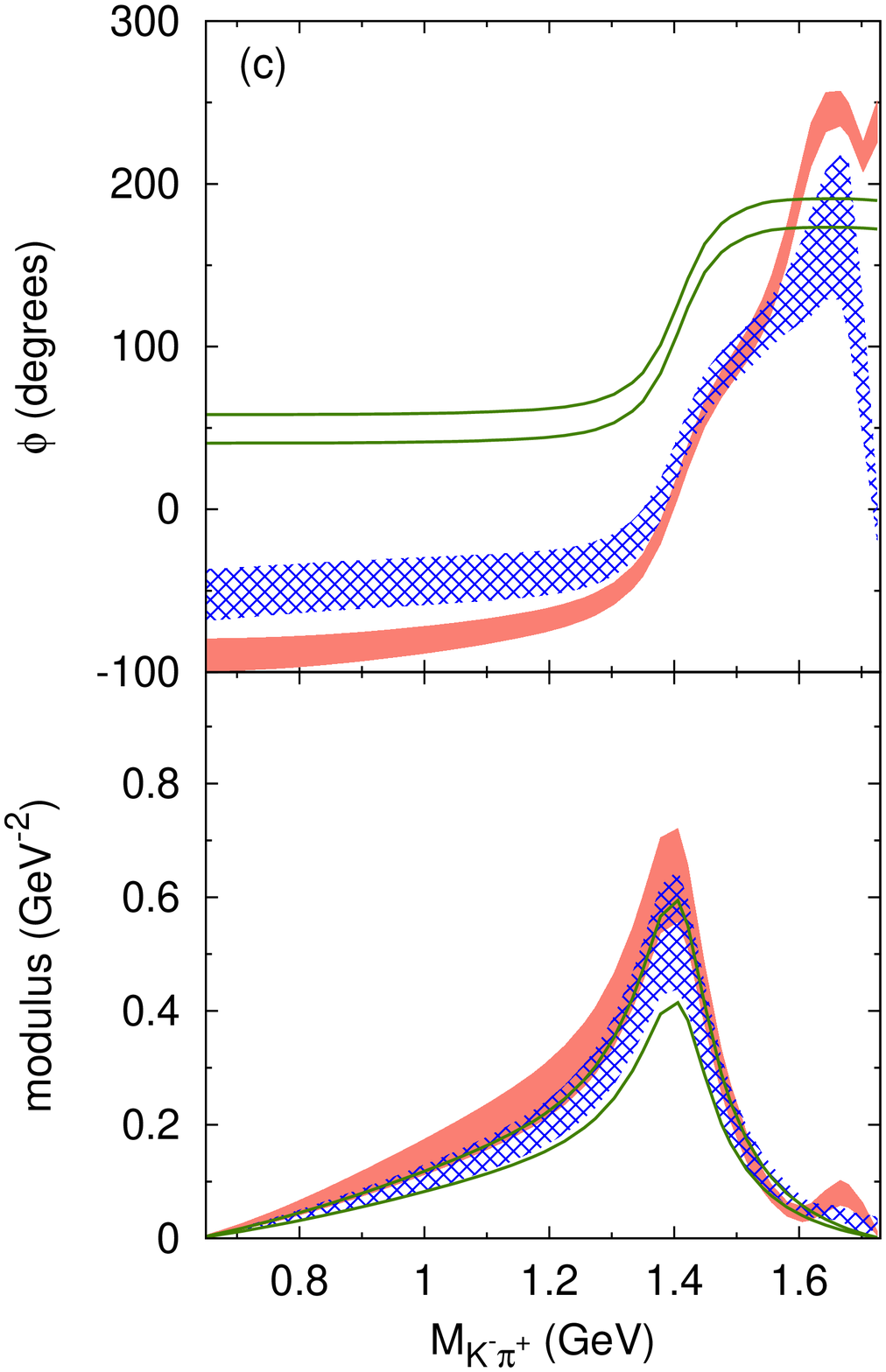}

\vspace{3mm}
\includegraphics[width=0.33\textwidth]{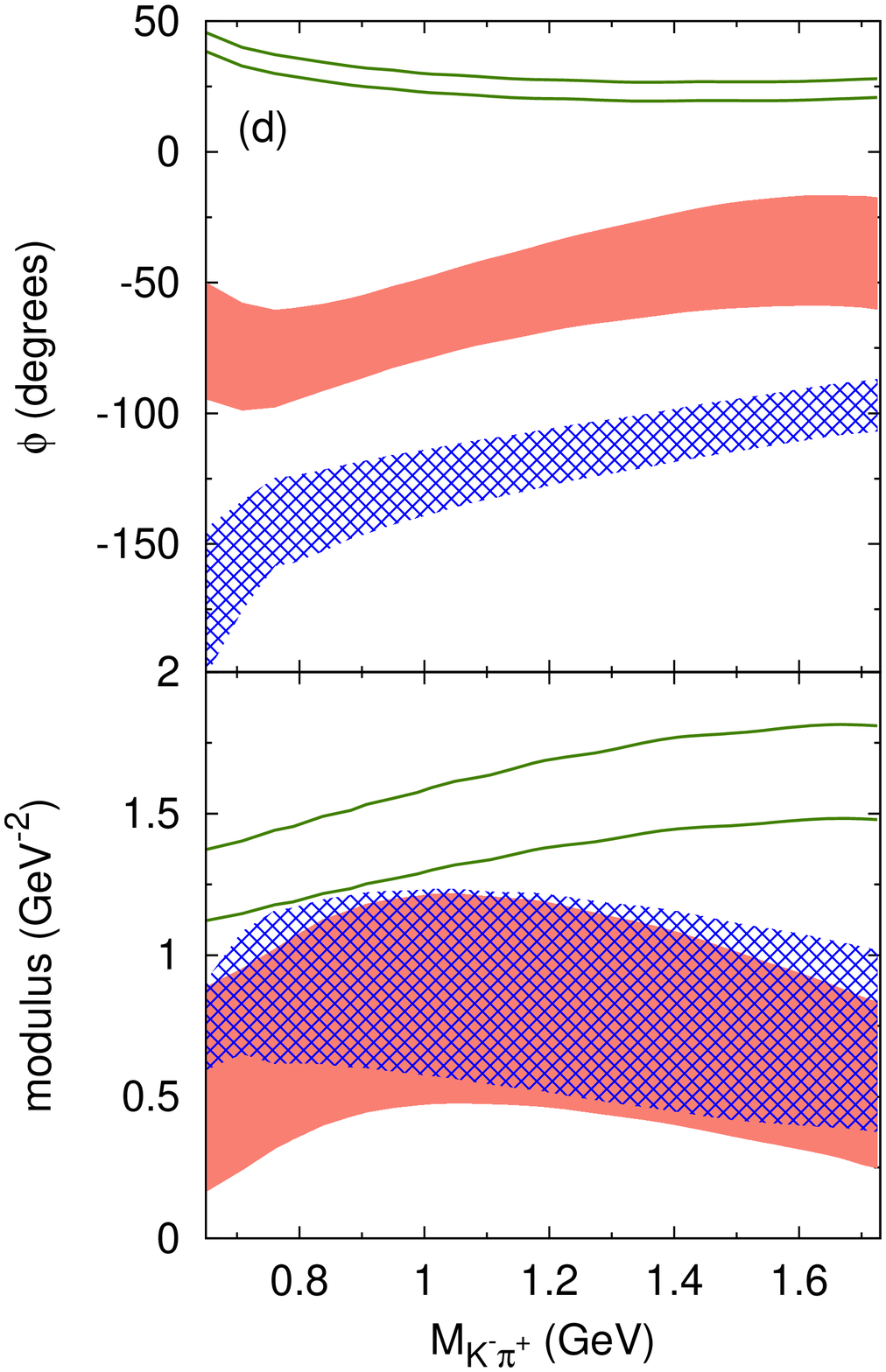}
\hspace{-3mm}
\includegraphics[width=0.33\textwidth]{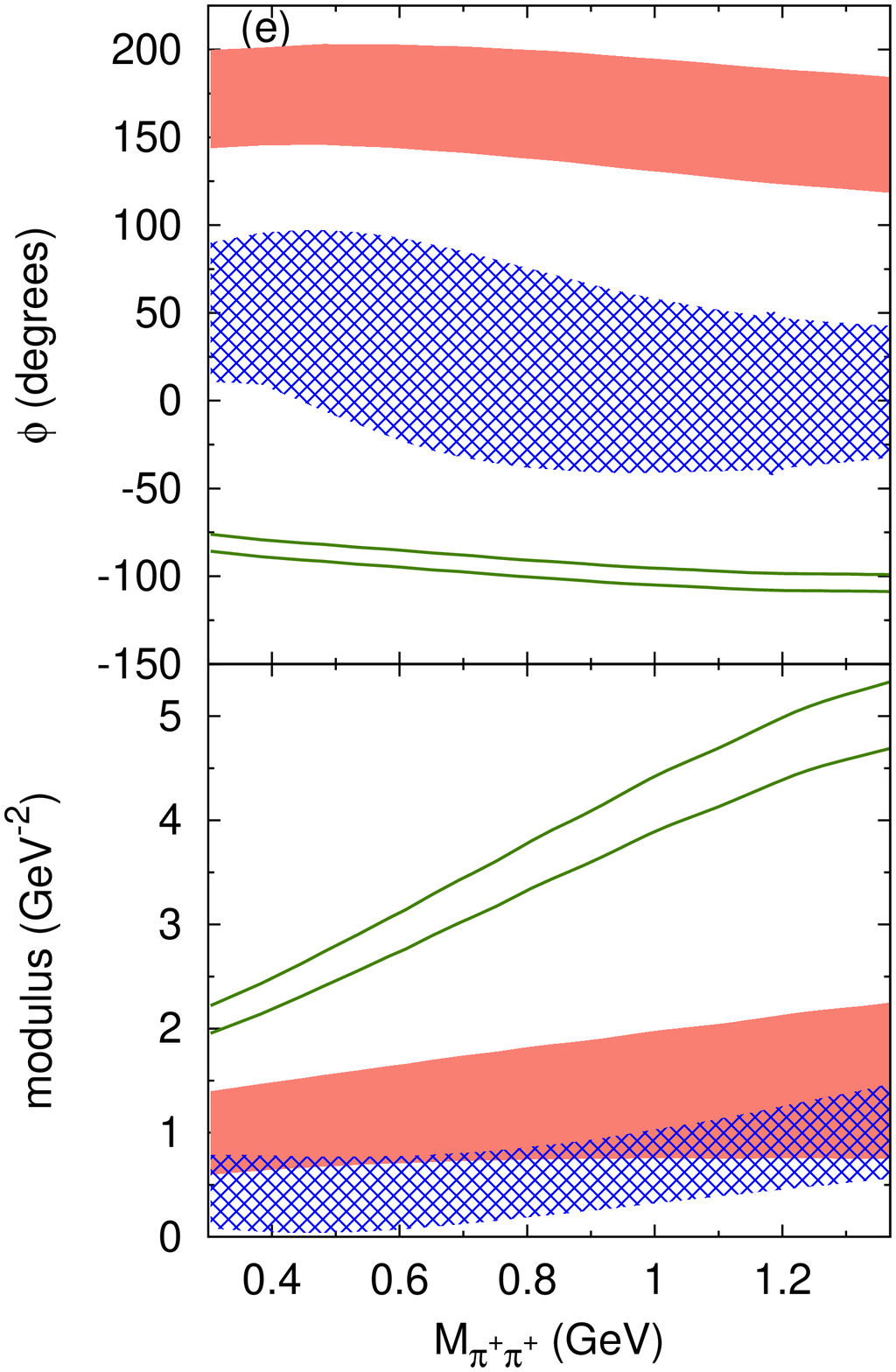}
\caption{\label{fig:amp-err} 
(Color online) 
Phase (upper) and modulus (lower) of the partial-wave amplitudes for
(a) $(K^-\pi^+)_S^{I=1/2}\pi^+$, 
(b) $(K^-\pi^+)_P^{I=1/2}\pi^+$,
(c) $(K^-\pi^+)_D^{I=1/2}\pi^+$,
(d) $(K^-\pi^+)_S^{I=3/2}\pi^+$, 
(e) $(\pi^+\pi^+)_S^{I=2}K^-$.
The red solid, the blue cross-hatched, and the green bordered
bands are from the Full, Z, and Isobar models, respectively;
the band widths represents the errors. 
For the definition of the amplitudes, see the text.
In the panel (a), we also show the $(K^-\pi^+)_S\pi^+$ amplitude from
the MIPWA of the E791 Collaboration~\cite{e791}
as the black squares with error bars.
}
\end{figure}
Next we present partial-wave decay amplitudes defined by
Eq.~(\ref{eq:decay-amp}) at $\hat q\cdot \hat p_c = 1$
($\bm{q}$: relative momentum of the $ab$-pair in their CM frame)
and the summation $\sum_{{\cal R},{\cal R}'}$ being
replaced by 
$\sum_{{\cal R}',{\cal R}}\delta_{s_{\cal R},L}
\delta_{s_{\cal R'},L}
\delta_{t_{\cal R},I}\delta_{t_{\cal R'},I}$
where $\{L,I\}$ specify the partial wave.
These amplitudes correspond to Eq.~(3) of Ref.~\cite{e791}
at $\cos\theta=1$, and in the same reference, 
the E791 Collaboration presented their amplitudes.
Thus we can compare our amplitudes with those from the E791 MIPWA.
We denote a partial wave by ``$(ab)^I_{L}c$'' in which
a two-pseudoscalar-meson pair $ab$ has the total angular momentum $L$ 
and the total isospin $I$.
When $ab$ is written with a charge state, $(ab)^I_{L}$ is understood to
be the total isospin $I$ state projected onto the particular charge state,
e.g., $(K^-\pi^+)^{I=1/2}_{S}\pi^+$.
We note that a partial-wave amplitude of a coupled-channel model generally
contains all the partial waves considered in the model
as intermediate states.
Therefore, we refer to an amplitude with
$(ab)^I_{L}c$ in the final state as the $(ab)^I_{L}c$ amplitude.
The partial-wave amplitudes for 
$(K^-\pi^+)_S^{I=1/2}\pi^+$,
$(K^-\pi^+)_P^{I=1/2}\pi^+$,
$(K^-\pi^+)_D^{I=1/2}\pi^+$, 
$(K^-\pi^+)_S^{I=3/2}\pi^+$, and 
$(\pi^+\pi^+)_S^{I=2}K^-$ 
are presented in Figs.~\ref{fig:amp-err}(a)-(e), respectively.
The partial-wave amplitudes from the Full, Z, and Isobar models are shown 
with their error bands as a function of 
$M_{ab}\equiv\sqrt{(E_a+E_b)^2-(\bm{p}_a+\bm{p}_b)^2}$.
As a whole, the Full and Z models are similar in the amplitudes
while the Isobar model is rather different, particularly in 
$(K^-\pi^+)_S^{I=1/2}\pi^+$,
$(K^-\pi^+)_S^{I=3/2}\pi^+$, and 
$(\pi^+\pi^+)_S^{I=2}K^-$.
Because the models maintain the Watson theorem
when the hadronic rescattering is absent
(see the note in Sec.~\ref{sec:setup}),
the phases (the upper panels of Fig.~\ref{fig:amp-err})
from the Isobar model in the elastic region are essentially the
same as those in Figs.~\ref{fig:pik} and \ref{fig:pipi-pik} 
up to overall constant shifts.
The difference in the $M_{K^-\pi^+}$-dependence of 
the phases between the Isobar
model and the Full (or Z) model is purely the effect of the hadronic
rescattering. 
To put it the other way around, 
the Isobar model does not have a degree of freedom to change the
phases using the rescattering in order to obtain an optimal fit, 
which might have led to the rather different solution.

In Fig.~\ref{fig:amp-err}(a), we also show the 
$(K^-\pi^+)_S\pi^+$ amplitude from the E791 MIPWA~\cite{e791},
denoted by 
$[(K^-\pi^+)_S\pi^+]^{\rm\scriptscriptstyle E791}_{\rm\scriptscriptstyle MIPWA}$ 
hereafter.
The modulus in the figure is, in their notation, 
$c\times F^0_D(\sqrt{s})$, and numerical values for 
$c$ and $F^0_D(\sqrt{s})$ are tabulated in Table~III of Ref.~\cite{e791}.
Interestingly, the $M_{K^-\pi^+}$-dependence of 
the phases from the Full and Z models are in a very good
agreement with those of the MIPWA for $M_{K^-\pi^+}\ltap 1.5$~GeV.
The Full model also agrees with the modulus of
$[(K^-\pi^+)_S\pi^+]^{\rm\scriptscriptstyle E791}_{\rm\scriptscriptstyle MIPWA}$ 
in the elastic region.
Because
$[(K^-\pi^+)_S\pi^+]^{\rm\scriptscriptstyle E791}_{\rm\scriptscriptstyle MIPWA}$ 
implicitly contains all the partial-wave amplitudes other than 
$(K^-\pi^+)_P^{I=1/2}\pi^+$ and $(K^-\pi^+)_D^{I=1/2}\pi^+$, 
for a more meaningful comparison, 
we might need to compare it with
the coherent sum of 
the $(K^-\pi^+)_S^{I=1/2}\pi^+$,
$(K^-\pi^+)_S^{I=3/2}\pi^+$ and $(\pi^+\pi^+)_S^{I=2}K^-$ amplitudes of
our models.
However, because the $(\pi^+\pi^+)_S^{I=2}K^-$ amplitude depends on
$M_{\pi^+\pi^+}$ while the others on $M_{K^-\pi^+}$,
they cannot be simply summed to obtain the counterpart of 
$[(K^-\pi^+)_S\pi^+]^{\rm\scriptscriptstyle E791}_{\rm\scriptscriptstyle MIPWA}$. 
Still, we confirmed that
the $M_{K^-\pi^+}$-dependence of the phase of the 
$(K^-\pi^+)_S^{I=1/2}\pi^+$ amplitude of the Full and Z models does not 
significantly change even after 
the $(K^-\pi^+)_S^{I=3/2}\pi^+$ amplitude is coherently added.
Thus, our coupled-channel models explain
the gap between the phase of 
$[(K^-\pi^+)_S\pi^+]^{\rm\scriptscriptstyle E791}_{\rm\scriptscriptstyle MIPWA}$
and that of
the LASS $(K^-\pi^+)_S^{I=1/2}\pi^+$ amplitude
in a way qualitatively different from the previous explanation.
Edera et al.~\cite{pik_I=3/2_model} and the FOCUS $K$-matrix model
analysis filled the gap with a rather large (more than 100\%)
destructive interference between 
the $(K^-\pi^+)_S^{I=1/2}\pi^+$ and $(K^-\pi^+)_S^{I=3/2}\pi^+$ 
amplitudes, without an explicit consideration of the hadronic rescattering. 
Our coupled-channel models, on the other hand, fill the gap with the hadronic
rescattering, and have a moderately 
destructive interference between 
the $(K^-\pi^+)_S^{I=1/2}\pi^+$ and $(K^-\pi^+)_S^{I=3/2}\pi^+$
amplitudes (see Table~\ref{tab:FF}).

We notice in Fig.~\ref{fig:amp-err} that
the $(K^-\pi^+)_S^{I=3/2}\pi^+$ and $(\pi^+\pi^+)_S^{I=2}K^-$
partial-wave amplitudes have relatively large errors.
Because we analyzed the data of the single charge state,
it may be difficult to separate the different isospin states 
with a good precision. 
This situation would be improved by analyzing data of different charge
states, i.e., $D^+\to K^-\pi^+\pi^+$ and $D^+\to K_S^0\pi^+\pi^0$,
in a combined manner.
We also note that even though the phases of
the $(K^-\pi^+)_P^{I=1/2}\pi^+$ amplitude from the Isobar model
have very large errors for $M_{K^-\pi^+}\gtap 1.5$~GeV, 
this is simply because the absolute value of the amplitude is very small.

Now we look into 
the fraction of each channel's contribution (fit fraction).
In an isobar model that describes 
a $D^+$-meson decay as 
$D^+\to \sum_{R} Rc\to abc$
where $R$ is expressed by a Breit-Wigner function,
the fit fractions of different $Rc$ contributions
can be straightforwardly defined,
and are often presented in the previous analyses.
However, in a model like ours where the resonances are described as
poles in unitary scattering amplitudes, 
the fit fraction of a certain resonance contribution is not so
straightforwardly defined, because the scattering amplitude generally
contains more than a single resonance, as we have seen in
Tables~\ref{tab:pik-pole} and \ref{tab:pipi-pole}.
Furthermore, the amplitude also contains nonresonant contributions.
There is no unambiguous way to single out a certain resonance contribution.
Therefore, we present the fit fractions of contributions from different 
partial-wave amplitudes.
We calculate the $(ab)^I_{L}c$ partial width using the
partial-wave amplitudes presented in the above paragraph,
with the angular ($\hat q\cdot \hat p_c$) dependence being restored and 
the Bose symmetrization being taken into account as in 
Eq.~(\ref{eq:cyclic-sum}).
The fit fraction is then naturally defined as
the $(ab)^I_{L}c$ partial width divided by the total width.

\begin{table}[t]
\caption{\label{tab:FF} 
Fit fractions (\%) from each of the models fitted to
the $D^+\to K^-\pi^+\pi^+$ pseudodata.
See the text for the definition of the fit fraction.
The ``S-waves'' fit fraction is
the coherent sum of the 
$(K^-\pi^+)_S^{I=1/2}\pi^+$, $(K^-\pi^+)_S^{I=3/2}\pi^+$, and
$(\pi^+\pi^+)_S^{I=2}K^-$ fit fractions.
The numbers in the parentheses are not included in the ``Sum'' in the
last row.
The fit fractions from the E791 Isobar model~\cite{e791},
the FOCUS $K$-matrix model analysis~\cite{focus},
%the FOCUS MIPWA~\cite{focus2009}, 
and the CLEO QMIPWA~\cite{cleo}
are also presented.
The fit fractions in $\langle\rangle$ are obtained by the incoherent sum of
 different resonance contributions in the same partial wave, and the
 errors are added in quadrature.
}    
{\footnotesize
\begin{ruledtabular}
\begin{tabular}{lcccccc}
                           &\multirow{2}{*}{Full}    &\multirow{2}{*}{Z}               & \multirow{2}{*}{Isobar}         &E791~\cite{e791}       	      &FOCUS~\cite{focus}    &CLEO~\cite{cleo}		       \\
                           &	    &               &          &isobar      	      &$K$-matrix    &QMIPWA		       \\\hline
$(K^-\pi^+)_S^{I=1/2}\pi^+$&$ 95.9\pm   7.3$&$ 78.7\pm   4.3$&$ 68.9\pm   4.4$&$33.8\pm\langle 10.8\rangle$   &$207.25\pm 25.45     $&---			       \\
$(K^-\pi^+)_P^{I=1/2}\pi^+$&$ 15.3\pm   1.5$&$ 16.2\pm   2.1$&$ 13.7\pm   0.7$&$16.2\pm\langle 1.6\rangle$    &$\langle 15.99\pm 1.18\rangle$&$\langle 10.076\pm 0.47\rangle$\\
$(K^-\pi^+)_D^{I=1/2}\pi^+$&$  0.5\pm   0.1$&$  0.4\pm   0.1$&$  0.3\pm   0.1$&$0.6\pm 0.1$  		      &$0.39\pm 0.09        $&$0.204\pm 0.040$	       \\
$(K^-\pi^+)_S^{I=3/2}\pi^+$&$ 27.9\pm  27.5$&$ 34.2\pm  25.7$&$109.8\pm  27.8$&        ---       	      &$40.50\pm 9.63       $&---			       \\
$(\pi^+\pi^+)_S^{I=2}K^-$  &$ 21.1\pm  19.3$&$  5.1\pm   5.6$&$174.6\pm  28.9$&---			      &--                    &$15.5\pm 2.8$		       \\
Background &---&---&---                                                       &$17.2\pm 5.3$      	      &--                    &---			       \\
S-waves                    &($81.7\pm  0.8$)&($82.3\pm  0.8$)&($81.2\pm  0.7$)&($79.8\pm\langle 12.0\rangle$) &$(83.23\pm 1.50)$     &$\langle 97.1\pm 3.9\rangle$    \\\hline
Sum                        & 160.7          & 134.5	     &367.3           &66.3                           &264.13                & 122.9                         \\
%%
%                           &Z(without $\rho$)       E791 MIPWA     FOCUS MIPWA
%$(K^-\pi^+)_S^{I=1/2}\pi^+$&$116.1\pm   5.5$         ---	       ---			       
%$(K^-\pi^+)_P^{I=1/2}\pi^+$&$ 15.2\pm   0.8$ 	    <13.1\pm 0.6>     $\langle 14.11\pm 0.71\rangle$   
%$(K^-\pi^+)_D^{I=1/2}\pi^+$&$  0.4\pm   0.1$ 	    0.2\pm 0.1         $0.58\pm 0.1$ 	               
%$(K^-\pi^+)_S^{I=3/2}\pi^+$&$141.1\pm  49.4$ 	      ---	       ---
%$(\pi^+\pi^+)_S^{I=2}K^-$  &$135.3\pm  39.6$ 	      ---	       ---
%						      ---	       ---       
%S-waves                    &$ 83.1\pm   0.8$ 	    78.6\pm 1.4     $80.24\pm 1.38$
%Sum                        & 408.0	       	    91.9                94.93			       
\end{tabular}
\end{ruledtabular}
}
\end{table}
The fit fractions defined in the above paragraph are presented in
Table~\ref{tab:FF} for the Full, Z, and Isobar models.
The incoherent sum of the fit fractions in the last row
is not necessarily 100\% because of the
interferences between different contributions in the different rows of the table.
We can see that the Full and Z models agree fairly well within the
errors, although the errors are rather large for 
the $(K^-\pi^+)_S^{I=3/2}\pi^+$ and
$(\pi^+\pi^+)_S^{I=2}K^-$ fit fractions. 
On the other hand, the Isobar model has quite different
$(K^-\pi^+)_S^{I=3/2}\pi^+$ and
$(\pi^+\pi^+)_S^{I=2}K^-$ fit fractions
that are rather large,
leading to 
the large incoherent sum (367\%).
This indicates a very destructive interference between 
the $(K^-\pi^+)_S^{I=3/2}\pi^+$ and
$(\pi^+\pi^+)_S^{I=2}K^-$ partial waves.
These results are 
consistent with what we can expect from the partial-wave
amplitudes shown in Fig.~\ref{fig:amp-err}.
We also show the coherent sum of the 
$(K^-\pi^+)_S^{I=1/2}\pi^+$, $(K^-\pi^+)_S^{I=3/2}\pi^+$, and
$(\pi^+\pi^+)_S^{I=2}K^-$ fit fractions, 
labeled ``S-waves''.
Interestingly, 
all three models have consistent ``S-waves'' fit fractions
within the drastically reduced errors.
Therefore, the data used in our analysis can constrain
the ``S-waves'' fit fraction rather well,
while they cannot well constrain
the $(K^-\pi^+)_S^{I=3/2}\pi^+$ and
$(\pi^+\pi^+)_S^{I=2}K^-$ fit fractions individually.

In Table~\ref{tab:FF}, we also list fit fractions from 
the previous analyses done by the experimental groups for a
comparison with our result.
Although each of the experimental groups obtained several models through their
analyses, we do not try to exhaust all the models in the comparison.
Rather we pick up some cases that are particularly interesting to
compare with our results.
Here, we tabulated three analyses results in Table~\ref{tab:FF}: 
the E791 Isobar model~\cite{e791} on which our pseudodata are based,
the FOCUS $K$-matrix model analysis~\cite{focus},
and the CLEO QMIPWA~\cite{cleo}.
We note that the definition for the fit fraction used by the E791 in Ref.~\cite{e791}
is different from what used here; 
the partial-wave amplitudes are not Bose-symmetrized 
in calculating the fit fractions, while our and the CLEO's~\cite{cleo} fit fractions are from 
Bose-symmetrized amplitudes.
However, we actually tabulated in Table~\ref{tab:FF} 
the fit fractions for the E791 Isobar model
calculated by ourselves with our definition, and their errors are assumed to be the same
as those given in Ref.~\cite{e791}.
In the original papers, the experimental groups presented 
the fit fractions of each of resonances considered. 
In order to (roughly) compare their results with ours, we sum the
resonance contributions in the same partial wave incoherently,
and the errors are summed in quadrature.
The numbers enclosed by $\langle\rangle$ are obtained by the incoherent
(quadrature) sum.
For the $({K^-}\pi^+)_P^{I=1/2}\pi^+$ fit fraction
obtained by the incoherent sum,
the $\bar K^*(892)$ fit fraction
dominates, and thus the interference effect would not be so large.
A general trend in Table~\ref{tab:FF}
is that all the analyses listed,
and the E791~\cite{e791} and FOCUS~\cite{focus2009} MIPWA results 
(not listed on the table) as well,
are in fairly good agreement on the 
$(K^-\pi^+)_P^{I=1/2}\pi^+$,
$(K^-\pi^+)_D^{I=1/2}\pi^+$, and 
``S-waves'' fit fractions.
Although the ``S-waves'' fit fraction from the CLEO QMIPWA is somewhat larger than the others, 
it is probably because 
the $(\pi^+\pi^+)_S^{I=2}K^-$ and 
$\bar K^*_0(1430)$ fit fractions have been added incoherently.
The other fit fractions are rather largely dependent on each of the analyses.
For example, as we have mentioned, the FOCUS $K$-matrix model analysis
gives rather large 
$(K^-\pi^+)_S^{I=1/2}\pi^+$ and 
$(K^-\pi^+)_S^{I=3/2}\pi^+$ fit fractions that interfere
very destructively while 
the interference is moderate in the Full and Z models.

\begin{table}[t]
\caption{\label{tab:bareFF} 
{\it Bare} $D^+\to K^-\pi^+\pi^+, \bar{K}^0\pi^0\pi^+$ decay fit fractions (\%).
}    
\begin{ruledtabular}
\begin{tabular}{lccc}
                           &Full	    &Z               & Isobar                      	      \\\hline
$(\bar{K}\pi)_S^{I=1/2}\pi^+$  &$ 45.0\pm   9.2$&$ 39.8\pm   5.0$&---\\
$(\bar{K}\pi)_P^{I=1/2}\pi^+$  &$ 13.1\pm   2.2$&$  5.8\pm   1.1$&---\\
$(\bar{K}\pi)_D^{I=1/2}\pi^+$  &$  0.3\pm   0.2$&$  0.1\pm   0.1$&---\\
$(\pi^+\pi^0)_P^{I=1}\bar{K}^0$&$ 16.0\pm   0.9$&$ 42.7\pm   0.7$&---\\
$(\bar{K}\pi)_S^{I=3/2}\pi$    &$  8.7\pm  11.7$&$  7.7\pm   8.2$&---\\
$(\pi\pi)_S^{I=2}\bar{K}$      &$ 16.8\pm  19.1$&$  3.9\pm   4.4$&---\\
%
%                               &Z(without $\rho$)             	      \\\hline
%$(\bar{K}\pi)_S^{I=1/2}\pi^+$  &$ 12.4\pm   1.8$\\
%$(\bar{K}\pi)_P^{I=1/2}\pi^+$  &$  4.5\pm   0.7$\\
%$(\bar{K}\pi)_D^{I=1/2}\pi^+$  &$  0.2\pm   0.1$\\
%$(\pi^+\pi^0)_P^{I=1}\bar{K}^0$&---\\
%$(\bar{K}\pi)_S^{I=3/2}\pi$    &$ 36.2\pm  10.9$\\
%$(\pi\pi)_S^{I=2}\bar{K}$      &$ 46.7\pm  13.8$\\
%
\end{tabular}
\end{ruledtabular}
\end{table}
With our coupled-channel analysis, it is interesting to examine  
{\it bare} fit fractions defined as follows.
We first calculate a bare $(ab)^I_{L}c$ partial width
using the decay amplitude in which 
all contributions from the rescattering
processes [the second term of Eq.~(\ref{eq:dressed-g});
Fig.~\ref{fig:d-decay}(b)] are omitted. 
In our coupled-channel description of the $D^+\to K^-\pi^+\pi^+$ decay,
we consider not only $K^-\pi^+\pi^+$ but also $\bar{K}^0\pi^0\pi^+$ in 
the hadronic intermediate states.
Therefore, when calculating the bare partial width, 
we consider both of the states in the final state sum.
(Precisely speaking, $\bar{K}^0\eta'\pi^+$, $\bar{K}^0\bar{K}^0K^+$, and the
effective inelastic channels in the $\pi\bar{K}$ $p$- and $d$-waves 
are also included in the intermediate states; their partial widths are
rather small.)
Then,
the bare fit fraction for a given $(ab)^I_{L}c$
is defined by the bare $(ab)^I_{L}c$ partial width
divided by the sum of the bare partial widths for all the considered
$(ab)^I_{L}c$.
Thus the sum of all the bare fit fractions is 100\% by definition.
The result is presented in Table~\ref{tab:bareFF}.
We leave the column for the Isobar model blank.
This is because 
the $D^+\to {\cal R}c$ vertices of the Isobar model implicitly includes
the rescattering effect, and
we cannot eliminate the effect to extract the bare vertices.

A remarkable feature in Table~\ref{tab:bareFF} is the large
fit fraction of $(\pi^+\pi^0)_P^{I=1}\bar{K}^0$ in which $\rho(770)$
plays a major role.
This is particularly true for the Z model.
This fit fraction did not appear in Table~\ref{tab:FF}
because $(\pi^+\pi^0)_P^{I=1}\bar{K}^0$ can appear only in
intermediate states of the $D^+\to K^-\pi^+\pi^+$ decay,
and its contribution is genuinely a coupled-channel effect.
The previous $D^+\to K^-\pi^+\pi^+$ analyses cannot see this effect
because they did not explicitly consider three-body dynamics.
\begin{figure}[t]
\includegraphics[width=0.45\textwidth]{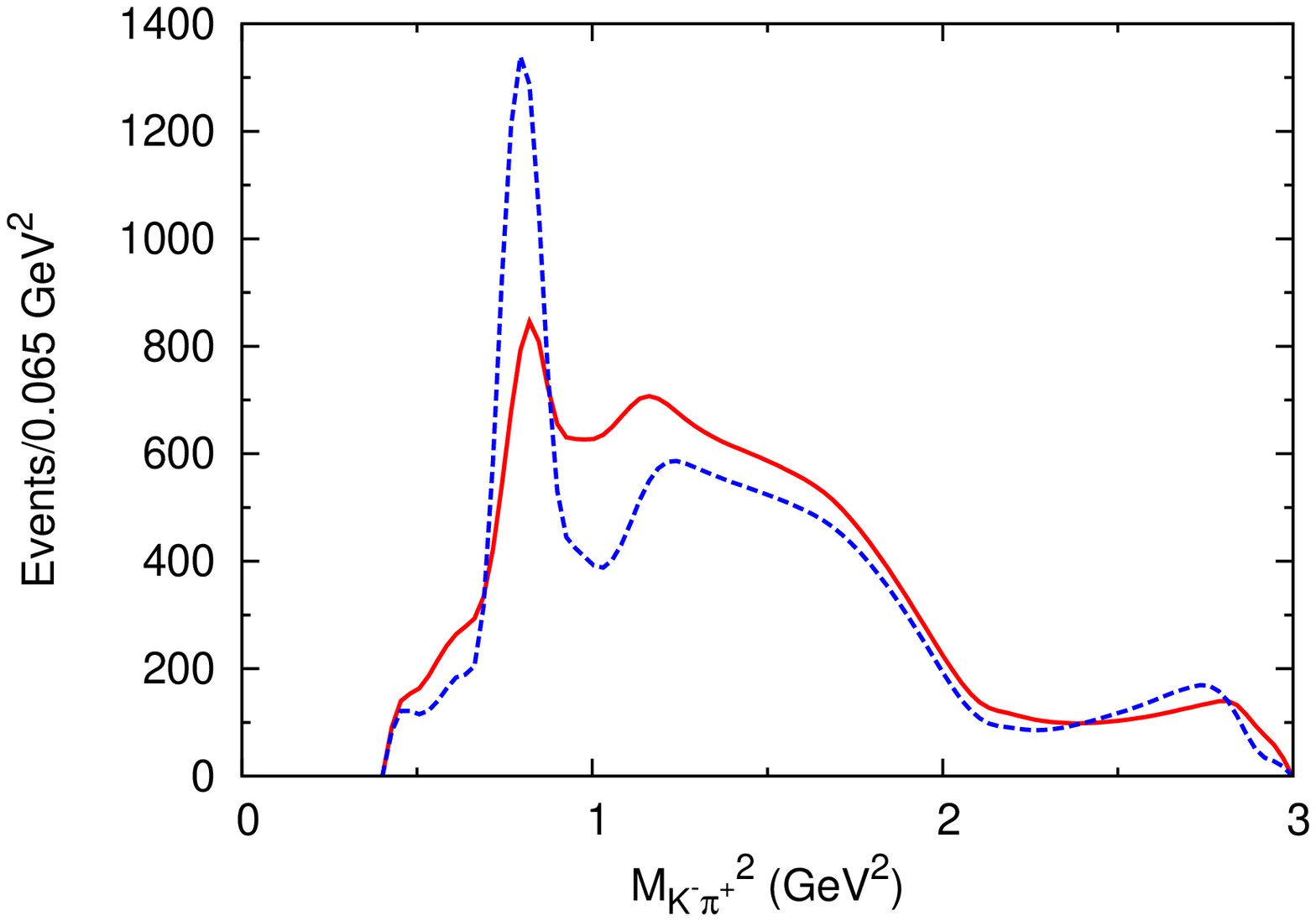}
\hspace{5mm}
\includegraphics[width=0.45\textwidth]{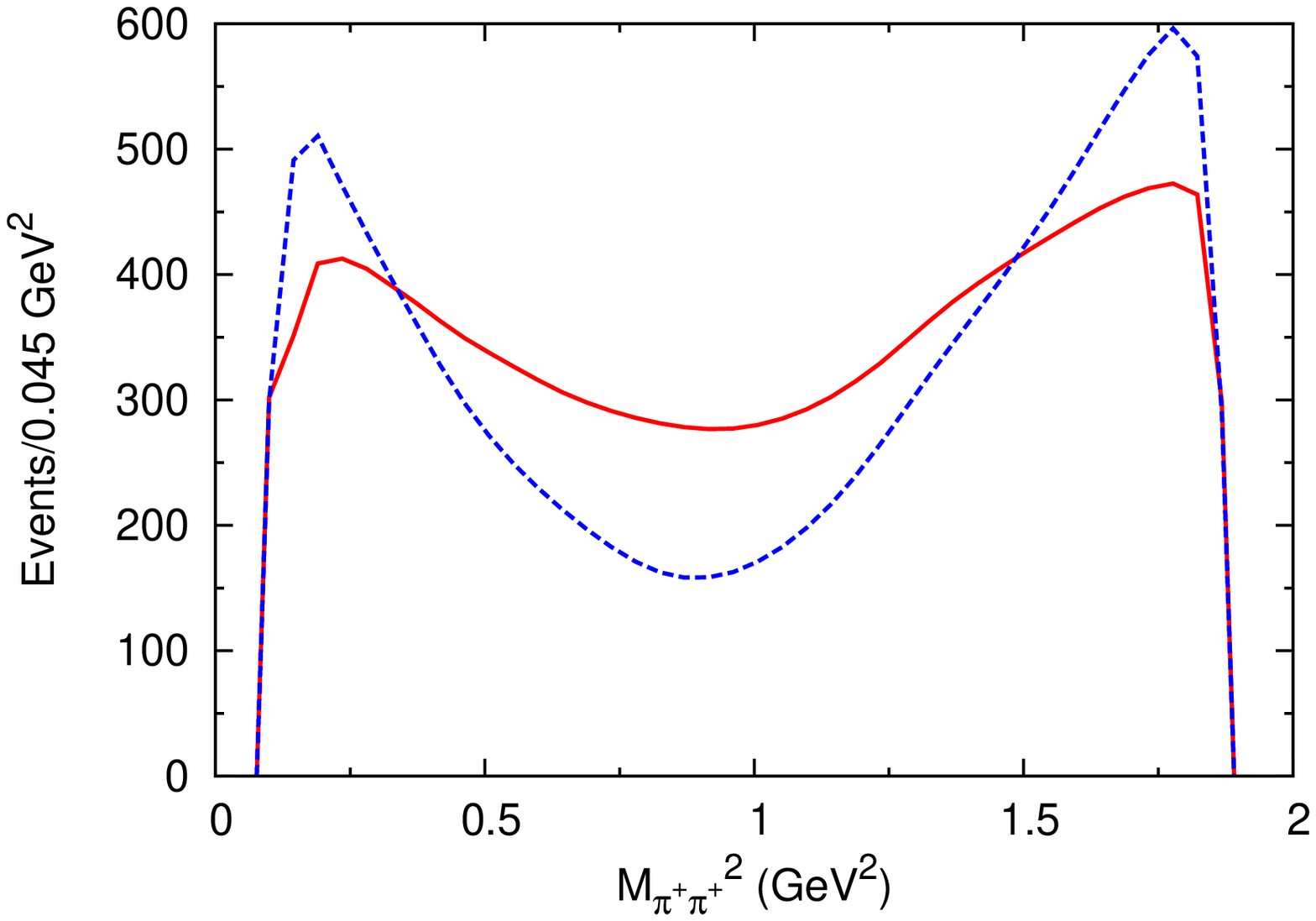}
\caption{\label{fig:proj-full-wo-rho}
(Color online) 
Contributions of 
$(\pi^+\pi^0)_P^{I=1}\bar{K}^0$ to
the $K^-\pi^+$ [$\pi^+\pi^+$] squared invariant mass spectrum
in the left [right] panel.
The Full model is shown by the red solid curve.
The blue dashed curve is also from the Full model but
couplings to
$(\pi^+\pi^0)_P^{I=1}\bar{K}^0$ are turned off.
}
\end{figure}
In order to see the contribution of $(\pi^+\pi^0)_P^{I=1}\bar{K}^0$
to the $D^+\to K^-\pi^+\pi^+$ decay more clearly, 
we show in Fig.~\ref{fig:proj-full-wo-rho}
the $K^-\pi^+$ ($\pi^+\pi^+$) squared invariant mass spectrum of the Full model
but couplings to $(\pi^+\pi^0)_P^{I=1}\bar{K}^0$ are turned
off.
The figures clearly show the large contribution from
$(\pi^+\pi^0)_P^{I=1}\bar{K}^0$. It increases 
the decay rate by $\sim$7\%, and significantly changes the shape
of the spectra.
We found a further enhanced contribution from 
$(\pi^+\pi^0)_P^{I=1}\bar{K}^0$ for the Z model;
it increases the decay rate by $\sim$ 30\%.
Actually, the dominance of the $(\pi^+\pi^0)_P^{I=1}\bar{K}^0$ fit
fraction ($\sim$85\%) was also found in a recent analysis of 
$D^+\to K_S^0\pi^+\pi^0$ done by the BESIII
Collaboration~\cite{bes3}.
The $D^+\to K^-\pi^+\pi^+$ and $D^+\to K_S^0\pi^+\pi^0$ decays
share the same hadronic dynamics, except for additional but much smaller 
doubly Cabibbo suppressed channels in the latter. 
Therefore, 
the large bare fit fraction of $(\pi^+\pi^0)_P^{I=1}\bar{K}^0$
for the $D^+\to K^-\pi^+\pi^+$ decay seems natural 
in order to understand the decay mechanisms for
$D^+\to K^-\pi^+\pi^+$ and $D^+\to K_S^0\pi^+\pi^0$ consistently. 
However, 
we have seen the rather large model dependence of
the contribution from $(\pi^+\pi^0)_P^{I=1}\bar{K}^0$.
This would be largely due to the fact that we analyzed
the $D^+\to K^-\pi^+\pi^+$ data that do not
contain this partial wave in the final state.
Because the $D^+\to K_S^0\pi^+\pi^0$ decay contains 
$(\pi^+\pi^0)_P^{I=1}\bar{K}^0$ in the final state,
a combined analysis of these two decay modes would significantly reduce
the uncertainty associated with $(\pi^+\pi^0)_P^{I=1}\bar{K}^0$.

\begin{figure}[t]
\includegraphics[width=0.45\textwidth]{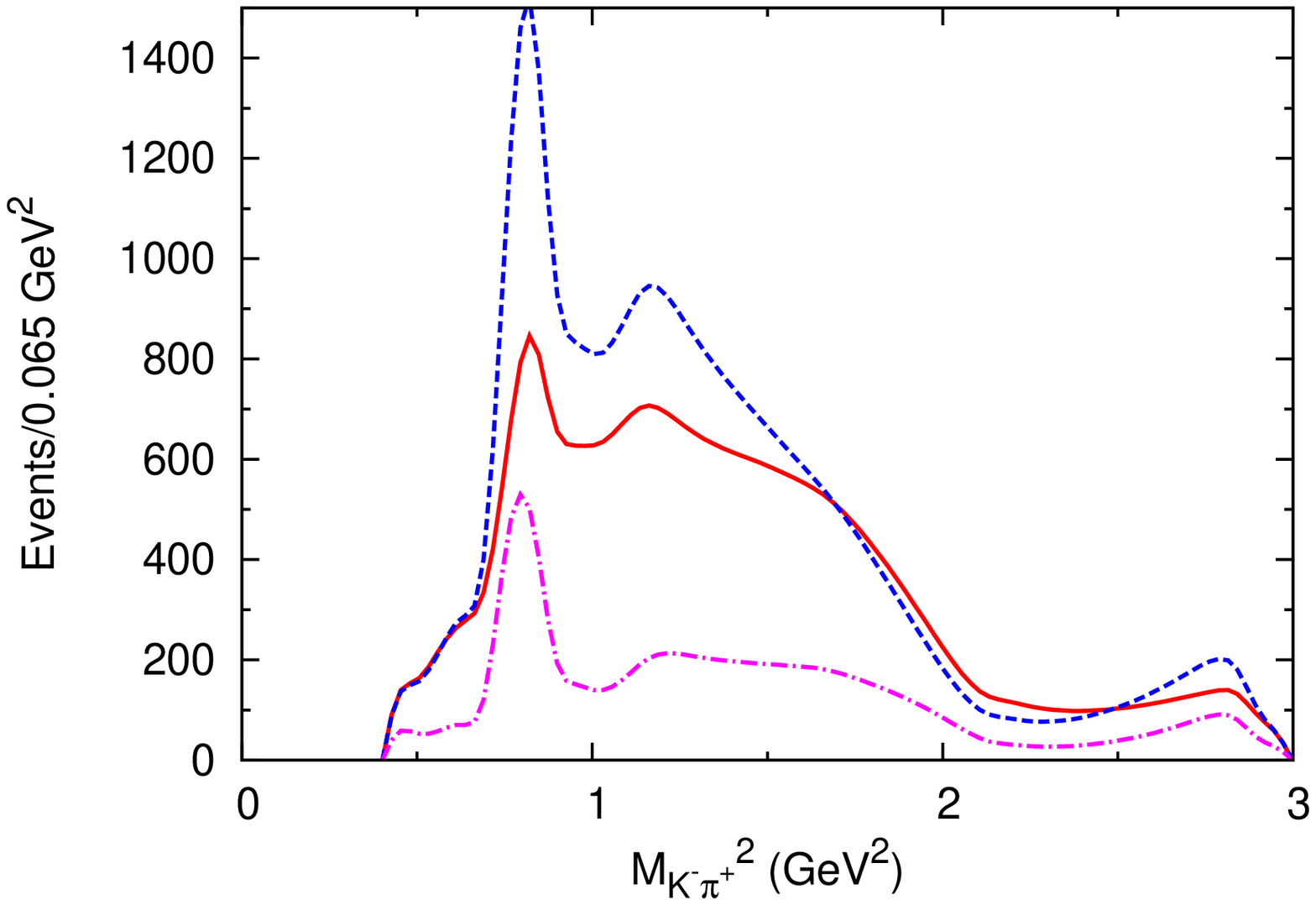}
\hspace{5mm}
\includegraphics[width=0.45\textwidth]{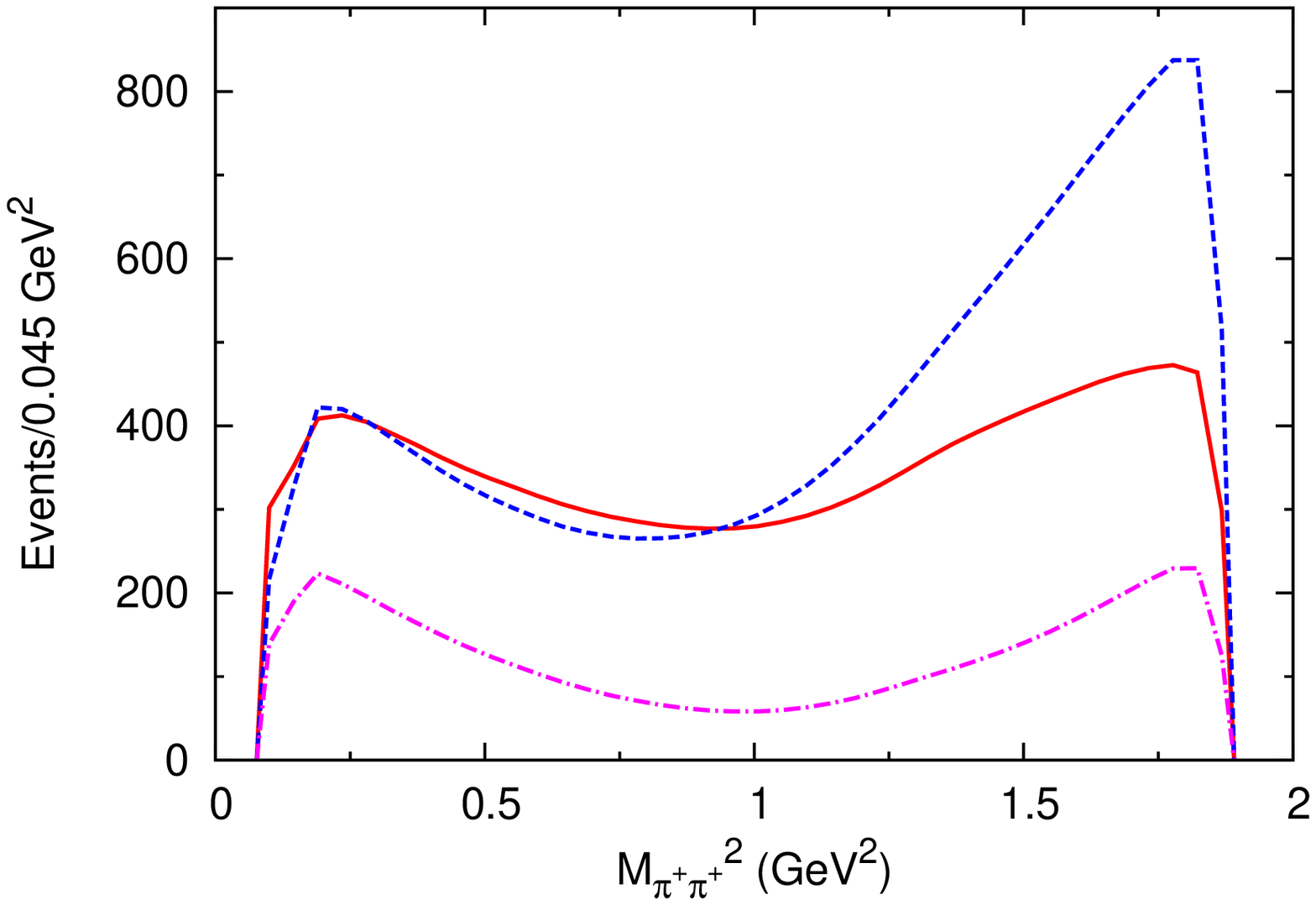}
\caption{\label{fig:3mf}
(Color online) 
Contributions of 
the three-meson force (Fig.~\ref{fig:vecpot}) to
the $K^-\pi^+$ [$\pi^+\pi^+$] squared invariant mass spectrum
in the left [right] panel.
The Full model is shown by the red solid curve.
The blue dashed curve is also from the Full model but
the three-meson force is turned off.
The magenta dot-dashed curve is from the Full model with
all the rescattering effects turned off.
}
\end{figure}
Now let us study how much the three-meson force contributes
to the $D^+\to K^-\pi^+\pi^+$ decay.
In Fig.~\ref{fig:3mf}, we compare 
the $K^-\pi^+$ ($\pi^+\pi^+$) squared invariant mass spectrum of the Full model
with those from the same model but the three-meson force being turned
off. 
The three-meson force suppresses
the decay width by $\sim$22\%, and 
change the spectrum shape significantly, as seen in Fig.~\ref{fig:3mf}.
The $K^-\pi^+$ spectrum is 
suppressed by the three-meson force
at the $K^*(892)$ peak region, and
the $\pi^+\pi^+$ spectrum is suppressed
at higher $M^2_{\pi^+ \pi^+}$.
We found that the effect of the diagram in Fig.~\ref{fig:vecpot}(a) 
connected to $(\pi^+\pi^0)_P^{I=1}\bar{K}^0$
is the
most important among the three-body-type diagrams we consider. 
In the same figure, we also show the spectrum from the Full model with
all the rescattering processes
[Fig.~\ref{fig:d-decay}(b); second term in Eq.~(\ref{eq:dressed-g})]
being turned off.
Effects of the rescattering mechanisms are quite large;
the decay width gets almost triplicated by the rescattering effect.
The spectra are rather different from the blue dashed curves in 
Fig.~\ref{fig:proj-full-wo-rho} where
$(\pi^+\pi^0)_P^{I=1}\bar{K}^0$ is turned off.
Therefore, the hadronic rescattering through partial waves other than 
$(\pi^+\pi^0)_P^{I=1}\bar{K}^0$ also gives a major contribution.

\begin{figure}[t]
\includegraphics[width=0.33\textwidth]{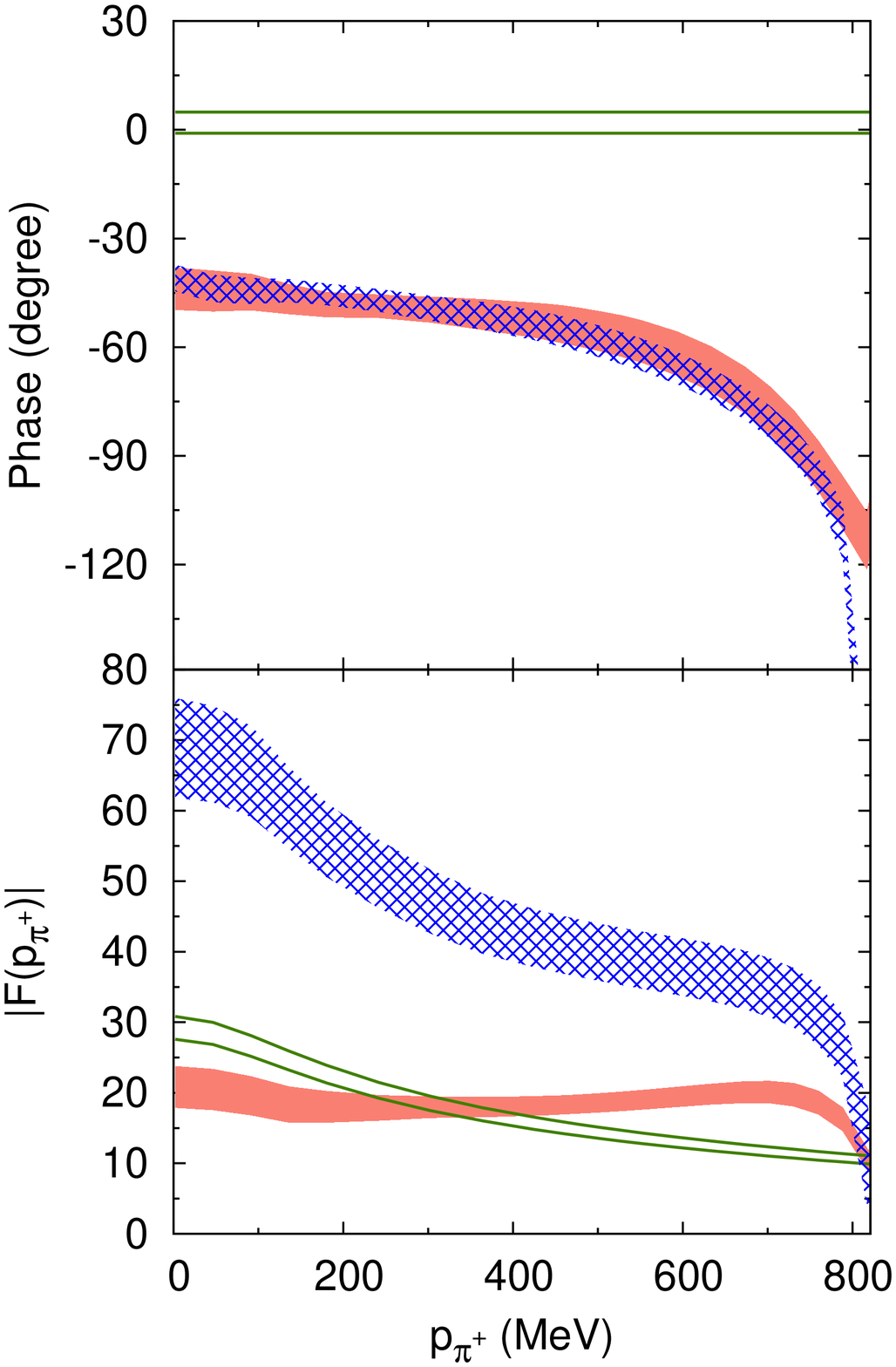}
\hspace{-2mm}
\includegraphics[width=0.33\textwidth]{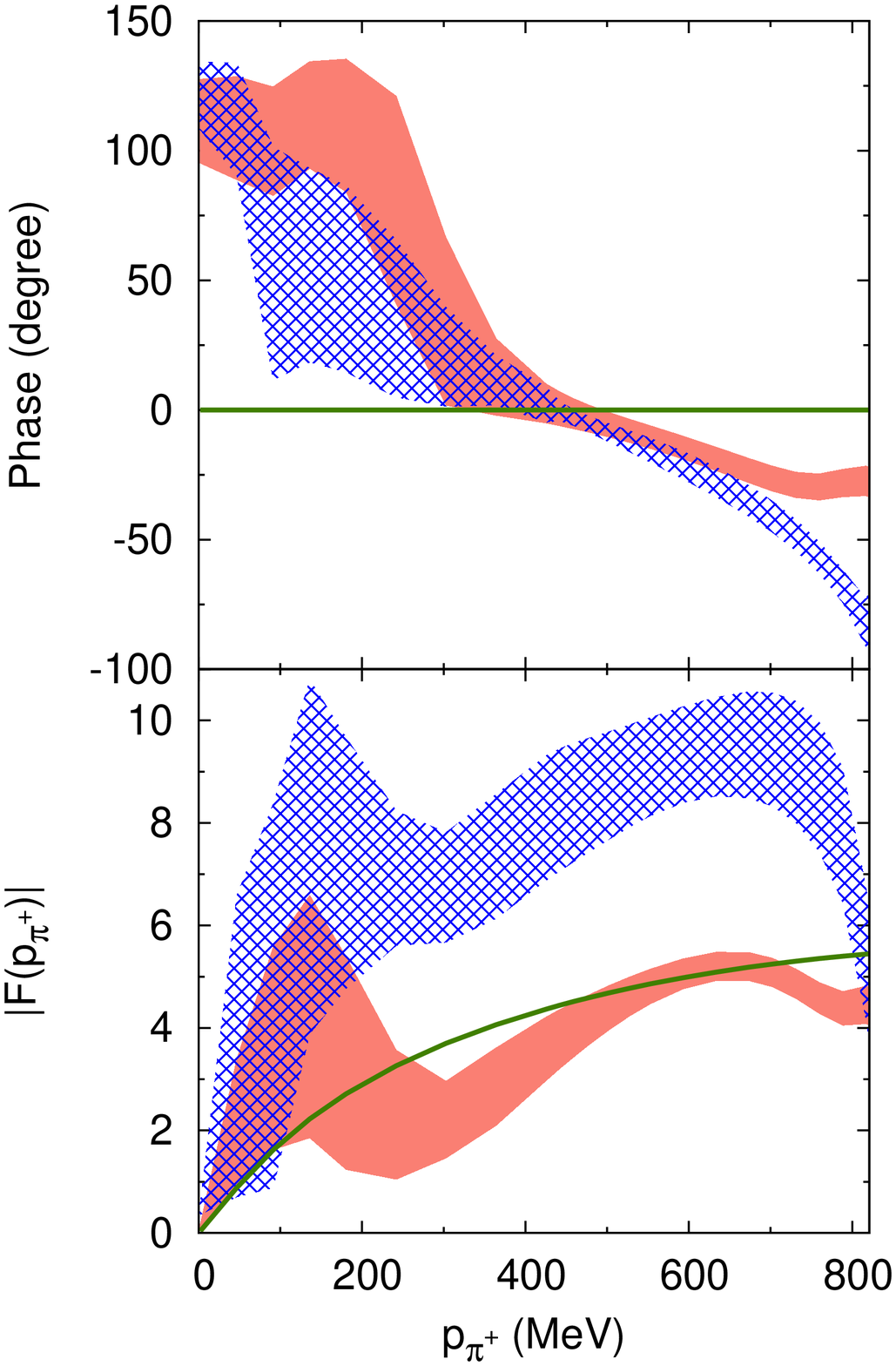}
\hspace{-2mm}
\includegraphics[width=0.33\textwidth]{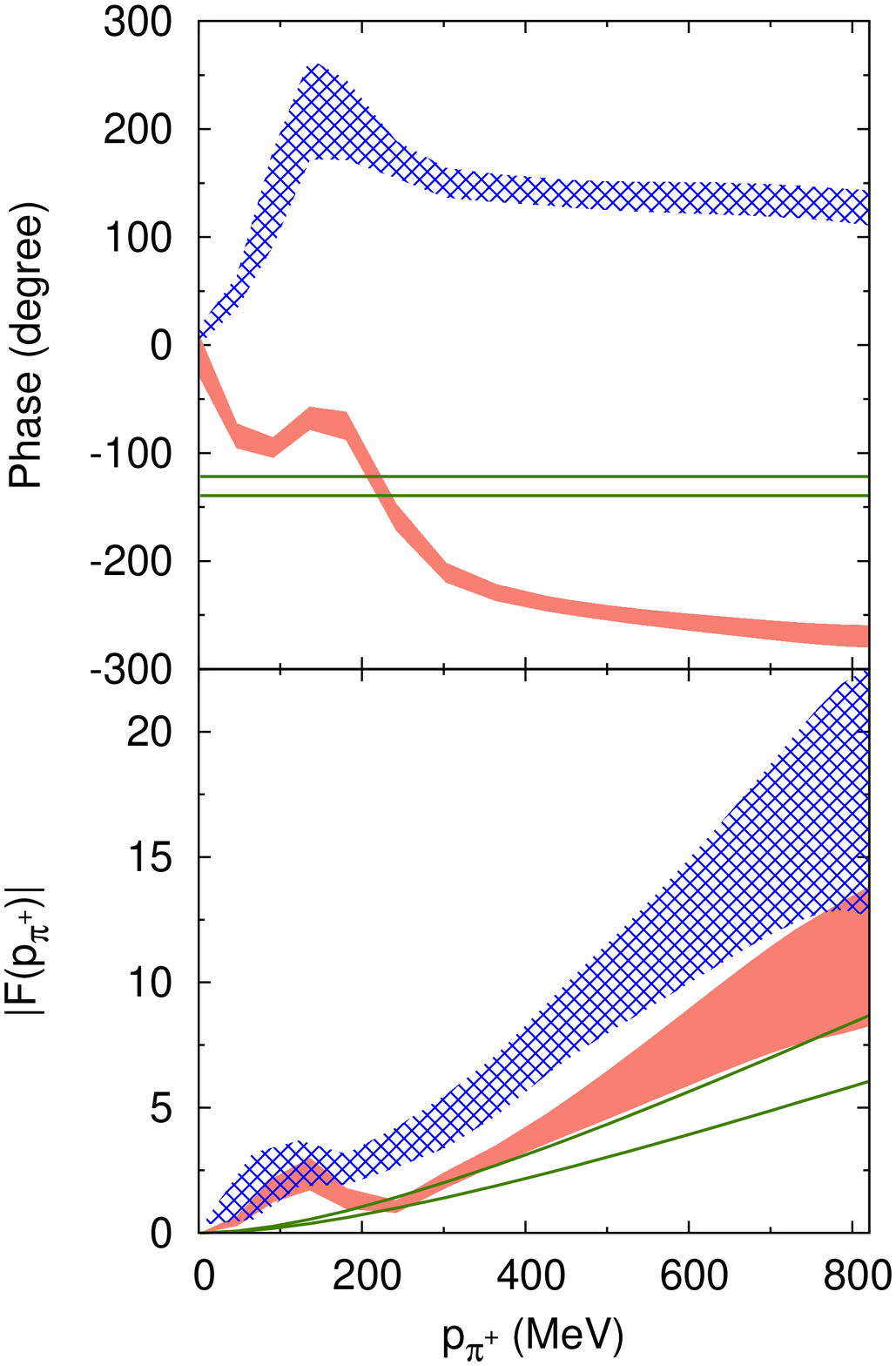}
\caption{\label{fig:Rc-vertex} 
(Color online) 
Phase (upper) and modulus (lower) of $D^+\to{\cal R}^{L,2I}_ic$ dressed vertex:
(Left) ${\cal R}^{L,2I}_ic=R^{01}_1\pi^+$; 
(Middle) ${\cal R}^{L,2I}_ic=R^{11}_1\pi^+$; 
(Right) ${\cal R}^{L,2I}_ic=R^{21}_1\pi^+$.
The red solid bands are from the Full model,
the blue cross-hatched bands from the Z model, 
and the green bordered bands
from the Isobar model; the band widths represents the errors. 
The unit of the modulus is arbitrary, but the relative magnitudes between
the different bands in the three panels are given by the model.
}
\end{figure}
In a Dalitz plot analysis with an isobar-type model,
it is assumed that a
$M^*\to {\cal R}c$ ($M^*$: heavy meson, meson resonance, etc.)
decay vertex implicitly contains effects of rescattering mechanisms
that are simulated by a complex coupling constant for the vertex.
It is interesting to examine the extent to which this assumption is
valid.
In Fig.~\ref{fig:Rc-vertex}, 
the upper [lower] panels give the phase [modulus] of 
dressed $D^+ \to {\cal R}c$ vertices 
defined in Eq.~(\ref{eq:dressed-g}) for 
the Full model (red solid bands)
and the Z model (blue cross-hatched bands),
as a function of the momentum of the unpaired meson, $p_c$.
For comparison, we also show in the same figure the $D^+ \to {\cal R}c$
vertices from the Isobar model by the green bordered bands.
The left, middle, and right panels are for 
${\cal R}^{L,2I}_ic=R^{01}_1\pi^+$,
${\cal R}^{L,2I}_ic=R^{11}_1\pi^+$, and  
${\cal R}^{L,2I}_ic=R^{21}_1\pi^+$, respectively.
$R^{01}_1$ works as a seed to develop $\bar\kappa$ and
$\bar K^*_0(1430)$ resonances, while 
$R^{11}_1$ and $R^{21}_1\pi^+$ will develop
$\bar K^*(892)$ and 
$\bar K^*_2(1430)$ resonances, respectively.
Even though all of the models have been fitted to the same pseudodata
of the Dalitz plot distribution, they are rather different in 
the $D^+ \to {\cal R}c$ vertices.
In particular,
the significant phase variations as a function of $p_c$ 
in the Full and Z models 
are purely due to the three-body
hadronic dynamics required to satisfy the three-body unitarity.
The constant phase assumed in the isobar-type models is not justified
from this viewpoint.

\section{Summary and Future Prospects}
\label{sec:conclusion}

In this work, we have performed a coupled-channel analysis of 
the pseudodata for the $D^+\to K^-\pi^+\pi^+$ Dalitz plot.
The pseudodata are generated from the isobar model of the E791
Collaboration~\cite{e791}, and are reasonably realistic.
As far as we know, this is the first coupled-channel analysis
of a realistic Dalitz plot distribution for a $D$-meson 
decay into a three-pseudoscalar-meson state.
We have demonstrated that it is indeed possible to analyze this kind of
high-quality data within a coupled-channel framework,
and found lots of interesting results that are beyond what can be obtained
with the conventional isobar-type model analyses.
Let us summarize below what we have done and found in this work.

In our build-up approach, 
we started with developing a two-pseudoscalar-meson interaction model.
With a suitable combination of contact interactions and bare
resonance-excitation mechanisms, our two-pseudoscalar-meson interaction
model successfully describes empirical $\pi\pi$ and $\pi \bar K$ scattering
amplitudes of low partial waves that are needed to analyze 
the $D^+\to K^-\pi^+\pi^+$ Dalitz plot. 
Poles associated with $\pi\pi$ and $\pi \bar K$ resonances have been extracted, and
most of them are in agreement with the PDG listings.
Then using the two-pseudoscalar-meson interactions as building blocks,
we developed a three-pseudoscalar-meson interaction model that describes
the FSI of the $D^+\to K^-\pi^+\pi^+$ decay.
The main driving force for the three-pseudoscalar-meson scattering
process is the Z-diagrams and the dressed ${\cal R}$-propagator that
appear as a necessary consequence of the three-body unitarity.
These mechanisms do not contain any adjustable parameters once the
two-pseudoscalar-meson interactions have been fixed using the $\pi\pi$
and $\pi \bar K$ scattering data.
In addition,
we considered mechanisms that are beyond simple iterations
of the two-body interactions, and thus may be called a three-meson force.
We introduced the three-meson force to 
an important channel in the FSI of 
the $D^+\to K^-\pi^+\pi^+$ decay, i.e., the vector-pseudoscalar channels.
Guided by the hidden local symmetry model~\cite{hls} that incorporates vector and
pseudoscalar mesons in a chiral Lagrangian,
we derived the vector-pseudoscalar interactions that work as a
three-meson force. 

In our analysis of the pseudodata for
the $D^+\to K^-\pi^+\pi^+$ Dalitz plot distribution,
we basically used three models: Full, Z, and Isobar models.
In the models,
we took account of all of the channels (partial waves) that had been
considered in the previous analyses of the same process.
This includes 
$(\bar{K}\pi)_S^{I=3/2}\pi$ ($(\pi\pi)_S^{I=2}\bar{K}$)
that is considered only in the FOCUS~\cite{focus} (CLEO~\cite{cleo}) analysis.
In addition, in the Full and Z models, we also considered
$(\pi^+\pi^0)_P^{I=1}\bar{K}^0$ where
$\rho(770)\bar{K}^0$ plays a dominant role.
This partial wave can contribute to the $D^+\to K^-\pi^+\pi^+$ decay
only through the rescattering,
and thus this contribution is a pure coupled-channel effect.
A distinct feature of our models is that all of the
two-pseudoscalar-meson resonances are included as poles of the
unitary scattering amplitudes that fit well the empirical $\pi\pi$ and
$\pi \bar K$ amplitudes. 
With an appropriately selected set of fitting parameters associated with
bare $D^+\to {\cal R}h$ ($h=\pi, \bar K$) vertices,
the Watson theorem is satisfied by the models when the rescattering is
turned off. 
The three-body unitarity is also taken care of within the Full and Z
models.
Unlike the previous analyses, we did not include
a flat background amplitude.

Our models fit the pseudodata with a reasonable precision.
As far as the $\chi^2$-value is concerned,
the Full and Z models are significantly better than
the Isobar model.
Although this may be partly due to the larger number of adjustable parameters in the fits,
it would also be because important mechanisms are considered in the better models.
Indeed, we showed that the Isobar model can be significantly improved by just
introducing the hadronic rescattering ($\chi^2$/d.o.f. is reduced by $\sim$29\%),
keeping the number of the fitting parameters unchanged [Z(without $\rho$) model].
The inclusion of the $\rho \bar K$ channel further
reduces $\chi^2$/d.o.f. by $\sim$47\%, and thus
the importance of this channel seems rather clear.
On the other hand, the three-meson force does not
improve $\chi^2$ although it gives a significant effect
on the FSI.
Another important point is that we achieved the good fits using 
two-pseudoscalar-meson amplitudes fixed by the empirical $\pi\pi$ and
$\pi\bar K$ amplitudes.
This is in sharp contrast with the previous analyses 
where the $p$-wave $\pi\bar K$ amplitude is given by a sum of the
Breit-Wigner functions that does not necessarily satisfy the Watson theorem.
Also, the previous analyses commonly adjusted parameters associated with 
$\bar K^*_0(1430)$ in their fits, and obtained widths significantly narrower
than those in the PDG listings. 
In our analysis,
on the other hand,
we were able to obtain the reasonable fits 
with $\bar K^*_0(1430)$ whose width is
close to those from the PDG.

We examined the partial-wave amplitudes and 
found that the Full and Z models have similar amplitudes,
while the Isobar model has significantly different amplitudes,
particularly for 
$({K}^-\pi^+)_S^{I=3/2}\pi^+$ and
$(\pi^+\pi^+)_S^{I=2}{K}^-$.
The $({K}^-\pi^+)_S^{I=3/2}\pi^+$ and $(\pi^+\pi^+)_S^{I=2}{K}^-$ amplitudes
have rather large errors
due to the fact that the fitting parameters associated
with these partial waves cannot be precisely
determined by the data used in this work. 
We compared the $({K}^-\pi^+)_S^{I=1/2}\pi^+$ amplitudes
from our models
with the $({K}^-\pi^+)_S\pi^+$
amplitude from the E791 MIPWA. 
We found a good agreement between the Full and Z models, and the
E791 MIPWA for $M_{K^-\pi^+}\ltap$~1.5~GeV.
We stress that
the hadronic rescattering plays an essential role to bring the phases of
our models into agreement with those from the E791 MIPWA. 
On the other hand, the Isobar model, that maintains the Watson theorem,
does not have a freedom to change the phases in the elastic region, 
ending up with a rather different solution.
The partial-wave amplitudes were used to calculate the fit fractions
that were compared with those from the previous analyses done by the
experimental groups.
We found a fairly good agreement between the analyses on the 
$(K^-\pi^+)_P^{I=1/2}\pi^+$ and $(K^-\pi^+)_D^{I=1/2}\pi^+$ fit
fractions.
For the $({K}^-\pi^+)_S^{I=1/2}\pi^+$,
$({K}^-\pi^+)_S^{I=3/2}\pi^+$ and $(\pi^+\pi^+)_S^{I=2}{K}^-$ fit
fractions, however, 
we found a rather large dependence on each of the analyses and also
large errors.
Even so, their coherent sum turned out to be
consistent among our three models and also
the $({K}^-\pi^+)_S\pi^+$ fit fractions from the E791 and FOCUS MIPWA
within greatly reduced errors.

With the coupled-channel framework, we were able to study the bare fit
fractions for which all the hadronic rescattering effects are absent.
This quantity could be useful to study the intrinsic quark-gluon
substructure of the $D$-meson.
Remarkably, $(\pi^+\pi^0)_P^{I=1}\bar{K}^0$, that does not
show up in the usual fit fraction, gives a large bare fit fraction.
This result may appear a bit surprising.
However, considering that 
the $D^+\to {K}^-\pi^+\pi^+$ and $D^+\to {K}^0_S\pi^0\pi^+$ decays
share the same hadronic dynamics to a large extent, 
this finding is actually
consistent with the recent BESIII analysis~\cite{bes3}
that found a dominant fit fraction ($\sim$85\%) of 
$(\pi^+\pi^0)_P^{I=1}\bar{K}^0$
in the $D^+\to {K}^0_S\pi^0\pi^+$ decay.

We further studied coupled-channel effects.
We found that 
the $D^+\to {K}^-\pi^+\pi^+$ decay width gets triplicated by
the rescattering effects in the Full model.
We also found that the phases of
the dressed $D^+\to {\cal R}\pi^+$ vertices have
rather large dependence on the unpaired $\pi^+$ momentum.
The phase variation is a consequence of the explicit treatment of the
three-body dynamics.
In the conventional isobar-type model analyses of heavy or excited meson
($M^*$) decay into three light mesons,
it is assumed that the
rescattering effects are small and/or 
the phases of the $M^*\to {\cal R}c$ vertices can be approximated by
constants.
Clearly, our analysis indicates that
neither of these assumptions are supported from this more microscopic
viewpoint, at least for the $D^+\to {K}^-\pi^+\pi^+$ decay.
With the above results as a basis, 
we can rather clearly conclude that
explicit treatment of the hadronic FSI is essential for extracting partial
wave amplitudes from Dalitz plots for $D^+\to {K}^-\pi^+\pi^+$
and probably also many other processes.

In future, we will apply our coupled-channel model to a combined
analysis of the $D^+\to {K}^-\pi^+\pi^+$ and $D^+\to {K}^0_S\pi^0\pi^+$
decays.
The strength of the coupled-channel framework is to describe different
processes in a unified manner. 
Because different aspects of the hadronic dynamics appear in different
processes, the combined analysis is a powerful method to understand the
hadron dynamics with smaller model-dependence.
This is why a combined analysis has become standard in the study of the
baryon spectroscopy.
For example, $\pi N, \gamma N\to \pi N, \eta N, K\Lambda, K\Sigma$
reaction data are analyzed with a coupled-channel model in a unified
manner to
extract the nucleon resonance properties in Ref.~\cite{knls13}.
This direction should also be pursued to better understand the hadronic
dynamics in heavy-meson decays and meson resonances.
Getting back to the future combined
analysis of $D^+\to {K}^-\pi^+\pi^+$ and ${K}^0_S\pi^0\pi^+$,
we expect to better understand
the role of $(\pi^+\pi^0)_P^{I=1}\bar{K}^0$
because contributions of this partial wave can be directly seen in 
the $D^+\to {K}^0_S\pi^0\pi^+$ Dalitz plot.
Although we found the very important contribution of
$(\pi^+\pi^0)_P^{I=1}\bar{K}^0$
to the $D^+\to {K}^-\pi^+\pi^+$ decay through the FSI,
this finding is based on the indirect information. 
Also in the combined analysis, we would be able to better extract
the $(\bar K \pi)_S^{I=3/2}\pi$ and $(\pi\pi)_S^{I=2}\bar K$ amplitudes
that have been determined with large uncertainties in this work.
By simultaneously analyzing the $D^+$ decays with differently charged final states,
it will be possible to better separate contributions from these different isospin states.
It will be interesting to
examine how the partial-wave amplitudes and fit fractions obtained in this work will be
in the combined analysis.
Finally, we will also study if a three-meson force plays an important
role in understanding the hadronic FSI of the two $D^+$-meson decays in a unified manner.

\begin{acknowledgments}
The author thanks Manoel Robilotta for stimulating discussions.
He thanks Masanori Hirai for useful discussions on the error analysis.
He also thanks the Yukawa Memorial Foundation for support during the
 early stage of this work in the form of a Yukawa Fellowship.
\end{acknowledgments}

%\clearpage
\appendix

\section{Three-meson force based on Hidden Local Symmetry Model}
\label{app:lag}

In this Appendix,
we first present a set of Lagrangians
from the hidden local symmetry (HLS) model~\cite{hls}.
Then we present expressions for potentials,
derived from the Lagrangians,
between vector-mesons and pseudoscalar-mesons.
These potentials work as a three-body-force in a
three-pseudoscalar-meson system.
Finally, we present the potentials in a partial-wave form that is useful
for numerical calculations.

\subsection{Lagrangians from the HLS model}
\label{app:lag1}

We use symbols $P$ and $V$
to denote octet pseudoscalar-mesons and nonet vector-mesons, respectively.
The pseudoscalar meson fields in the SU(3) matrix form are
\begin{equation}
P = {1\over 2} \sum_{a=1}^8  P_a \lambda_a
= {1\over \sqrt{2}}
\left(
\begin{array}{ccc}
\frac{1}{\sqrt{2}} \pi^0 + \frac{1}{\sqrt{6}} \eta & \pi^+ & K^+ \\
\pi^- & - \frac{1}{\sqrt{2}} \pi^0 + \frac{1}{\sqrt{6}} \eta & K^0 \\
K^- & \bar{K}^0 & - \frac{2}{\sqrt{6}} \eta
\end{array}
\right) \ ,
\label{eq:su3-meson}
\end{equation}
where $\lambda_a$ is a Gell-Mann matrix,
while the vector meson nonet is given by 
\begin{equation}
V = {1\over 2} \sum_{a=0}^8  V_a \lambda_a
=  {1\over \sqrt{2}}
\left(
\begin{array}{ccc}
\frac{1}{\sqrt{2}} \rho^0 + \frac{1}{\sqrt{2}} \omega & \rho^+ & K^{*+} \\
\rho^- & - \frac{1}{\sqrt{2}} \rho^0 + \frac{1}{\sqrt{2}} \omega & K^{*0} \\
K^{*-} & \bar{K}^{*0} & \phi
\end{array}
\right) \ ,
\label{eq:su3-vector}
\end{equation}
where $\lambda_0=\sqrt{2\over 3}\unit$ ($\unit$: unit matrix), and
the ideal mixing between the neutral vector mesons is assumed.
When $P$ and $V$ are enclosed in the trace symbol, they are 
fields of the SU(3) matrix form.
Otherwise, e.g., they are in brackets, they are understood to be
one of particles contained in the SU(3) matrix elements.
It is convenient to work with isospin states rather than the charge
states used in Eqs.~(\ref{eq:su3-meson}) and (\ref{eq:su3-vector}).
For the relation between the two sets of the basis,
we employ a convention where
the charge states
are the same as their isospin states ($\ket{II^z}$) with some exceptions that need
additional phases as follows:
\begin{eqnarray}
\ket{\rho^+} &=& - \ket{I=1,I^z=1}, \qquad
\ket{K^{*-}} = - \ket{I=1/2,I^z=-1/2}, \nonumber\\
\ket{\pi^+} &=& - \ket{I=1,I^z=1}, \qquad
\ket{K^{-}} = - \ket{I=1/2,I^z=-1/2} \ .
\end{eqnarray}
In what follows, 
$\ket{P}$ and $\ket{V}$ are understood to be isospin states
rather than charge states.
Also, we use curly symbols to denote creation or annihilation operators.
For example, ${\cal P}$ is the annihilation operator contained in 
the field $P$, and its normalization is $\bra{0} {\cal P}\ket{P_a}=\lambda_a/2$.

The mesonic interaction Lagrangians we use are
those from the HLS model~\cite{hls,hls2}.
The $VPP$ interactions are
(with the Bjorken-Drell convention for the metric)
\begin{equation}
\label{eq:vmm}
   {\cal L}_{VPP} =  2 i g 
{\rm Tr}\left[ V_\mu (\partial^\mu P P - P \partial^\mu P) \right] \ ,
\end{equation}
where the trace is taken in the SU(3) space.
The coupling $g$ is related to the $\rho\pi\pi$ coupling by
$g=g_{\rho\pi\pi}$, and we use $g_{\rho\pi\pi}=6.0$ determined from
the $\rho\to\pi\pi$ decay width.
The Yang-Mills type Lagrangian, from which we use $VVV$ interactions, is 
\begin{eqnarray}
\label{eq:ym}
 {\cal L}_{YM} = - {1\over 2} {\rm Tr}\left[ F_{\mu\nu}F^{\mu\nu}  \right] \ ,
\end{eqnarray}
with
\begin{eqnarray}
F_{\mu\nu} = \partial_\mu V_\nu - \partial_\nu V_\mu - i g [V_\mu,V_\nu]
 \ .
\end{eqnarray}
The $VVP$ interactions are given by
\begin{equation}
\label{eq:vvm}
   {\cal L}_{VVP} = g^2 C\epsilon^{\alpha\beta\gamma\delta}
{\rm Tr}\left[\partial_\alpha V_\beta \partial_\gamma V_\delta P \right] \ ,
\end{equation}
where we use the convention, $\epsilon^{0123}=+1$.
The coupling is $C=-3/(4\pi^2f_\pi)$ with $f_\pi$ being the pion decay
constant.
In our numerical calculations, we use $|g^2C/2|=g_{\omega\pi\rho}\sim 0.012$
from an analysis by Durso~\cite{durso} on the decay
$\omega\to\pi\rho\to\pi\gamma$.

\subsection{$VP\to V'P'$ potentials}

We consider a process
$V(p_V,\epsilon_V,I_V,I^z_V) P(p_P,I_P,I^z_P)$ $\to$ $V'(p_{V'},\epsilon_{V'},I_{V'},I^z_{V'})P'(p_{P'},I_{P'},I^z_{P'})$
where the variables in the parentheses specify each particle's state
such as the four-momentum ($p$), polarization ($\epsilon$), isospin ($I$) and
its $z$-component ($I^z$).
The 0th component of the four-momentum is related to the spatial part by
$p^0_x=\sqrt{\bm{p}^2_x + m^2_x}$ where $m_x$ is the mass for a
particle $x$.
The potential diagrammatically represented in Fig.~\ref{fig:vecpot}(a) is 
derived from the Lagrangians in Eqs.~(\ref{eq:vmm}) and (\ref{eq:ym})
following the unitary transformation method~\cite{sko}, and
is given by
\begin{eqnarray}
 V_{\rm Fig.\ref{fig:vecpot}(a)} &=& {g_{VV'V_{ex}} g_{V_{ex}PP'} \over q^2 - m^2_{V_{ex}}} 
(-1)^{I_{V}+2I_{V'}+I_{V_{ex}}-2I^z_{V}+I_{P'}+1}
\sqrt{(2I_{P'}+1)(2I_{V}+1)}  \nonumber \\
 &\times& \sum_{I} (-1)^{-I} W(I_{V} I_{V'} I_{P} I_{P'}; I_{V_{ex}} I)
(I_{V} I^z_{V} I_{P} I^z_{P}| I I^z)
(I_{V'} I^z_{V'} I_{P'} I^z_{P'}| I I^z) \nonumber \\
 &\times& \left\{
(p_V+p_{V'})\cdot (p_P+p_{P'}) \epsilon_V\cdot\epsilon^*_{V'}
-(p_V+q)\cdot\epsilon^*_{V'} (p_P+p_{P'})\cdot\epsilon_V \right.
\nonumber \\
&&  + \left. (q-p_V)\cdot\epsilon_V (p_P+p_{P'})\cdot\epsilon^*_{V'}
\right\} \ ,
\label{eq:vvv-pot}
\end{eqnarray}
where $q=p_V-p_{V'}=p_{P'}-p_P$, and
$m_{V_{ex}}$ and $I_{V_{ex}}$ are
the exchanged vector-meson mass and isospin, respectively.
We have used
the Racah and Clebsch-Gordan coefficients denoted by 
$W(j_1 j_2 J j_3; j_{12} j_{23})$ and
$(j_1 m_1 j_2 m_2 | JM)$, respectively.
The propagator for the exchanged vector-meson is more explicitly written
as
\begin{eqnarray}
{1\over q^2 - m^2_{V_{ex}}} = 
{1\over 2}
\left[
{1\over (p_V-p_{V'})^2 - m^2_{V_{ex}}} +
{1\over (p_{P'}-p_{P})^2 - m^2_{V_{ex}}} 
\right]  \ ,
\end{eqnarray}
as specified by the unitary transformation~\cite{sko}.
The effective coupling $g_{VV'V_{ex}}$ is given by 
\begin{eqnarray}
g_{VV'V_{ex}} &=& {g\over \sqrt{2}}
{{\rm Tr}\left[ \langle {\cal V} \rangle [\langle {\cal V'}\rangle,\langle {\cal V}_{ex}\rangle] \right]
\over
(I_{V'}I^z_{V'}I_{V_{ex}}I^z_{V_{ex}}|I_VI^z_V)}\ , 
\end{eqnarray}
with 
$\langle {\cal V} \rangle=\langle 0| {\cal V} | V \rangle$,
$\langle {\cal V'} \rangle=\langle V'| {\cal V} | 0 \rangle$ and 
$\langle {\cal V}_{ex} \rangle=\langle V_{ex}| {\cal V} | 0 \rangle$.
The effective couplings $g_{VV'V_{ex}}$ as well as 
$g_{V_{ex}PP'}$ and $g_{VV_{ex}P'}$ appearing below 
are independent of isospin $z$-components.
$g_{V_{ex}PP'}$ is given by
\begin{eqnarray}
g_{V_{ex}PP'} &=& {g\over \sqrt{2}}
{{\rm Tr}\left[\langle{\cal V}_{ex}\rangle [\langle{\cal P'}\rangle,\langle{\cal P}\rangle]\right]
\over
(I_{V_{ex}}I^z_{V_{ex}}I_{P}I^z_{P}|I_{P'}I^z_{P'})}\ ,
\end{eqnarray}
with 
$\langle {\cal V}_{ex} \rangle=\langle 0 | {\cal V} | V_{ex} \rangle$,
$\langle {\cal P }\rangle=\langle 0| {\cal P} | P \rangle$ and 
$\langle {\cal P'} \rangle=\langle P'| {\cal P} | 0 \rangle$.

Another potential diagrammatically represented by Fig.~\ref{fig:vecpot}(b) is 
derived from the Lagrangian in Eq.~(\ref{eq:vvm}) and is given by
\begin{eqnarray}
 V_{\rm Fig.\ref{fig:vecpot}(b)} &=& {g_{VV_{ex}P'} g_{V'V_{ex}P} \over q^2 - m^2_{V_{ex}}} 
\sqrt{(2I_{V}+1)(2I_{V'}+1)}  \nonumber \\
 &\times& \sum_{I} W(I_{V'} I_{P} I_{P'} I_{V}; I_{V_{ex}} I)
(I_{V} I^z_{V} I_{P} I^z_{P}| I I^z)
(I_{V'} I^z_{V'} I_{P'} I^z_{P'}| I I^z) \nonumber\\
 &\times& \left\{
p_{V'}\cdot p_{P'} (\epsilon^*_{V'}\cdot p_V  \epsilon_V\cdot p_P -
\epsilon^*_{V'}\cdot\epsilon_V p_V\cdot p_P)
+ p_V\cdot p_{V'} (\epsilon^*_{V'}\cdot\epsilon_V p_P\cdot p_{P'}
- \epsilon^*_{V'}\cdot p_{P'} \epsilon_V\cdot p_P) \right. \nonumber \\
&& \left. 
+ \epsilon_V\cdot p_{V'}  (\epsilon^*_{V'}\cdot p_{P'} p_V\cdot p_P - 
\epsilon^*_{V'}\cdot p_V p_P\cdot p_{P'})
\right\} \ ,
\label{eq:vvp-pot}
\end{eqnarray}
where the propagator for the exchanged vector-meson is
\begin{eqnarray}
{1\over q^2 - m^2_{V_{ex}}} = 
{1\over 2}
\left[
{1\over (p_V-p_{P'})^2 - m^2_{V_{ex}}} +
{1\over (p_{V'}-p_{P})^2 - m^2_{V_{ex}}} 
\right]  \ .
\end{eqnarray}
The effective coupling $g_{VV_{ex}P'}$ is given by 
\begin{eqnarray}
g_{VV_{ex}P'} &=& 2 g_{\omega\pi\rho}
{{\rm Tr}\left[
\langle{\cal P'}\rangle (\langle{\cal V}\rangle\langle{\cal V}_{ex}\rangle + \langle{\cal V}_{ex}\rangle \langle{\cal V}\rangle) \right]
\over 
(I_{V_{ex}}I^z_{V_{ex}}I_{P'}I^z_{P'}|I_VI^z_V)}\ ,
\end{eqnarray}
with
$\langle {\cal P'} \rangle=\langle P'| {\cal P} | 0 \rangle$,
$\langle {\cal V} \rangle=\langle 0 | {\cal V} | V \rangle$ and
$\langle {\cal V}_{ex} \rangle=\langle V_{ex} | {\cal V} |0 \rangle$.
$g_{V'V_{ex}P}$ is obtained by exchanging labels $V\to V'$
and $P'\to P$ in the rhs of Eq.~(A14),
and the meaning of the matrix elements are 
$\langle {\cal V'} \rangle=\langle V'| {\cal V} | 0 \rangle$,
$\langle {\cal P} \rangle=\langle 0 | {\cal P} | P \rangle$ and
$\langle {\cal V}_{ex} \rangle=\langle 0 | {\cal V} |V_{ex} \rangle$.

Following Ref.~\cite{pirho2}, we multiply the following form factor
to the potentials of Eqs.~(\ref{eq:vvv-pot}) and
(\ref{eq:vvp-pot}):
\begin{eqnarray}
F_{\rm 3MF}(\bm{q}) = \left( {\Lambda_{\rm 3MF}^2 - m^2_{V_{ex}} \over \Lambda_{\rm 3MF}^2 +
	     \bm{q}^2}\right)^2 \ ,
\label{eq:ff-3mf}
\end{eqnarray}
where $\Lambda_{\rm 3MF}$ is a cutoff to be determined by fitting data.
We use the same cutoff value for all the potentials
of Eqs.~(\ref{eq:vvv-pot}) and (\ref{eq:vvp-pot}).
We checked numerical values of the potentials by comparing 
our calculation with Fig.~8 of Ref.~\cite{pirho2}.

\subsection{Partial-wave expansion}

In order to implement the $VP\to V'P'$ potentials 
of Eqs.~(\ref{eq:vvv-pot}) and (\ref{eq:vvp-pot})
into the scattering equation of
Eq.~(\ref{eq:pw-tcr}), we need to expand the potentials 
in terms of the partial-wave representation as,
\begin{eqnarray}
&&\langle
 V'(p_{V'},\epsilon_{V'},I_{V'},I^z_{V'})P'(p_{P'},I_{P'},I^z_{P'})|
V_{\rm Fig.\ref{fig:vecpot}}|
V(p_V,\epsilon_V,I_V,I^z_V) P(p_P,I_P,I^z_P)\rangle
\nonumber\\
&&\qquad
= \sum_{TJJ^zll^zl'l^{\prime z}}
\inp{I_VI^z_VI_PI^z_P}{T\, I^z_V\!+\!I^z_P}
\inp{I_{V'}I^z_{V'}I_{P'}I^z_{P'}}{T\, I^z_V\!+\!I^z_P}
\nonumber\\
&&\qquad\qquad
\times
\inp{ll^z 1 \epsilon_V}{JJ^z}
\inp{l'l^{\prime\, z} 1\epsilon_{V'}}{JJ^z}
Y_{l'l^{\prime z}}(\hat{p}_{V'})
Y^*_{ll^{z}}(\hat{p}_{V})
V^{{\rm Fig.\ref{fig:vecpot}}\ JPT}_{(P'V')_{l'},(PV)_{l}} 
(|\bm p_{V'}|,|\bm p_V|) \ ,
\end{eqnarray}
where $JPT$ are the total angular momentum, parity, and total isospin,
respectively, and $l$ ($l'$) is the orbital angular momentum of the
relative motion of $VP$ ($V'P'$).
Inverting this equation, we obtain
\begin{eqnarray}
&&V^{{\rm Fig.\ref{fig:vecpot}}\ JPT}_{(P'V')_{l'},(PV)_{l}} 
(|\bm p_{V'}|,|\bm p_V|) 
=
\sum_{I^z_V,I^z_P,I^z_{V'},I^z_{P'}}
\inp{I_VI^z_VI_PI^z_P}{T\, T^z}
\inp{I_{V'}I^z_{V'}I_{P'}I^z_{P'}}{T\, T^z}
\sqrt{4\pi\over 2l'+1}
\nonumber \\
&&\qquad
\times
\sum_{\epsilon_V,\epsilon_{V'}} 
(-1)^{l+l'+\epsilon_V+\epsilon_{V'}}
\inp{1 \epsilon_V J -\epsilon_V}{l0}
\inp{1 \epsilon_{V'} J -\epsilon_V}{l' \epsilon_{V'}-\epsilon_V}
\int d\Omega_{\hat{p}_{V'}}
Y^*_{l',\epsilon_{V}-\epsilon_{V'}}(\hat{p}_{V'})
\nonumber \\
&&\qquad\times
\langle
 V'(p_{V'},\epsilon_{V'},I_{V'},I^z_{V'})P'(p_{P'},I_{P'},I^z_{P'})|
V_{\rm Fig.\ref{fig:vecpot}}|
V(p_V,\epsilon_V,I_V,I^z_V) P(p_P,I_P,I^z_P)\rangle \ ,
\label{eq:partialv}
\end{eqnarray}
where $\bm{p}_V$ is taken along the $z$-axis.
This partial-wave form of the potentials is plugged in
Eq.~(\ref{eq:addition}).

\section{Model parameters}
\label{sec:para}

\begin{table}[t]
\caption{\label{tab:kpi} 
Model parameters for the $\pi\bar{K}$ partial-wave scattering with the angular
 momentum $L$ and the isospin $I$.
The parameters are defined in
Eqs.~(\ref{eq:cont-ptl}), (\ref{eq:vf-cont}), 
(\ref{eq:pw-2body-v}), and (\ref{eq:pipi-vertex}).
For each partial wave specified by \{$L$,$I$\},
masses ($m_{R_i}$), couplings ($g_{h_1h_2,R_i}$), 
and cutoffs ($c_{h_1h_2,R_i}$) are for the $i$-th bare
$R$ states, $R_i$, and $h_1$ and $h_2$ are particles in a
two-pseudoscalar-meson channel.
Couplings ($h_{h_1h_2,h_1h_2}$) 
and cutoffs ($b_{h_1h_2}$) are for the contact interactions.
Masses ($m_{1}$, $m_{2}$) are for two ``particles'' in the
effective inelastic channel, except for 
the \{$L$,$I$\}=\{0,1/2\} wave for which
($m_{1}$, $m_{2}$)=($m_K$, $m_{\eta'}$).
The superscripts, $LI$, of the parameters are suppressed for
simplicity. 
The masses and cutoffs are given in the unit of MeV.
The hyphens indicate unused parameters.
}    
\begin{ruledtabular}
\begin{tabular}{lcccc}
$R~\{L,I\}$&$\bar{K}^*_0$ \{0, 1/2\} & \{0, 3/2\} &$\bar{K}^*_1$ \{1, 1/2\} &$\bar{K}^*_2$ \{2, 1/2\} \\\hline
$m_{R_1}$                  & 1391 &      --- & 1081     & 3070 \\
$g_{\pi\bar{K},R_1}$       & $-$6.28&      --- &  0.52    &  0.18\\ 	 
$c_{\pi\bar{K},R_1}$       &  609 &      --- & 1973     & 1954 \\ 
$g_{h_1h_2,R_1}$           &  4.30&      --- & $-$0.00    &  0.97\\	 
$c_{h_1h_2,R_1}$           & 1966 &      --- & 1706     & 1035 \\
$m_{R_2}$ 	           & 1767 &      --- & 1580     &  --- \\	 
$g_{\pi\bar{K},R_2}$       &  8.22&      --- &  1.84    &  --- \\ 	 
$c_{\pi\bar{K},R_2}$       &  395 &      --- &  395     &  --- \\ 
$g_{h_1h_2,R_2}$           & $-$4.87&      --- & $-$3.00    &  --- \\	 
$c_{h_1h_2,R_2}$           &  395 &      --- &  411     &  --- \\
$m_{R_3}$ 	           &  --- &      --- & 1750     &  --- \\	 
$g_{\pi\bar{K},R_3}$       &  --- &      --- &  0.23    &  --- \\ 	 
$c_{\pi\bar{K},R_3}$       &  --- &      --- & 1316     &  --- \\ 
$g_{h_1h_2,R_3}$           &  --- &      --- &  2.65    &  --- \\	 
$c_{h_1h_2,R_3}$           &  --- &      --- &  395     &  --- \\
$h_{\pi\bar{K},\pi\bar{K}}$&  1.31&      0.45&      --- &  --- \\	 
$b_{\pi\bar{K}}$           & 710  &      1973&      --- &  --- \\
$m_{1}$ 	           & 494  &      --- &       591&   100\\	 
$m_{2}$ 	           & 958  &      --- &       662&  1049\\	 
\end{tabular}
\end{ruledtabular}
\end{table}

\begin{table}[t]
\caption{\label{tab:pipi} 
Model parameters for $\pi\pi$ partial-wave scatterings.
The coupling $h'$ is defined in Eq.~(\ref{eq:vf-cont2}).
The other features are the same as those in 
Table~\ref{tab:kpi}.
}
\begin{ruledtabular}
\begin{tabular}{lcccc}
$R~\{L,I\}$&$f_0$ \{0, 0\} & \{0, 2\} &$\rho$ \{1, 1\} &$f_2$ \{2, 0\} \\\hline
$m_{R_1}$          & 1166 &      --- &      850 &      1561 \\
$g_{\pi\pi,R_1}$   &  5.97&      --- &      1.02&      $-$0.32\\
$c_{\pi\pi,R_1}$   & 1162 &      --- &      805 &       962 \\
$g_{K\bar{K},R_1}$ & $-$2.19&      --- &     $-$0.18&       0.19\\ 
$c_{K\bar{K},R_1}$ & 1973 &      --- &      395 &      1216 \\  
$m_{R_2}$ 	   & 1627 &      --- &     1551 &      ---  \\
$g_{\pi\pi,R_2}$   & $-$5.23&      --- &      0.49&      ---  \\
$c_{\pi\pi,R_2}$   & 1973 &      --- &     1973 &      ---  \\
$g_{K\bar{K},R_2}$ & 11.99&      --- &      3.74&      ---  \\ 
$c_{K\bar{K},R_2}$ &  533 &      --- &      395 &      ---  \\
$h_{\pi\pi,\pi\pi}$&  0.47&     0.11 &      --- &      ---  \\	 
$h'$               &  --- &     0.21 &      --- &      ---  \\	 
$b_{\pi\pi}$       &   897&       913&      --- &      ---  \\
\end{tabular}
\end{ruledtabular}
\end{table}

\begin{table}[t]
\vspace*{-3mm}
\caption{\label{tab:param}
Parameters determined by fitting $D^+\to K^-\pi^+\pi^+$ Dalitz plot pseudodata;
$D^+\to h {\cal R}^{L,2I}_i$ ($h=\pi,\bar{K}$)
bare coupling ($C_{h {\cal R}^{L,2I}_i}$) and
phase ($\phi_{h {\cal R}^{L,2I}}$) in degrees,
as defined in Eq.~(\ref{eq:bare_mstar}).
$R^{L,2I}_i$ stands for $i$-th bare $R$ state with the spin $L$ and
isospin $I$.
$r^{L,2I}_{ab}$ 
stands for the spurious state 
(see Sec.~\ref{sec:three-meson} for the definition)
that decays into two pseudoscalar
mesons ($ab$) with the orbital angular momentum $L$
and total isospin $I$.
The total spin, parity, total isospin, and $h{\cal R}$ relative orbital
angular momentum 
are $J$=0, $P$=+1, $T$=3/2, and $l$=$L$, respectively
for all the parameters, and thus labels of $JPT$ and $l$ are suppressed
 in the table.
The second, third, fourth, and fifth columns show the parameters for the Full,
Z, Isobar, and Z(without $\rho$) models, respectively.
The hyphens indicate unused parameters.
The cutoff for the three-meson force in the unit of MeV
is denoted by $\Lambda_{\rm 3MF}$ [see Eq.~(\ref{eq:ff-3mf}) for definition].
}
\begin{ruledtabular}
\begin{tabular}{lcccc}
                                      & Full & Z &   Isobar&Z(without $\rho$)    \\\hline
$\phi_{\pi R^{01}  }$          &$   -72. \pm    5.$&$   -80. \pm    4.$&$     2. \pm    3.$&$   -24. \pm    6.$\\
   $C_{\pi R^{01}_1}$          &$  10.54 \pm  0.57$&$  24.32 \pm  2.94$&$  10.83 \pm  0.60$&$   7.81 \pm  0.79$\\
   $C_{\pi R^{01}_2}$          &$   0.00 \pm  0.28$&$   0.00 \pm  0.87$&$   8.37 \pm  0.59$&$  12.34 \pm  1.16$\\
   $C_{\pi r^{01}_{\pi\bar K}}$&$   0.00 \pm  0.01$&$   0.00 \pm  0.03$&$   0.00 \pm  0.00$&$   0.00 \pm  0.00$\\
$\phi_{\pi R^{11}  }$          &    0  (fixed)     &    0  (fixed)     &    0  (fixed)     &    0  (fixed)     \\
   $C_{\pi R^{11}_1}$          &    1  (fixed)     &    1  (fixed)     &    1  (fixed)     &    1  (fixed)     \\
   $C_{\pi R^{11}_2}$          &$   1.48 \pm  0.31$&$   4.08 \pm  0.83$&$   0.03 \pm  0.26$&$   0.19 \pm  0.24$\\
   $C_{\pi R^{11}_3}$          &$   1.34 \pm  0.46$&$   6.55 \pm  1.28$&$   0.03 \pm  0.43$&$   0.39 \pm  0.50$\\
$\phi_{\pi R^{21}  }$          &$    82. \pm   13.$&$   136. \pm   24.$&$  -131. \pm    9.$&$   129. \pm    6.$\\
   $C_{\pi R^{21}_1}$          &$   0.27 \pm  0.08$&$   0.37 \pm  0.10$&$   0.21 \pm  0.04$&$   0.34 \pm  0.04$\\
$\phi_{\bar{K} R^{12}  }$      &$   -44. \pm    8.$&$   178. \pm    7.$& ---  & --- \\ 	 
   $C_{\bar{K} R^{12}_1}$      &$   0.00 \pm  0.00$&$  19.37 \pm  3.12$& ---  & --- \\ 	 
   $C_{\bar{K} R^{12}_2}$      &$  17.27 \pm  1.48$&$   0.00 \pm  0.84$& ---  & --- \\ 	 
$\phi_{\pi r^{03}_{\pi\bar K}}$&$   121. \pm   19.$&$    62. \pm   16.$&$  -133. \pm    4.$&$  -170. \pm    8.$\\
   $C_{\pi r^{03}_{\pi\bar K}}$&$   0.90 \pm  0.52$&$   2.06 \pm  1.12$&$   2.61 \pm  0.26$&$   3.00 \pm  0.36$\\
$\phi_{\bar Kr^{04}_{\pi\pi}}$ &$   173. \pm   30.$&$    36. \pm   42.$&$   -78. \pm    5.$&$  -114. \pm    7.$\\
   $C_{\bar Kr^{04}_{\pi\pi}}$ &$   1.78 \pm  0.79$&$   2.09 \pm  1.20$&$   4.67 \pm  0.30$&$   4.97 \pm  0.48$\\
$\Lambda_{\rm 3MF}$            &$  2563. \pm  722.$&  --- &  --- &  ---\\
\end{tabular}
\end{ruledtabular}
\end{table}

\end{document}